\documentclass[a4paper,fleqn,usenatbib]{mnras}

\usepackage{newtxtext,newtxmath}

\usepackage[T1]{fontenc}
\usepackage{ae,aecompl}

\usepackage{graphicx}	
\usepackage{amsmath}	
\usepackage{amssymb}	
\usepackage{threeparttable}

\title[The physical nature of 22 New VVV-GC candidates]{Analysis of the physical nature of 22 New VVV Survey Globular Cluster candidates in the Milky Way Bulge}

\author[T. Palma et al.]{
Tali Palma,$^{1,2}$\thanks{E-mail: tali@oac.unc.edu.ar}
Dante Minniti,$^{3,4,5}$
Javier Alonso-Garc\'ia,$^{6,4}$
Juliana Crestani,$^{7}$
\newauthor Henryka Netzel,$^{8}$
Juan J. Clari\'a,$^{1,2}$
Roberto K. Saito,$^{9}$
Bruno Dias,$^{3,10}$
\newauthor Jos\'e G. Fern\'andez-Trincado,$^{11,12,13}$
Roberto Kammers,$^{9}$
Douglas Geisler,$^{13,14,15}$
\newauthor Mat\'ias G\'omez,$^{3}$
Maren Hempel$^{3}$
and Joyce Pullen$^{3}$
\\ \\
$^{1}$Universidad Nacional de C\'ordoba, Observatorio Astron\'omico de C\'ordoba, Laprida 854, B. Observatorio, C\'ordoba, 5000, Argentina\\
$^{2}$Consejo Nacional de Investigaciones Cient\'ificas y T\'ecnicas (CONICET), Godoy Cruz 2290, Buenos Aires, Argentina\\
$^{3}$Departamento de F\'isica, Facultad de Ciencias Exactas, Universidad Andr\'es Bello, Av. Fern\'andez Concha 700, Las Condes, Santiago, Chile \\
$^{4}$Instituto Milenio de Astrof\'isica, Santiago, Chile \\
$^{5}$Vatican Observatory, V00120 Vatican City State, Italy \\
$^{6}$Centro de Astronom\'{i}a (CITEVA), Universidad de Antofagasta, Av. Angamos 601, Antofagasta, Chile \\
$^{7}$Departamento de Astronomia, Universidade Federal do Rio Grande do Sul, Porto Alegre, Brazil \\
$^{8}$Nicolaus Copernicus Astronomical Center, Warsaw, Poland \\
$^{9}$Departamento de F\'{i}sica, Universidade Federal de Santa Catarina, Trindade 88040-900, Florian\'opolis, SC, Brazil \\
$^{10}$European Southern Observatory, Alonso de Cordova 3107, Vitacura, Santiago, Chile \\
$^{11}$Instituto de Astronom\'ia y Ciencias Planetarias, Universidad de Atacama, Copayapu 485, Copiap\'o, Chile \\
$^{12}$Institut Utinam, CNRS UMR 6213, Universit\'e Bourgogne-Franche-Comt\'e, OSU THETA Franche-Comt\'e, Observatoire de Besan\c{c}on, \\ BP 1615, F-25010 Besan\c{c}on Cedex, France \\
$^{13}$Departamento de Astronom\'ia, Casilla 160-C, Universidad de Concepci\'on, Concepci\'on, Chile \\
$^{14}$Departamento de F\'isica y Astronom\'ia, Universidad de La Serena, Av. Ra\'ul Bitr\'an S/N, La Serena, Chile \\
$^{15}$Instituto de Investigaci\'on Multidisciplinario en Ciencia y Tecnolog\'ia, Universidad de La Serena, Av. Ra\'ul Bitr\'an S/N, La Serena, Chile \\
}

\date{Accepted XXX. Received YYY; in original form ZZZ}

\pubyear{2019}

\begin{document}
\label{firstpage}
\pagerange{\pageref{firstpage}--\pageref{lastpage}}
\maketitle

\begin{abstract}
In order to characterize 22 new globular cluster (GC) candidates in the Galactic bulge, we present their colour-magnitude diagrams (CMDs) and $K_s$-band luminosity functions (LFs) using the near-infrared VVV database as well as Gaia-DR2 proper motion dataset. CMDs were obtained, on one hand, after properly decontaminating the observed diagrams from background/foreground disc stars and other sources. On the other hand, CMDs were also obtained based upon star selection in proper motion diagrams. Taking into account our deep CMDs and LFs analyses, we find that 17 out of 22 new GC candidates may be real and should therefore be followed-up, while 5 candidates were discarded from the original sample. We also search for RR\,Lyrae and Mira variable stars in the fields of these new GC candidates. In particular, we confirm that Minni\,40 may be a real cluster. If confirmed by further follow-up analysis, it would be the closest GC to the Galactic centre in projected angular distance, located only $0.5$\,deg away from it.
We consider that it is very difficult to confirm the physical reality of these small, poorly-populated bulge GCs so in many cases alternative techniques are needed to corroborate our findings.
\end{abstract}




\section{Introduction}

Globular clusters (GCs) are among the oldest objects known in the Milky Way (MW). 
They are believed to be the survivors of a larger population of primordial Galactic star clusters, most  of which have been destroyed by different dynamical processes \citep{fall77,fall85}.  The most important destruction processes are dynamical friction, tidal disruption  by the interaction with the MW potential, i.e., strong gravitational shocks \citep[e.g.,][for instance]{minniti18a} and evaporation that depend on the cluster masses, positions, and orbital parameters \citep[e.g.,][]{Baumgardt03,lamers05,trenti10,rossi15, wang16}. This is important to investigate in order to compare with other galaxies like the Magellanic clouds where large numbers of clusters can be studied. In particular, dynamical friction and tidal disruption are more severe in the inner regions of the MW. The GCs were presumably dragged and destroyed preferentially in the inner regions of the Galactic bulge, deep in the potential well, where the stellar density is very high. The present distribution of GCs is fairly concentrated towards the Galactic centre, following a steep spatial density law $ \propto r^{-3.5}$  \citep[e.g.,][]{zinn85,minniti95}.  Recent observational studies indicate that a few field stars in the inner Galaxy share chemical anomalies patterns similar to those found in GC stars \citep{recio17,trincado17,schiavon17,trincado19a,trincado19b,trincado19c}, suggestive that these are ancient fossil remnants of destroyed stellar clusters.
This is why we have labelled the Galactic bulge the ``elephant graveyard'' \citep{minniti17a}, and are now conducting a comprehensive search for the remnants of these GCs.
Such debris may be expected to be present as short streams, or as the remaining small cores of larger objects \citep{minniti18a}.
However, finding new GCs is very tricky in these regions due to the effects of reddening, both absolute and differential, and to high stellar density \citep{valenti07,alonsogarcia12}. The best way to recognize them is to use near-IR (NIR) observations because the bright red giants peak their emission and the interstellar clouds become more transparent in this spectral region. Another way to spot them is through high resolution NIR spectroscopy surveys like APOGEE \citep{majewski17}.  \\

The attempts to identify hidden Galactic GCs have lately been succesful, i.e. 37 new GCs were discovered only in the past couple of years, namely: Sagittarius\,II \citep{laevens15,mutlu18}, Kim\,3 \citep{kim16}, FSR\,1716 = VVV-GC05 (rediscovered by \citealt{minniti17b,contreras18}), Minni\,01 to 22 \citep{minniti17c}, DES-1 \citep{luque16}, 
Gaia\,2 \citep{koposov17}, RLGC\,1 and RLGC\,2 \citep{ryu18},
Camargo\,1102 \citep{bica18,camargo18}, Camargo\,1103 to  1109 \citep{camargo18,camargo19}, and FSR\,1758 \citep{barba19}.
While some of them are located in the MW halo, most of the new ones (33 out of 37) are located towards the Galactic bulge. 
Indeed, taking into account the newly confirmed GCs and candidates, the number of known GCs in the Galactic bulge has almost doubled in the last few years. The discovery of these new GCs represents an advance in this field of research, since it is important for future studies on the age and chemical composition of the oldest stars, the formation and evolution of the MW, the dynamics of stellar systems, the distance scale, the interstellar medium, the stellar evolution, and the Galactic structure, among other issues. \\

\begin{figure*}
\includegraphics[width=0.9\textwidth]{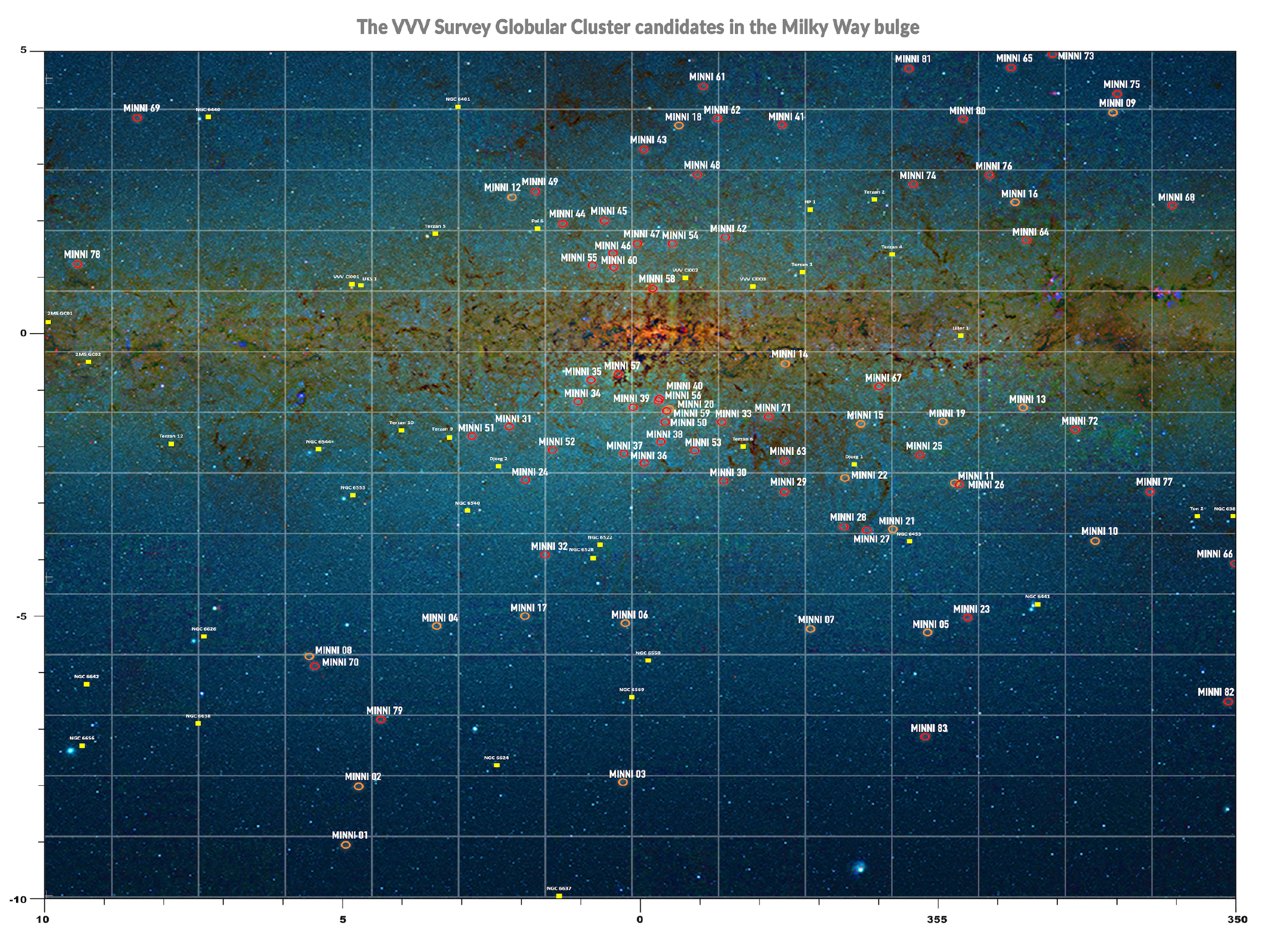}
\caption{VVV NIR image of the MW bulge (https://www.eso.org/public/news/eso1242) and the currently known population of GCs. Yellow squares represent the catalogued GCs, orange circles the 22 recently discovered candidates \citep{minniti17c}, and red circles the new GC candidates discovered and among those the selected sample studied in this work.
\label{fig1}}
\end{figure*}

Based on a search for overdensities in the red giant (RG) stars selected using Wesenheit CMDs \citep{minniti17c}, we recently reported the discovery of 22 new GC candidates in the MW. Later on we extended the search for more GC candidates using different GC tracers, such as the concentration of RR\,Lyrae and Type 2 Cepheid variable stars in the bulge fields \citep{minniti17a}. 
The clustering of horizontal branch (HB) stars traces GCs of all metallicities since metal-rich GCs will predominantly have red clump (RC) stars and the metal-poor ones will predominantly have a blue HB without a RC, while the concentration of RR\,Lyrae stars should indicate mostly metal-poor GCs. 
However, because the bulge fields are very crowded and reddened regions, it is important to secure the confirmation of bonafide GCs using deep CMDs, for example. The NIR VVV Survey PSF photometry is now publicly available \citep{alonsogarcia18}, allowing us to make a deep photometric study of many of the new bulge GC candidates. Here we present the deep NIR CMDs of 22 newly published candidates, together with their CMDs based on Gaia-DR2 proper motion (PM) studies, with the aim to confirm the best GC candidates and discard the bad ones. In Figure \ref{fig1} we superimposed, on the MW image, the known catalogued GCs (yellow squares), the 22 recently discovered candidates \citep{minniti17c} in orange circles, and the 22 new GC candidates (red circles) studied in this work among the total sample of newly discovered candidates. \\

\section{Near-IR CMDs of Globular Cluster Candidates}

In order to find the missing GCs, one needs to explore the complex region of the Galactic bulge, but the high stellar density and extinction of these regions make the search for new GCs confusing. The  VISTA Infrared Camera (VIRCAM) at the 4\,m VISTA telescope at the ESO Cerro Paranal Observatory \citep{emerson10} is the ideal instrument for such a search, because it is optimized for wide field NIR observations.
The VVV Survey has been mapping the MW bulge using this instrument since the year 2010 \citep{minniti10,saito12}.
The extended VVVX Survey is expected to last until approximately the year 2020, completing a total of about 400 observing nights.
We use observations from the VVV Survey reduced at the Cambridge Astronomical Survey Unit (CASU) with the VIRCAM pipeline v1.3. 
The PSF photometry used to build the deep NIR CMDs was performed with DoPhot \citep{schechter93}, following \cite{alonsogarcia15,alonsogarcia18}. \\

Given the photometric depth of the VVV images (the limiting magnitude in the specified tiles spans up to $K_s = 18$ mag.) and their spatial resolution ($0.34$ arcsec/pixel), the VVV imaging are prime data tools to search for stellar overdensities even at a distance as far as 16 kpc, hence including regions across the whole MW bulge. However, detecting new GCs in these regions is still complicated.
Indeed, the observed GC candidate overdensities are only a few sigma at best above the field star density. The searches for new GCs were made using two complementary techniques:
1. Seeking overdensities of RC stars that are numerous in metal-rich GCs. 
The first VVV GC, located in a very dense and reddened region of the bulge \citep{minniti11}, was discovered using this technique.
2. Searching for tight groups of RR\,Lyrae variable stars, which are characteristic of metal-poor GCs. The first VVV GC discovered using this technique is VVV-GC-005 \citep{minniti17b}, also known as FSR\,1716 \citep{froebrich07}, located in a very dense and reddened region of the MW plane \citep{minniti17b,contreras18}. \\

\begin{figure*}
\centering
\includegraphics[scale=.2]{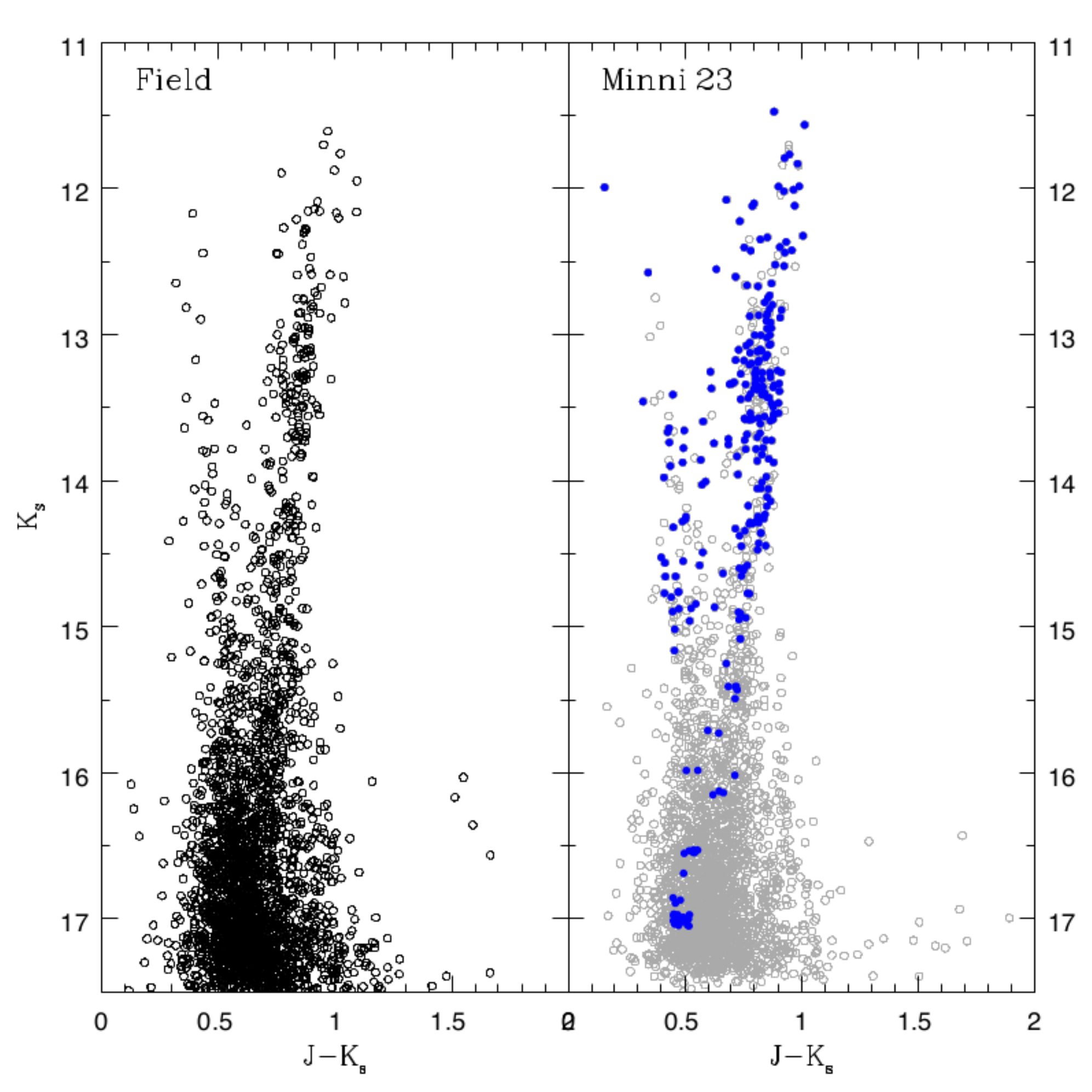}
\includegraphics[scale=.2]{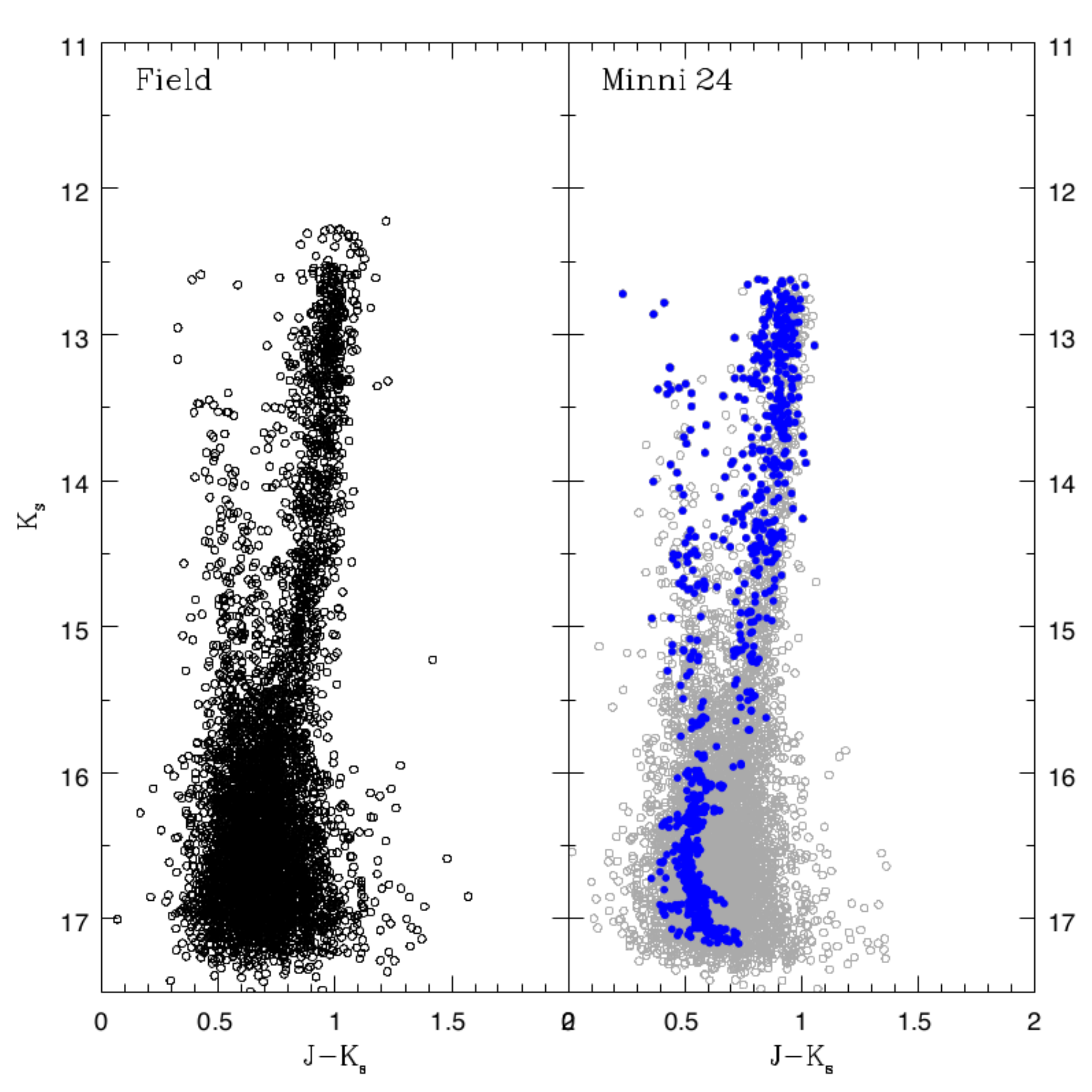}
\includegraphics[scale=.2]{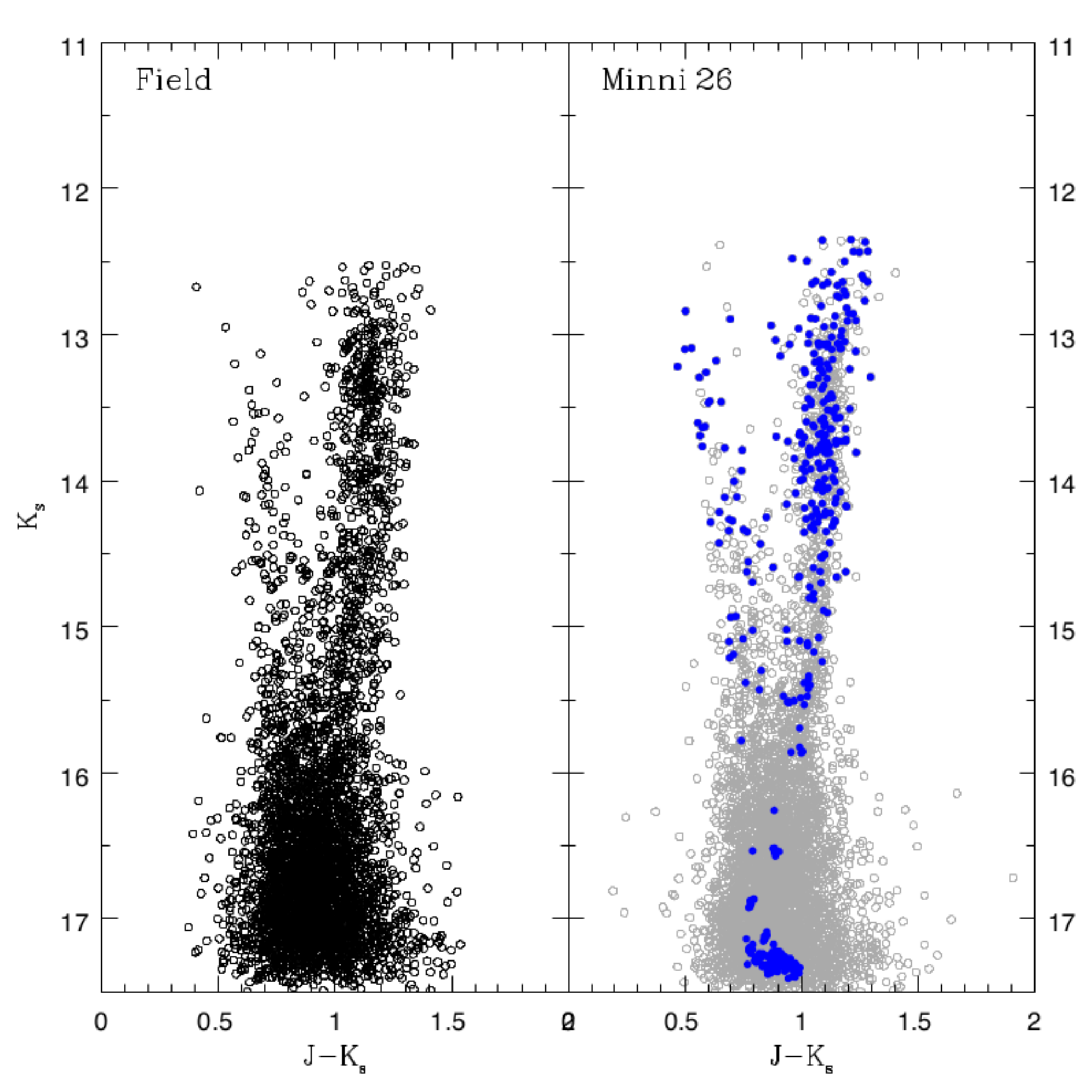}
\includegraphics[scale=.2]{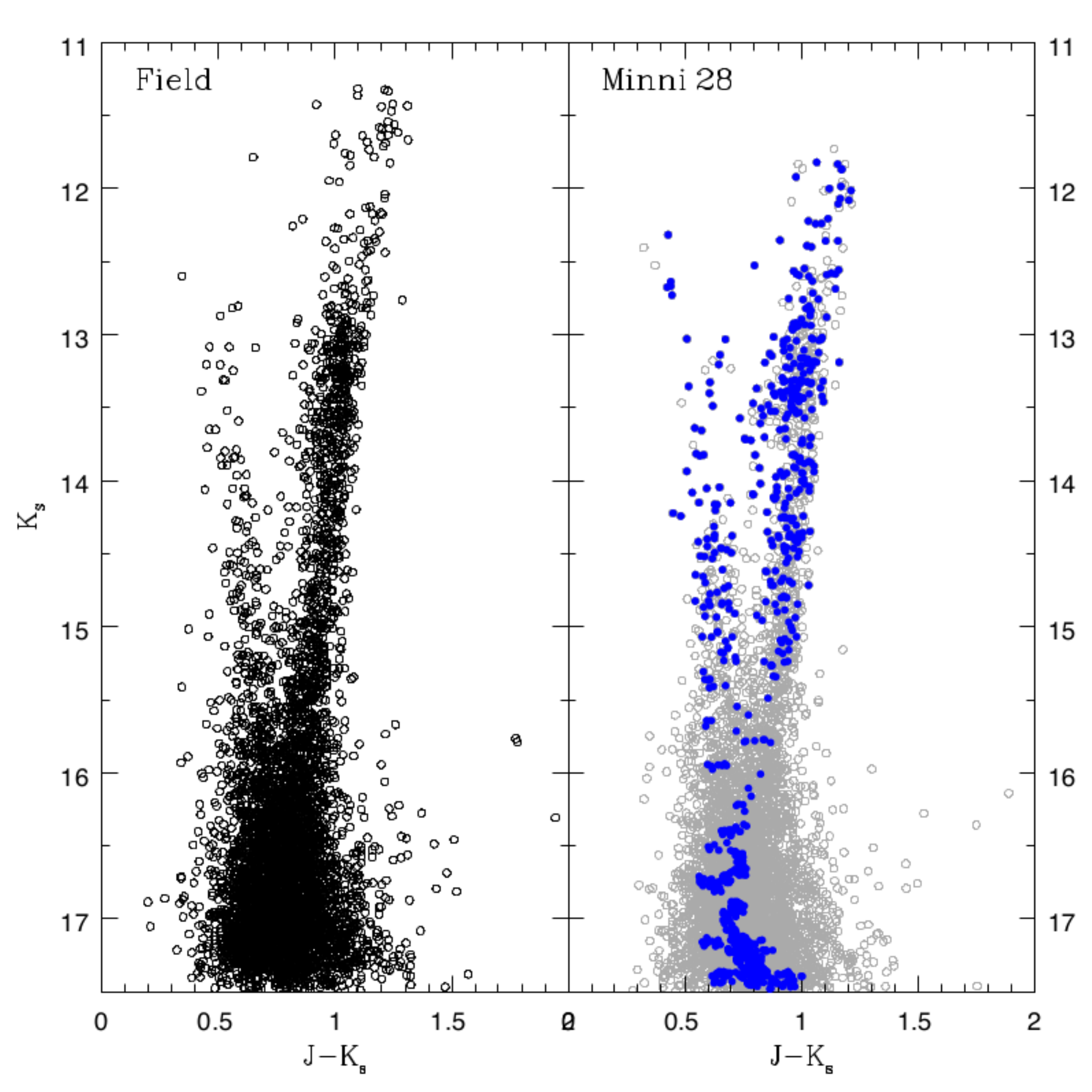}
\includegraphics[scale=.2]{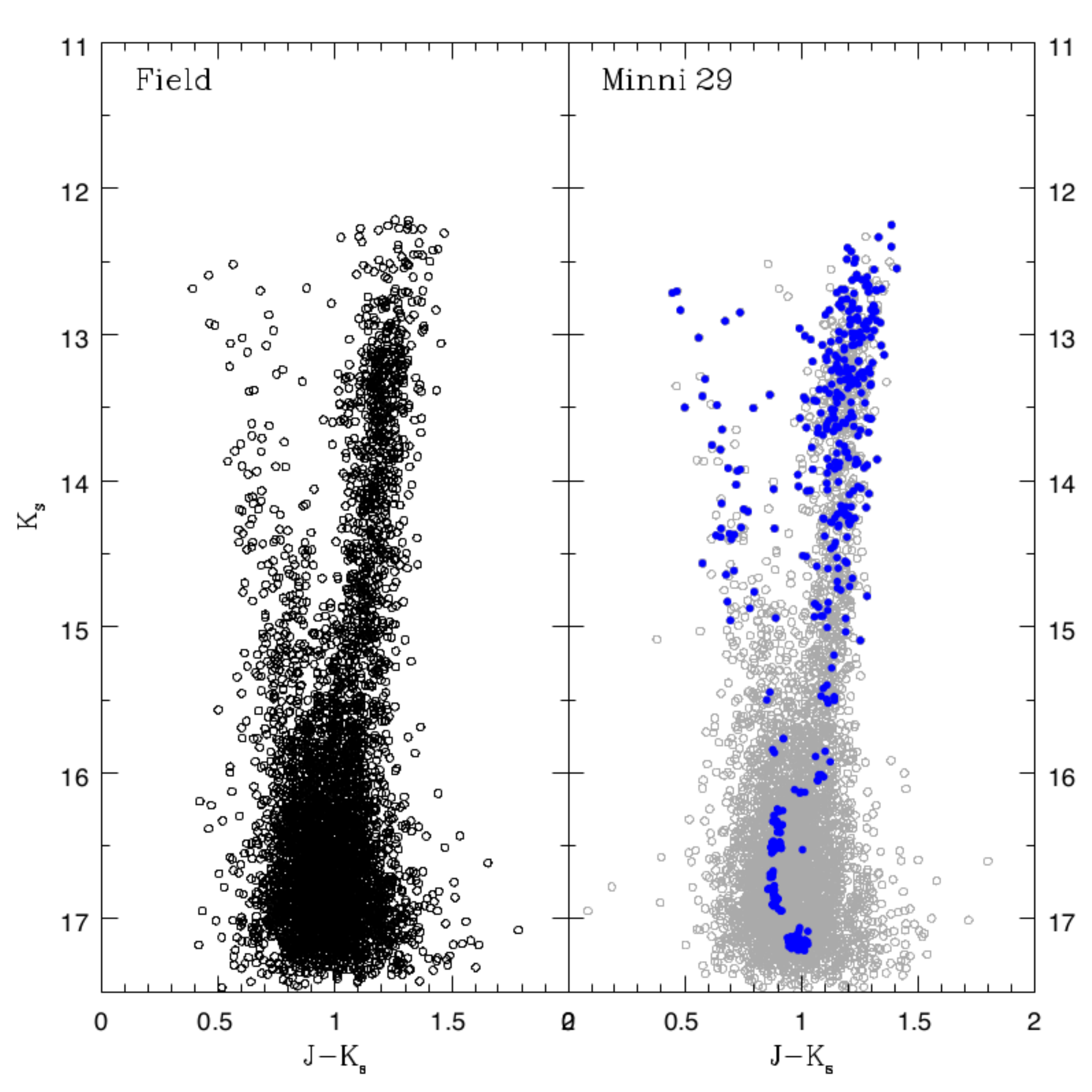}
\includegraphics[scale=.2]{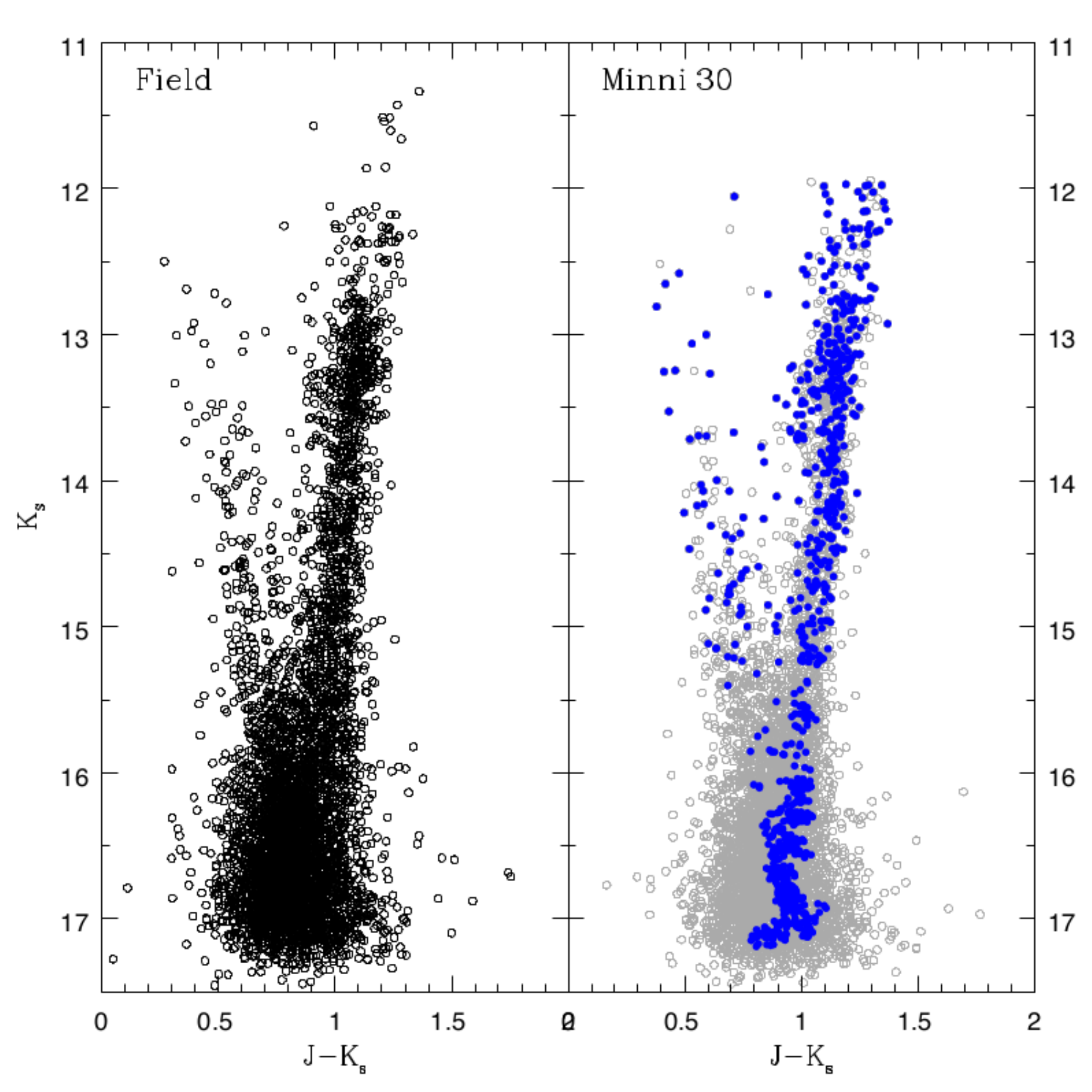}
\includegraphics[scale=.2]{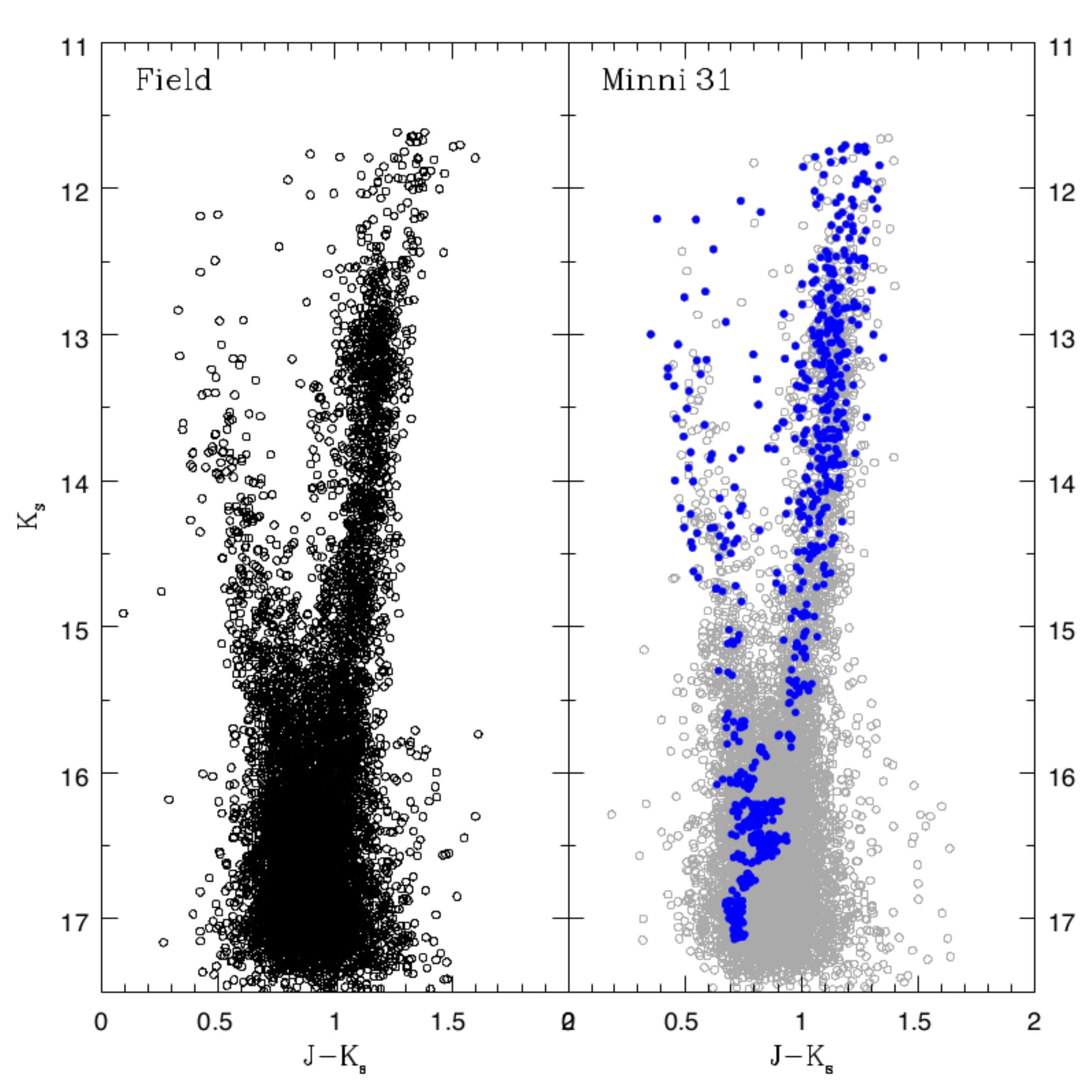}
\includegraphics[scale=.2]{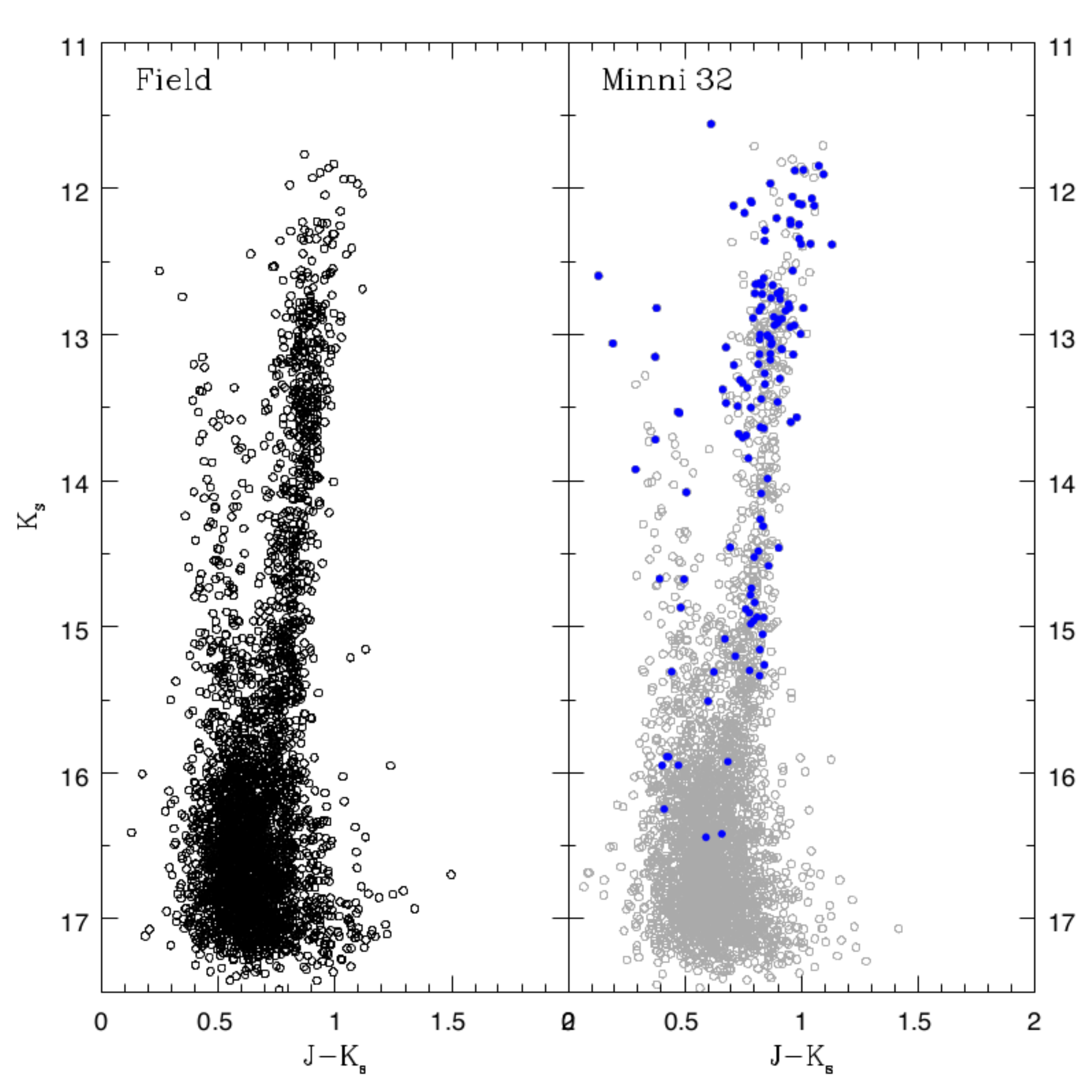}
\includegraphics[scale=.2]{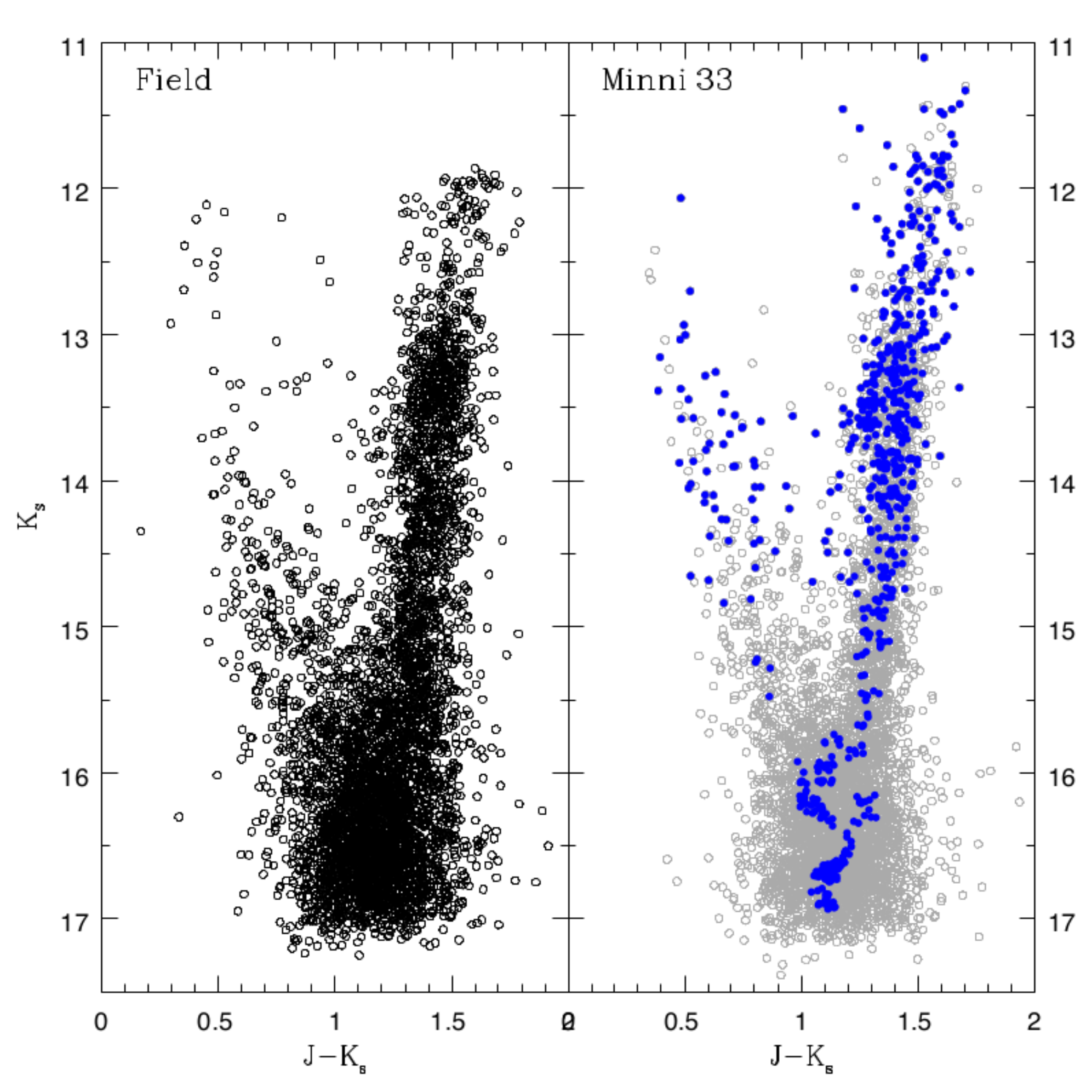}
\includegraphics[scale=.2]{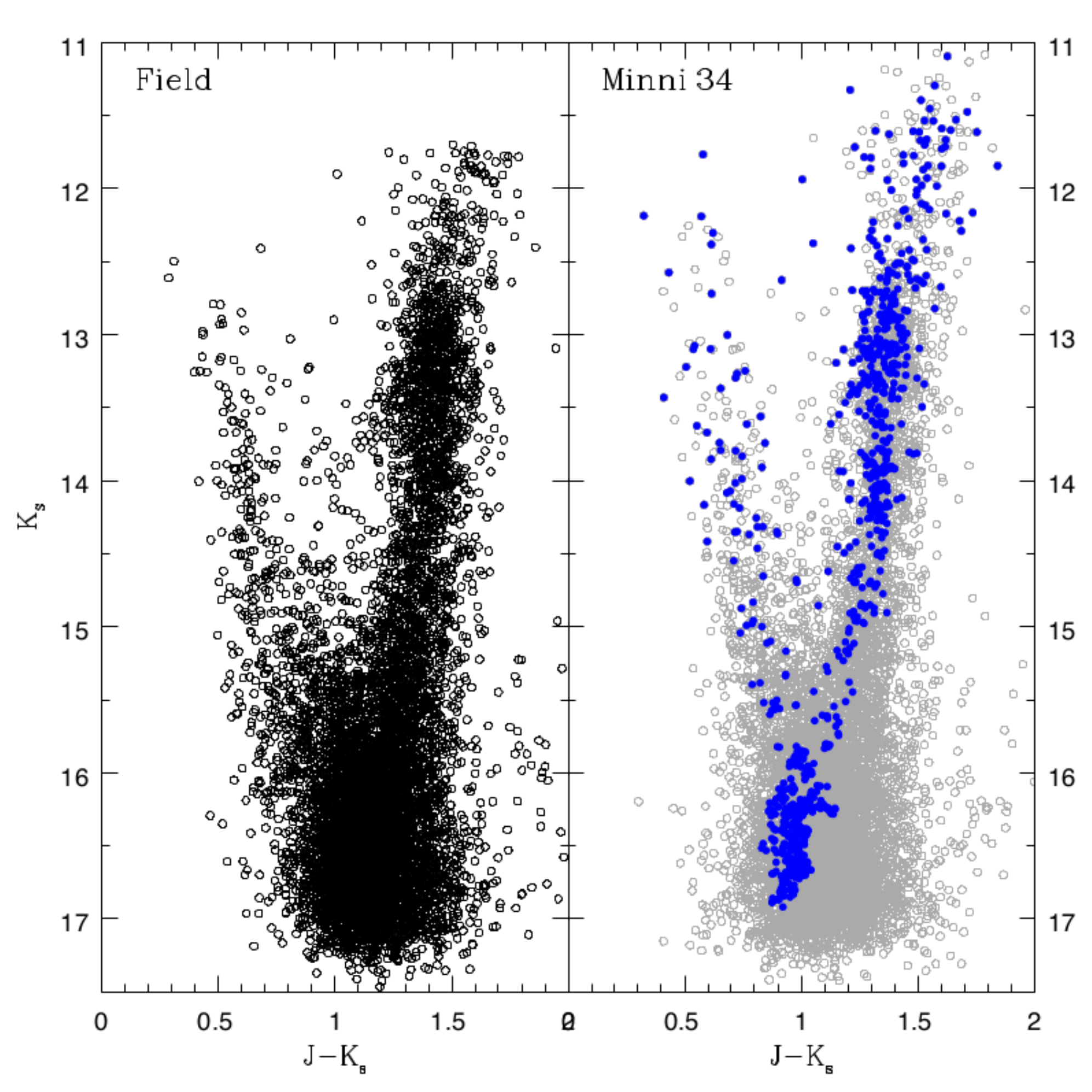}
\includegraphics[scale=.2]{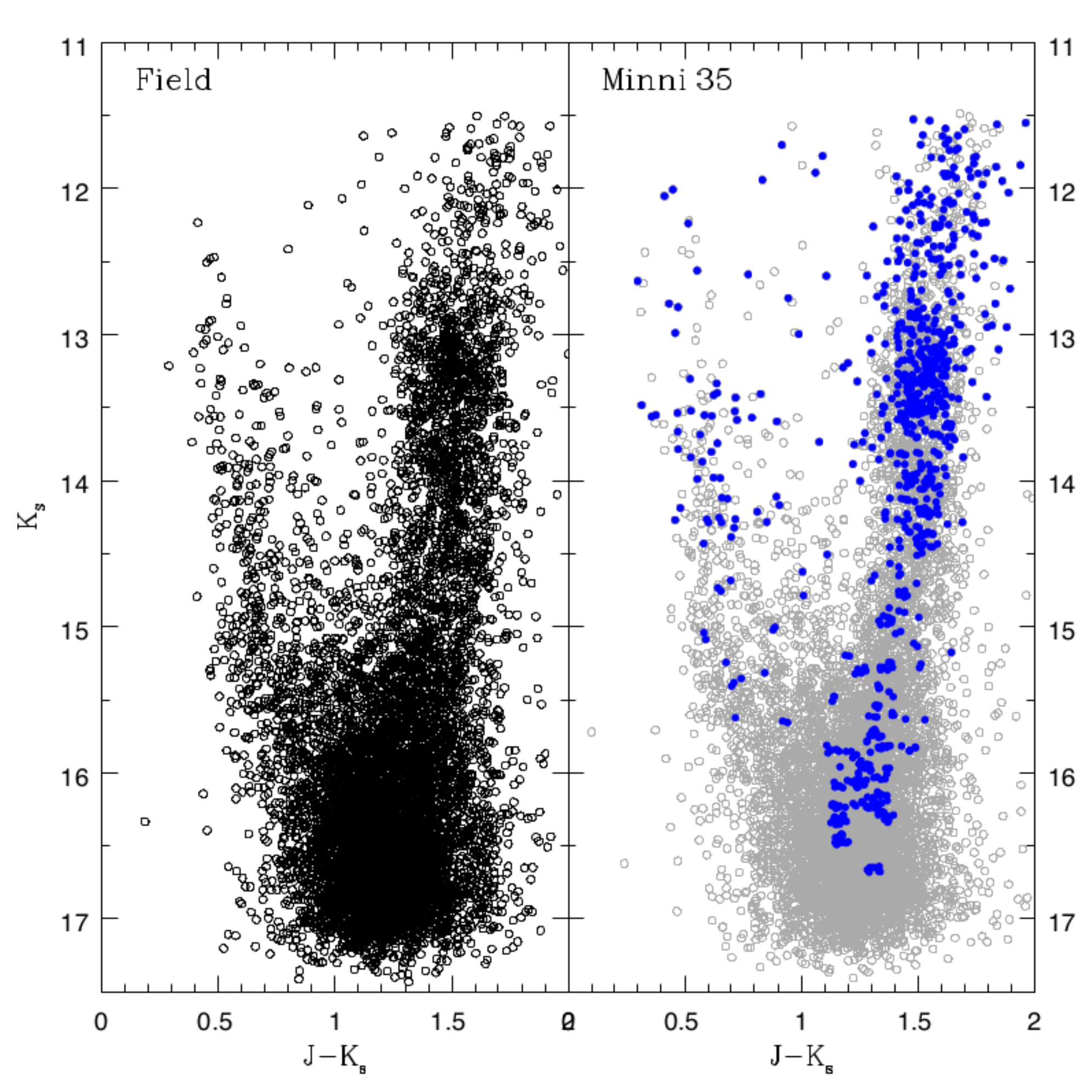}
\includegraphics[scale=.2]{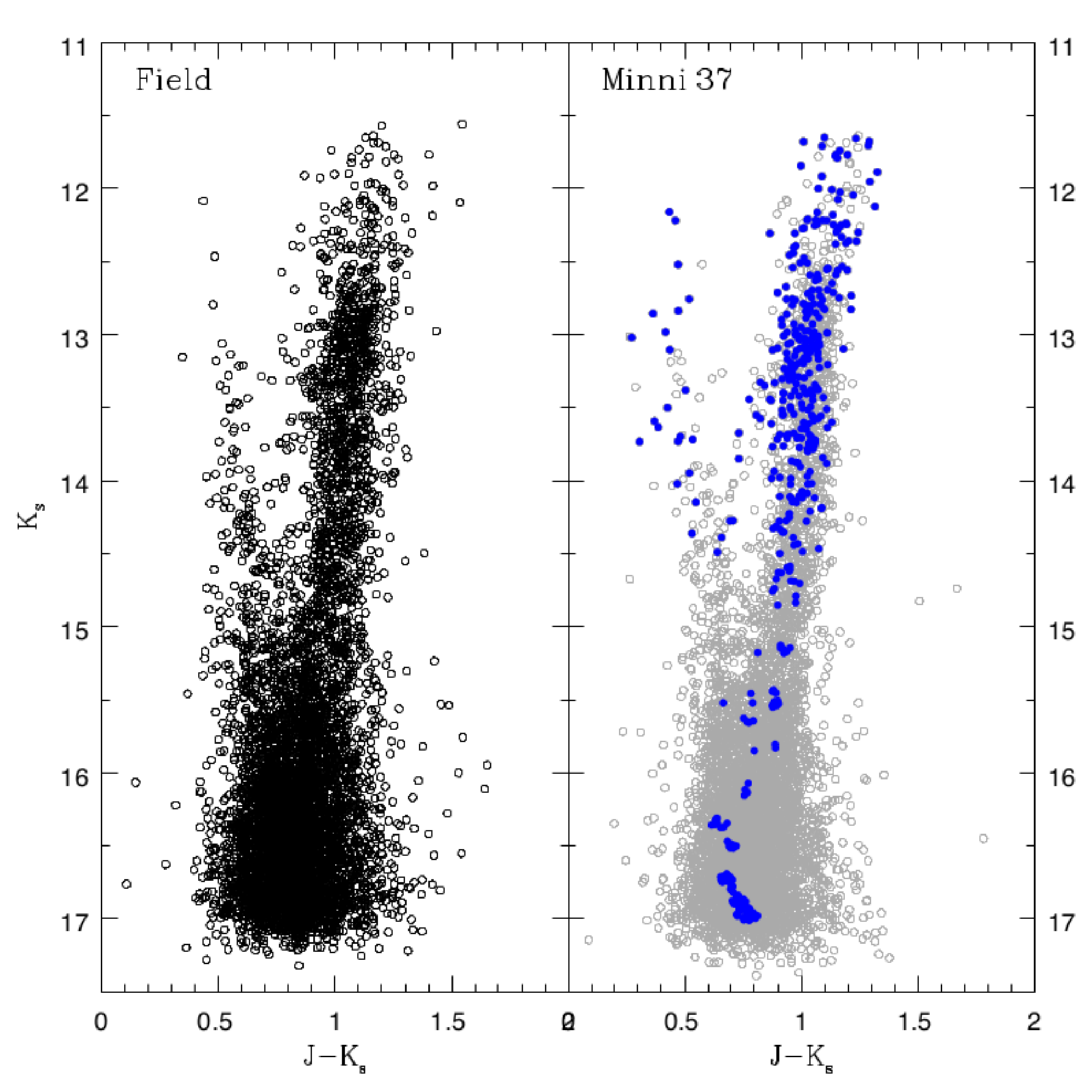}
\includegraphics[scale=.2]{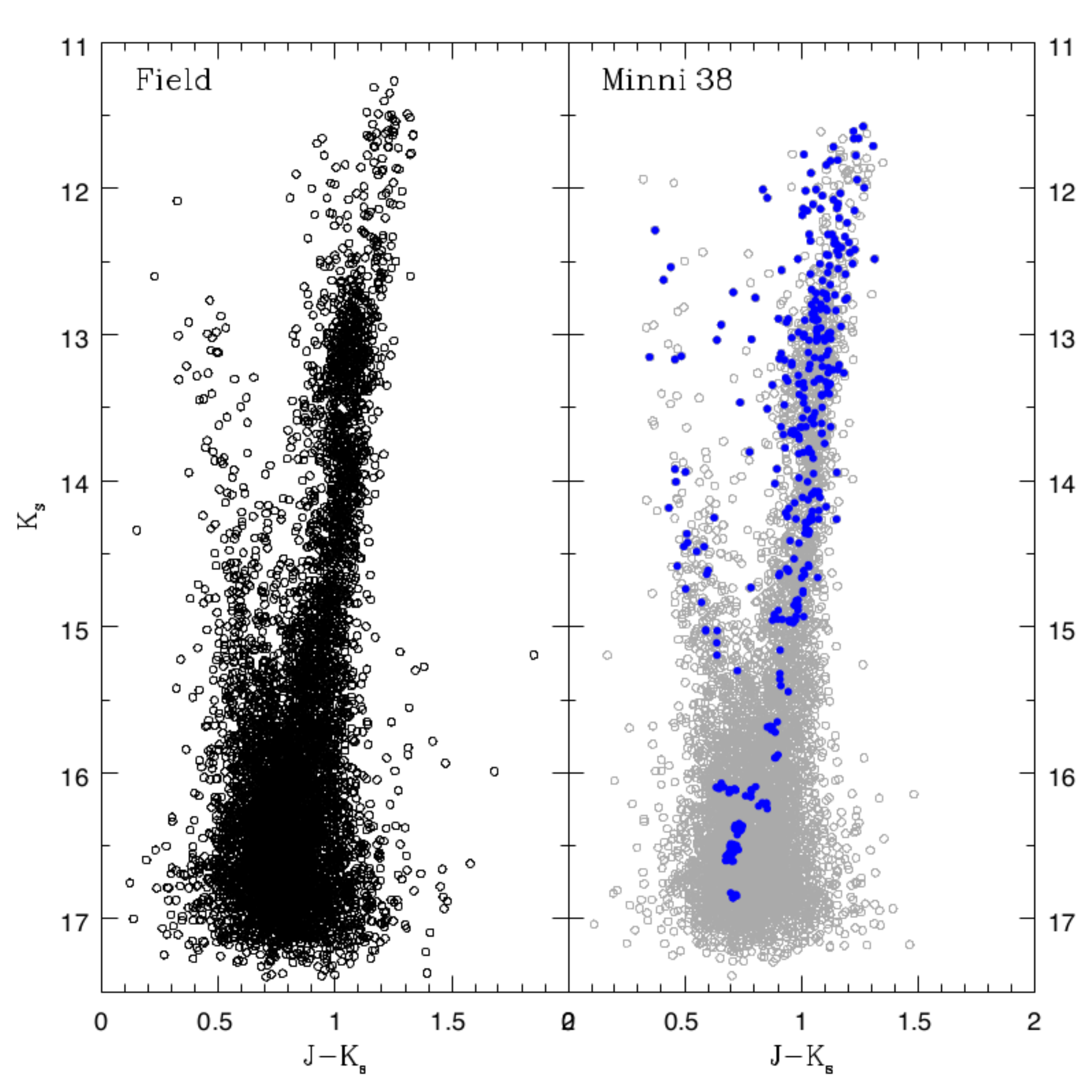}
\includegraphics[scale=.2]{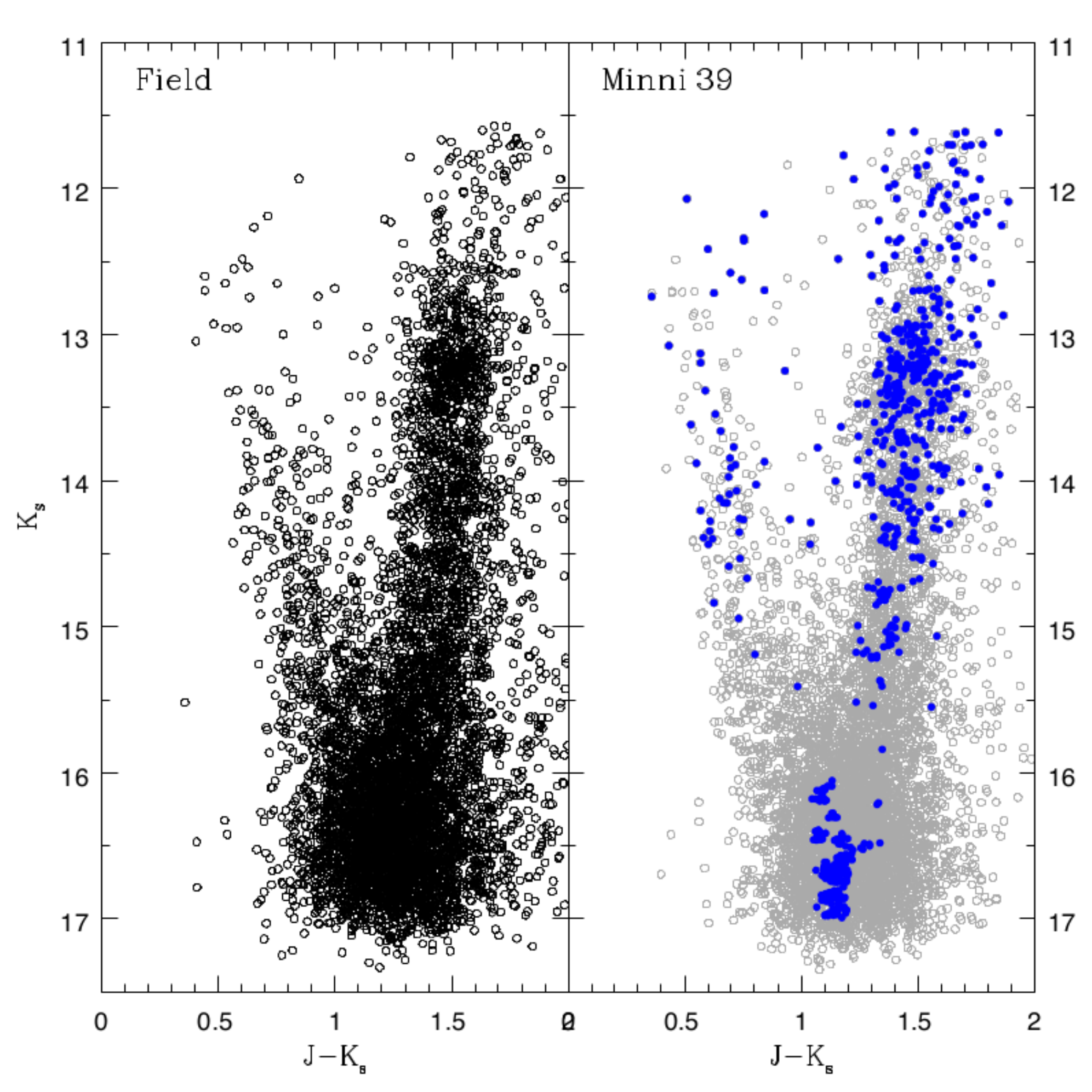}
\includegraphics[scale=.2]{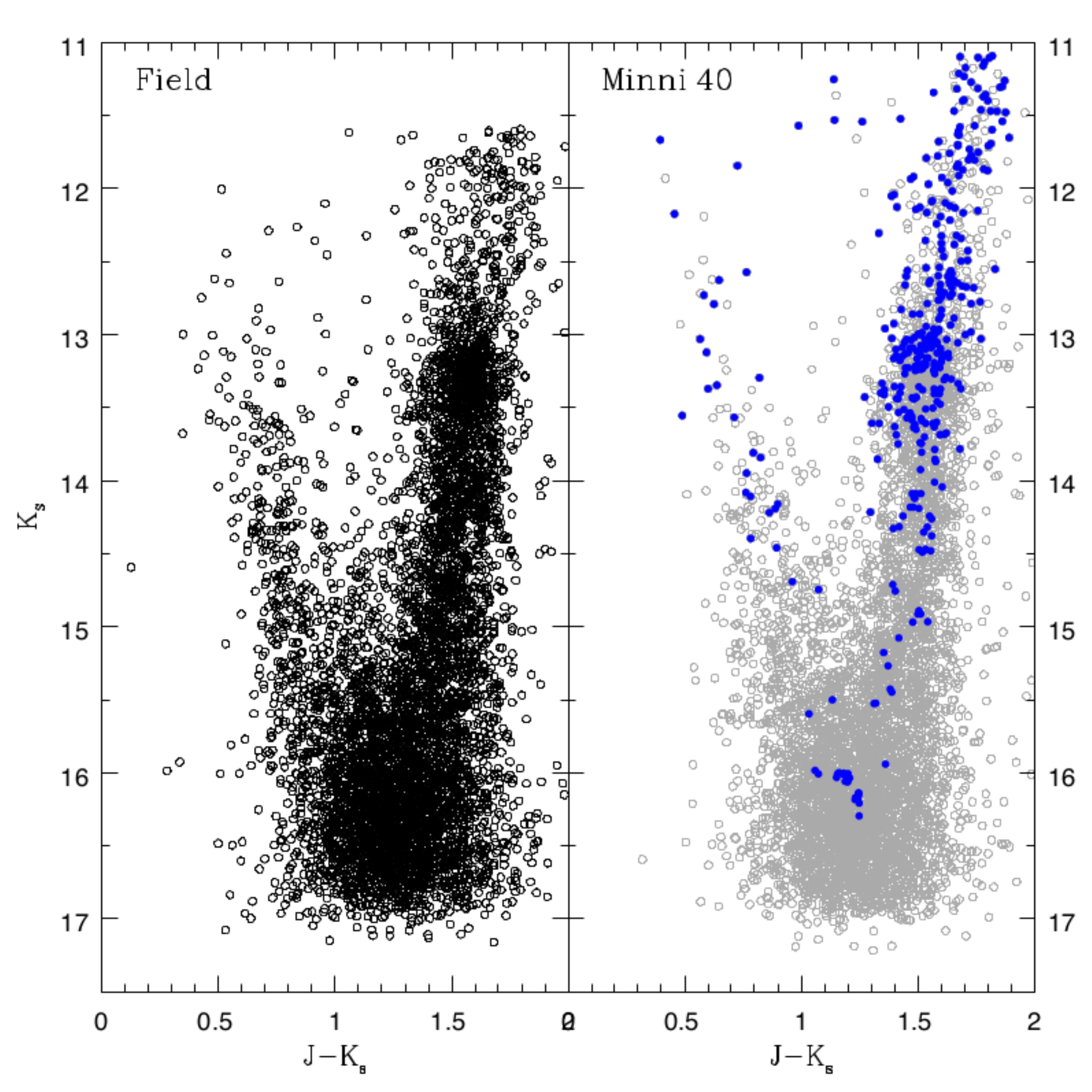}
\includegraphics[scale=.2]{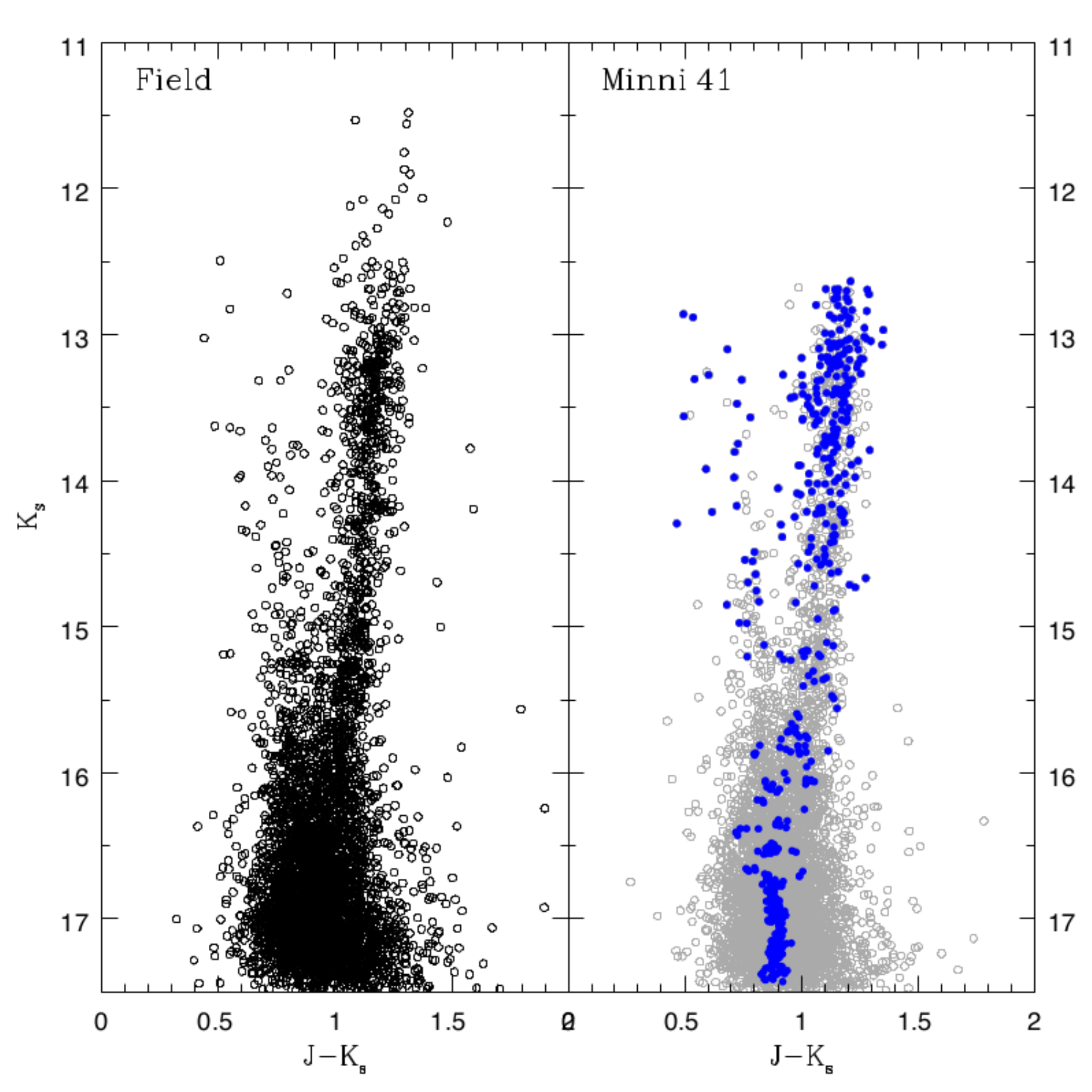}
\includegraphics[scale=.2]{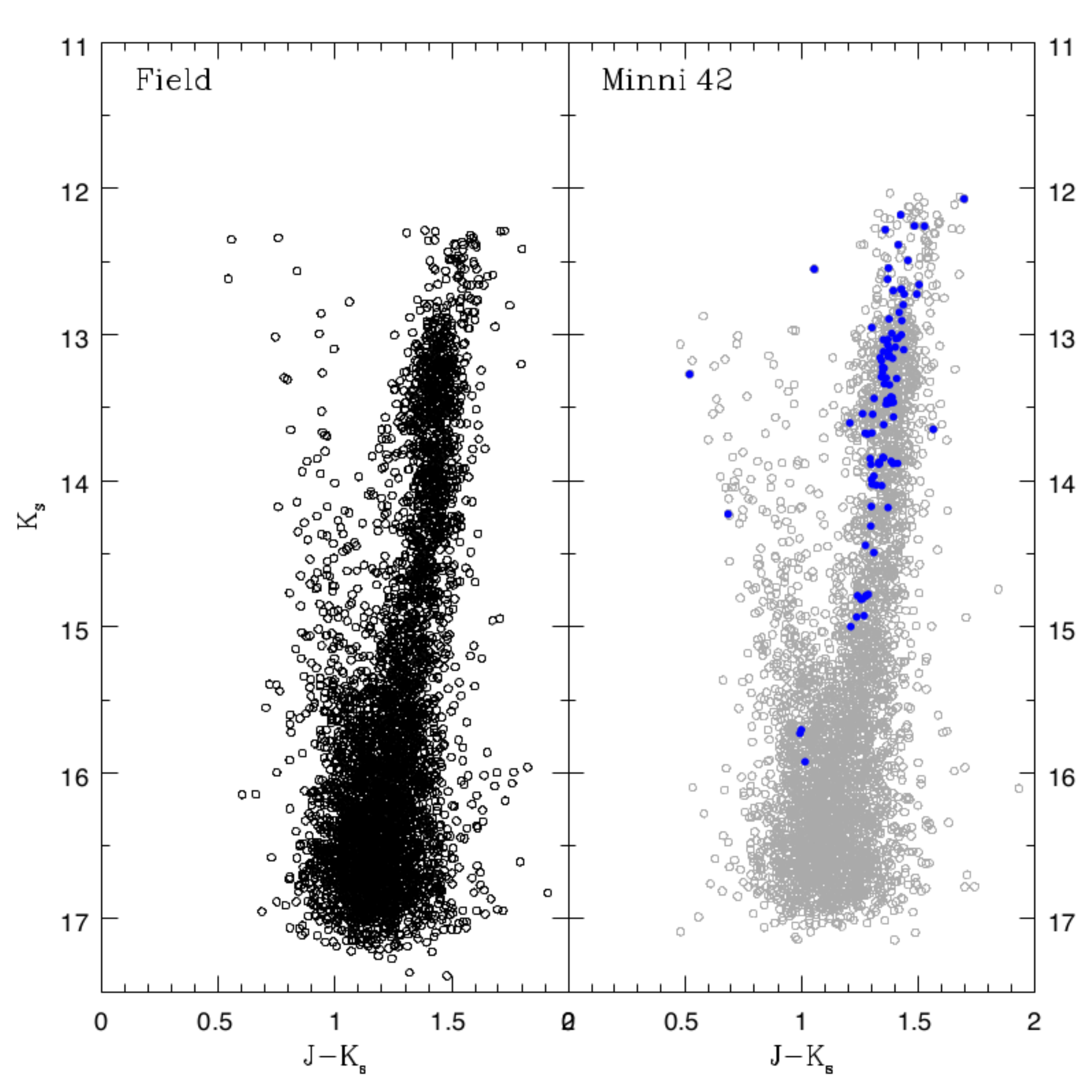}
\caption{CMDs of the regions within 2 arcmin for the seventeen candidates deemed to be possible real GCs. Left panels show the GC candidate fields. In the right panels, the background comparison field CMDs are shown in grey circles while CMDs decontaminated after applying a statistical procedure are shown in blue filled circles. 
\label{fig2}  }
\end{figure*}

\begin{figure*}
\centering
\includegraphics[scale=.2]{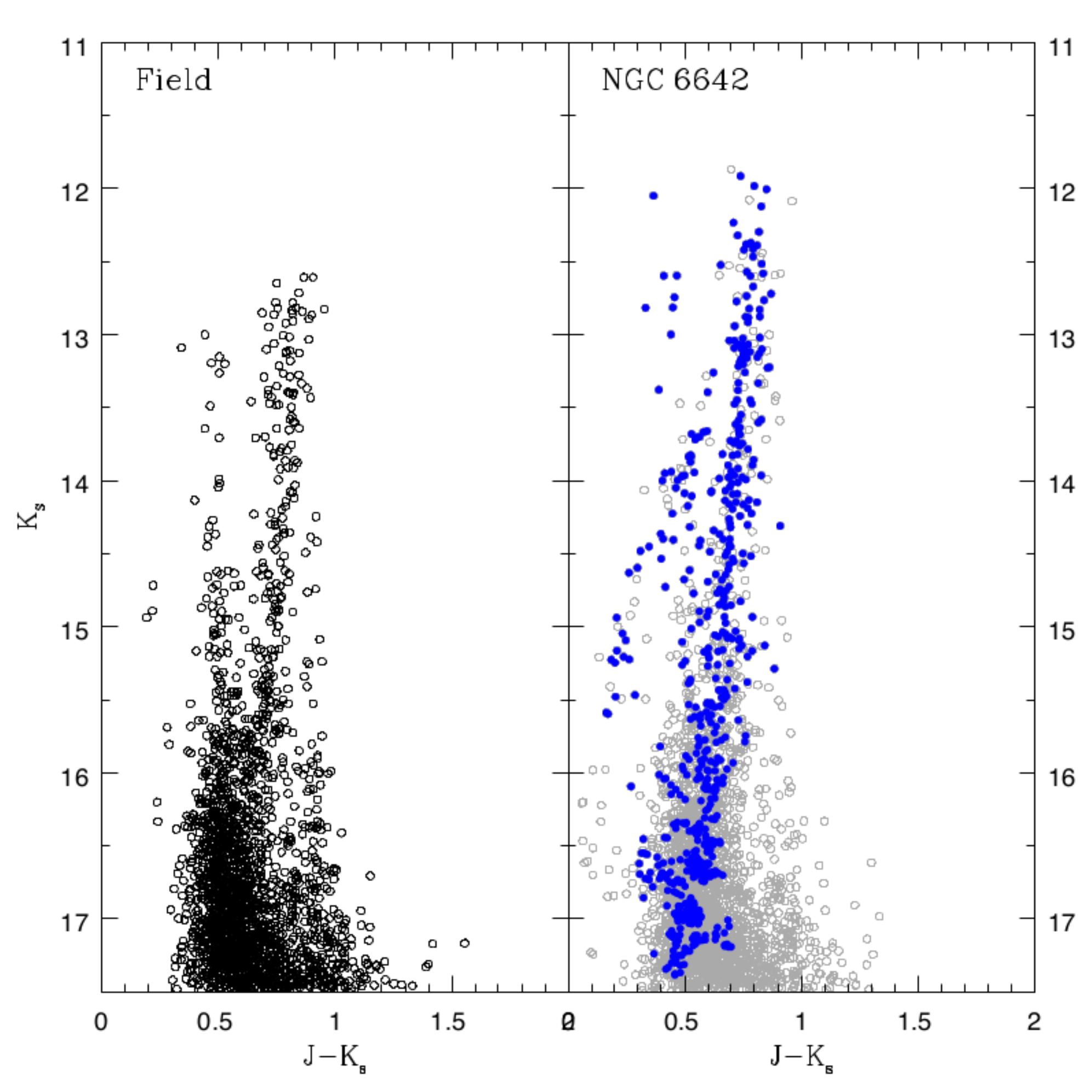}
\includegraphics[scale=.2]{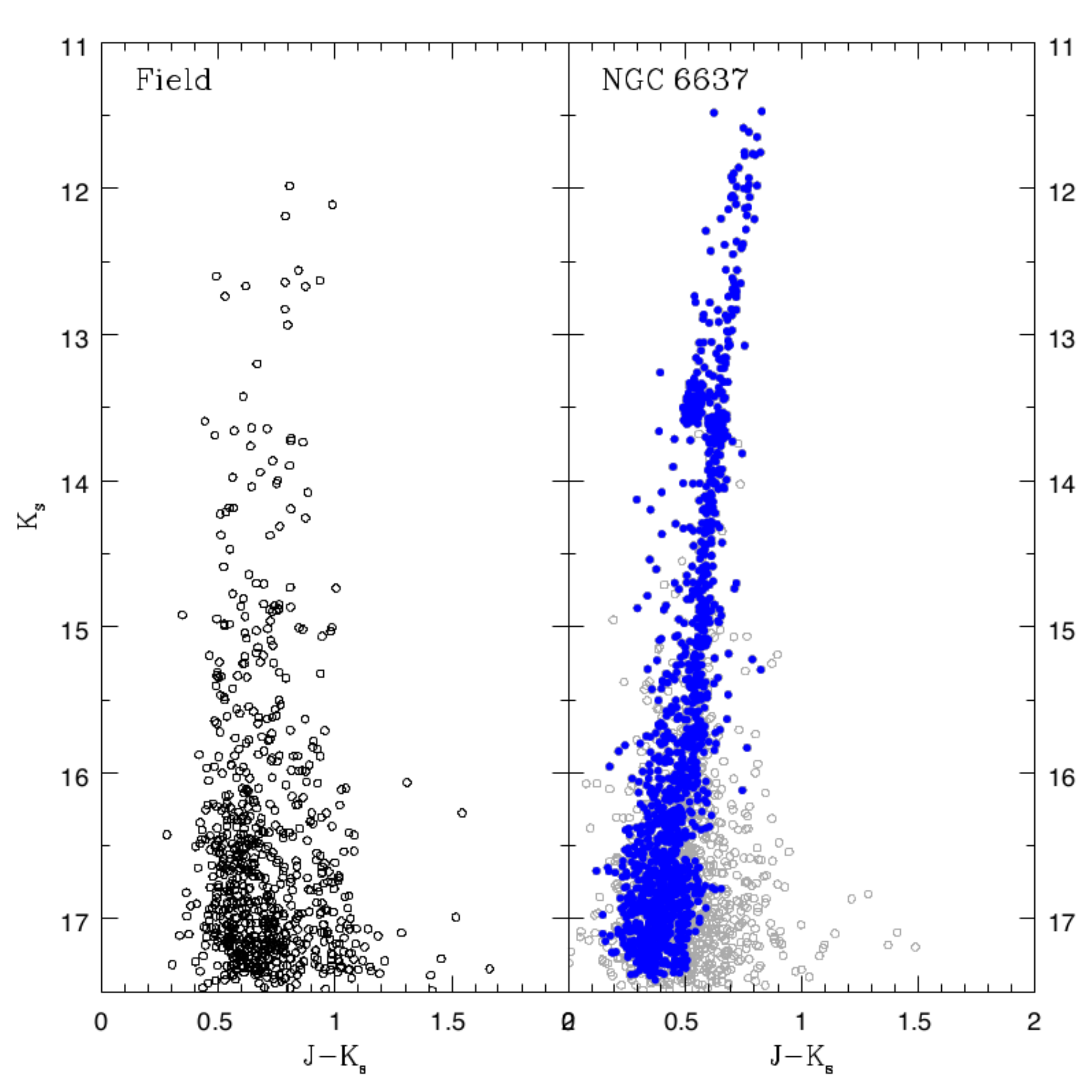}
\includegraphics[scale=.2]{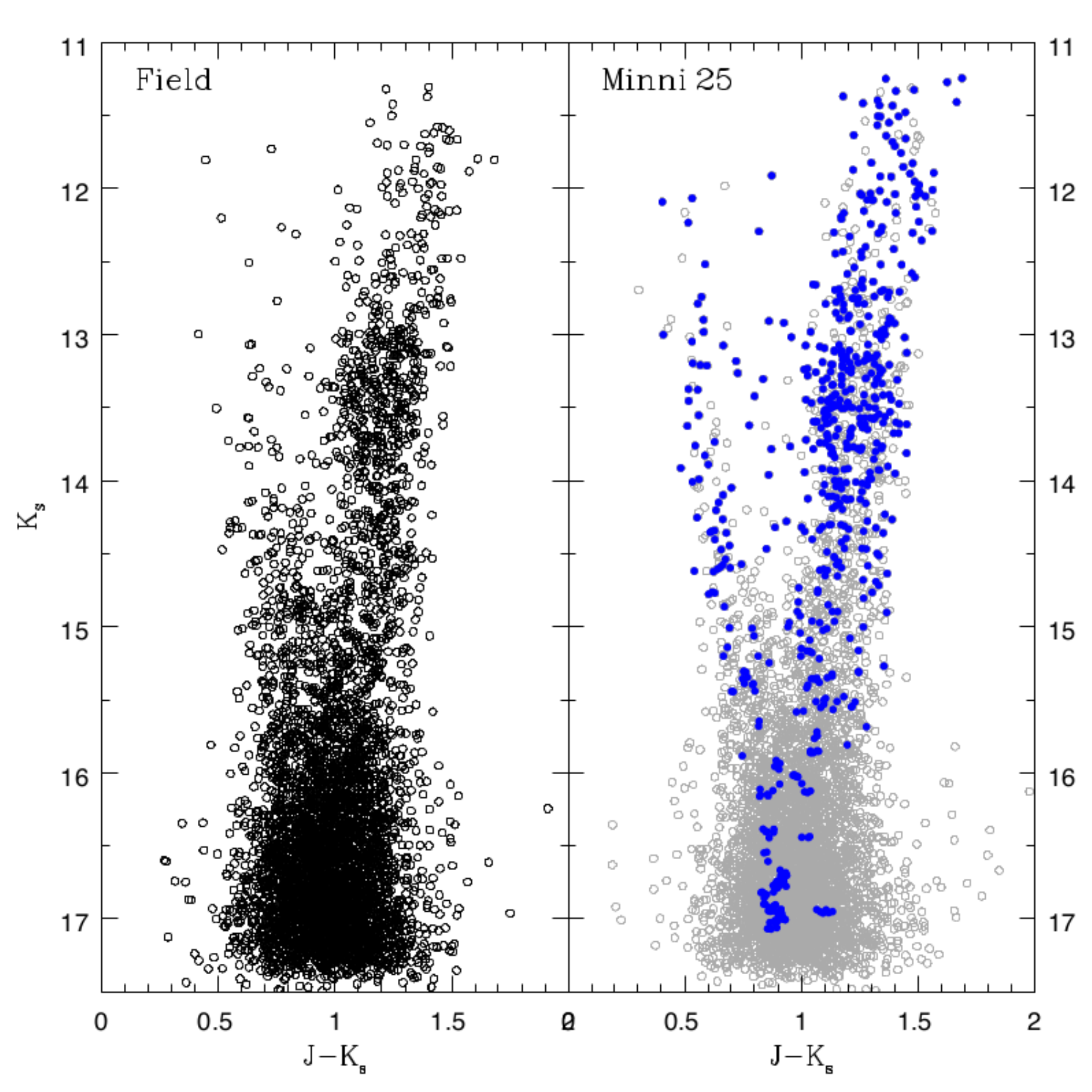}
\includegraphics[scale=.2]{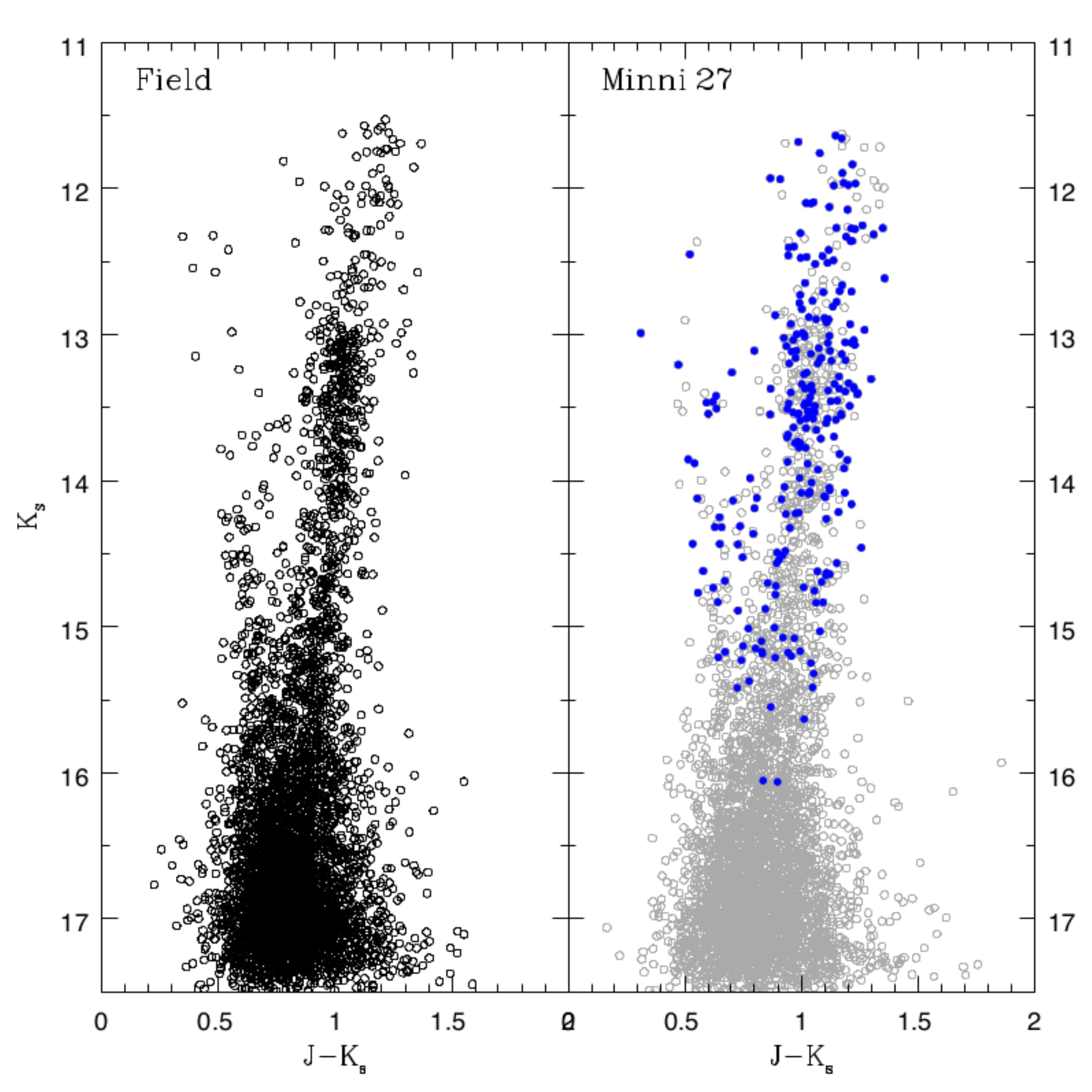}
\includegraphics[scale=.2]{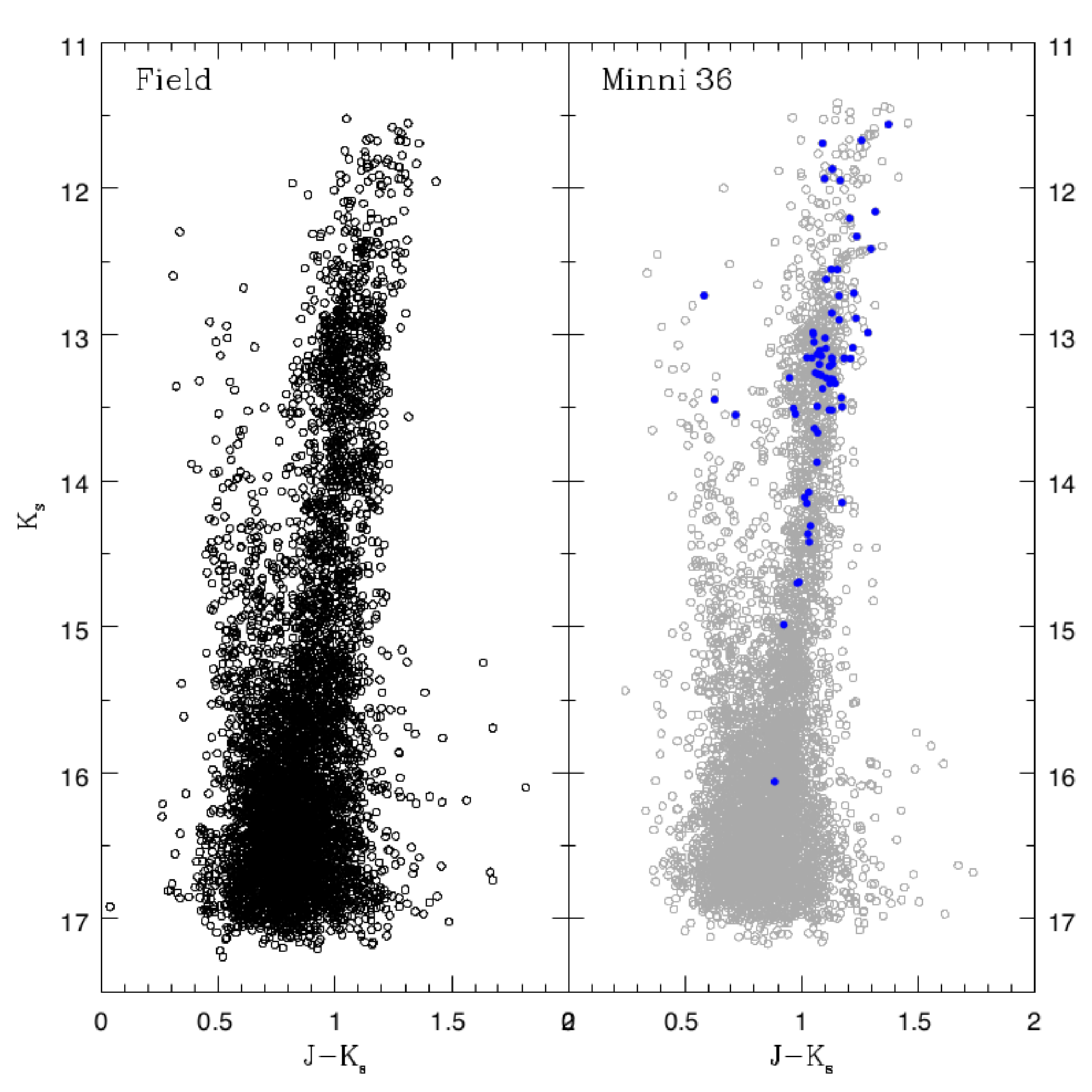}
\includegraphics[scale=.2]{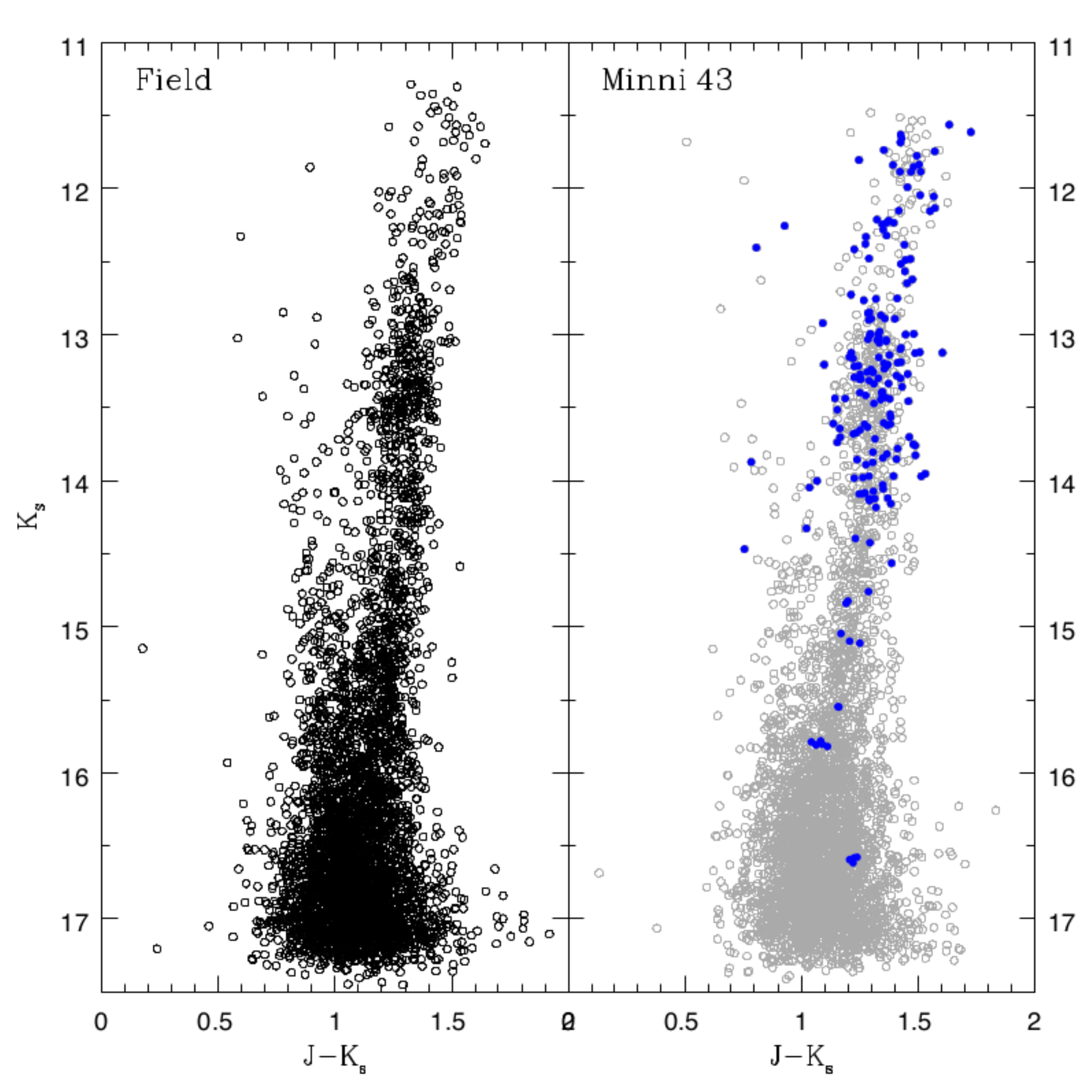}
\includegraphics[scale=.2]{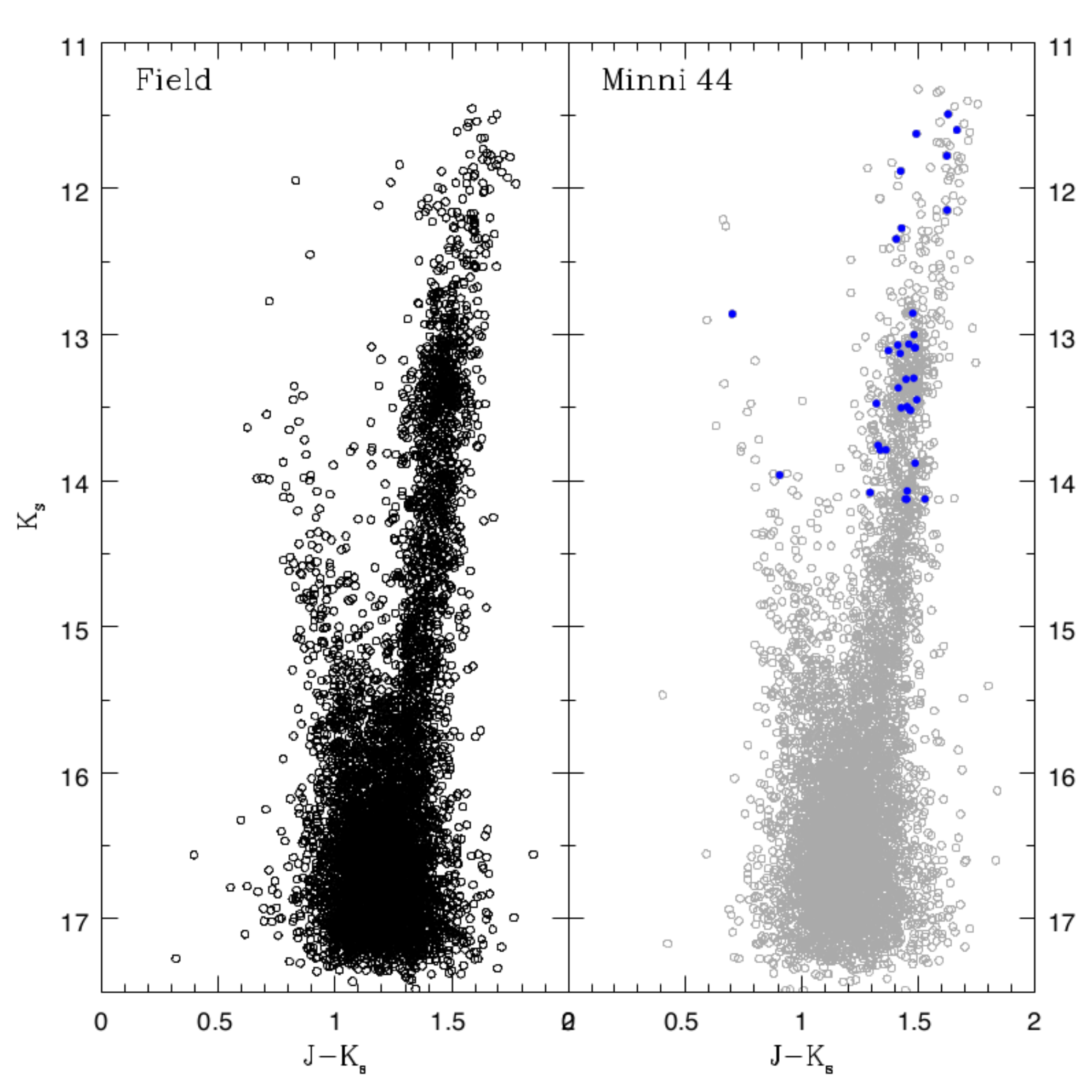}
\caption{CMDs of the regions within 2 arcmin of the GC candidate. The reference clusters NGC\,6642 and NGC\,6637 are shown at the left of the top panels. The remaining CMDs show the candidates deemed not to be real GCs.
\label{fig3}  }
\end{figure*}

Twenty two new GC candidates discovered using NIR VVV Survey photometry are presented in Table \ref{tab1}. Using catalogues from OGLE and VVV, those candidates were recognized by searching for RR Lyrae overdensities and for type 2 Cepheids to trace the unknown metal-poor GCs in the bulge \citep{minniti17a}.  This table lists in succession the GC identification, the equatorial coordinates (J2000), the tile ID, the numbers of RR\,Lyrae and Mira stars within 5 arcmin of each cluster centre, the corresponding absorption in the $K_s$ band obtained with the BEAM calculator which is based on the VVV data \citep{gonzalez12} and has a typical error of $\sigma _{A_{K_s}} \sim 0.10$ mag, our outcome analysis, the GC used as comparison, and the estimated heliocentric distances in kpc. The CMDs of the GC candidate fields were centred on the coordinates given by \cite{minniti17a}. Those CMDs were decontaminated by a statistical procedure \citep[see][]{palma16,minniti17b,minniti18b} in order to discriminate the background/foreground fields from the observed cluster CMDs (Figures \ref{fig2}-\ref{fig3}, black circles on the left panels). Briefly, the procedure consists in selecting about 4 to 6 circular background fields per cluster, of the same area and about 15 arcmin away, on average from its centre (Figures \ref{fig2}-\ref{fig3}, grey circles on the right panels), and having similar reddenings and stellar densities. The stars that fall in the same intervals of colour and magnitude of the background CMDs are subtracted from the cluster CMD. The resulting decontaminated CMDs are shown in the right panels of Figures \ref{fig2}-\ref{fig3} using blue filled circles.  A couple of known, bona fide, well-populated bulge GCs (NGC\,6642 and NGC\,6637) are also shown for comparison purposes in the top panels of Figure \ref{fig3}. These clusters were selected because of their well defined CMDs and well studied nature as a canonical metal-poor ($[Fe/H]=-1.19\pm 0.14$) and metal-rich ($[Fe/H]=-0.59\pm 0.07$) GCs \citep{nataf13}. 
The CMDs for the new GC candidates reveal populated red giant branches (RGBs) in the GCs, which in some cases appear tighter than those of their respective comparison regions (Minni\,30 and Minni\,34, for example) and in other cases they do not (as Minni\,24, for instance). Figure \ref{fig3} show that some of the candidates, however, do not have well defined RGBs and are indistinguishable from the typical Galactic bulge population, so they could be mere windows in the dust distribution towards the bulge, or simply asterisms \citep[see e.g.,][]{monibidin11}. For some of the clusters, we can clearly see a well defined red clump (RC) in the cluster RGBs, which appears to be more compact than the typical bulge field RC. In some of the statistically subtracted CMDs there is still some remnant from foreground disc sequence. However, if the statistical subtraction is run inversely (Field - GC region) we find that the decontaminated CMDs have no stars left. \\

We measured differentially the heliocentric distances of the possibly real GC candidates with respect to the selected comparison clusters (NGC\,6624 and NGC\,6637), which were also observed and analyzed in the same way in the VVV Survey. We adopted for the comparison clusters the following distances given by Harris (2010): $D_{NGC6624} =6.8$  kpc and  $D_{NGC6637} =8.8$  kpc, with a typical error of $\sigma_D \sim 1.5$ kpc.\\

\begin{figure*}
\includegraphics[scale=.17]{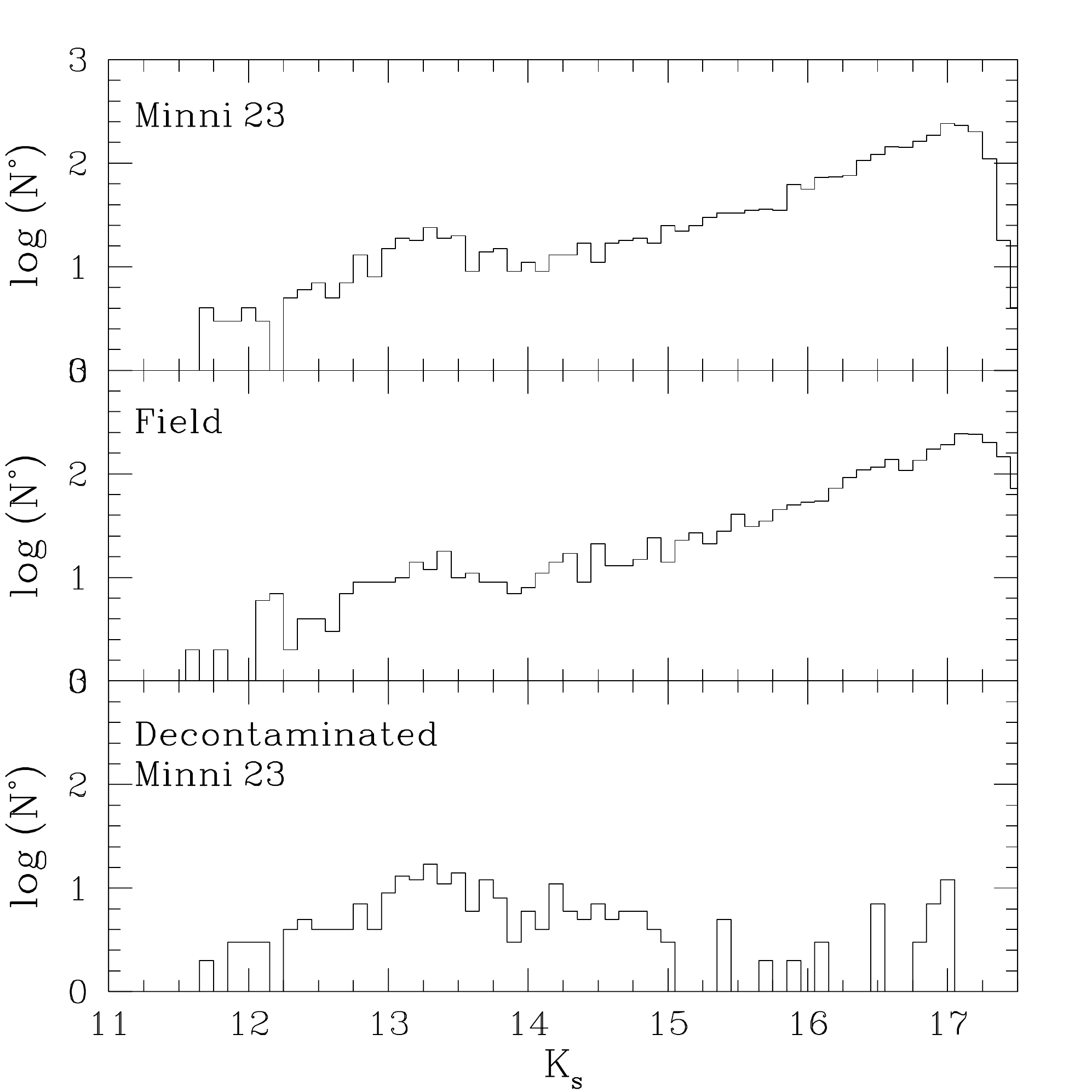}
\includegraphics[scale=.17]{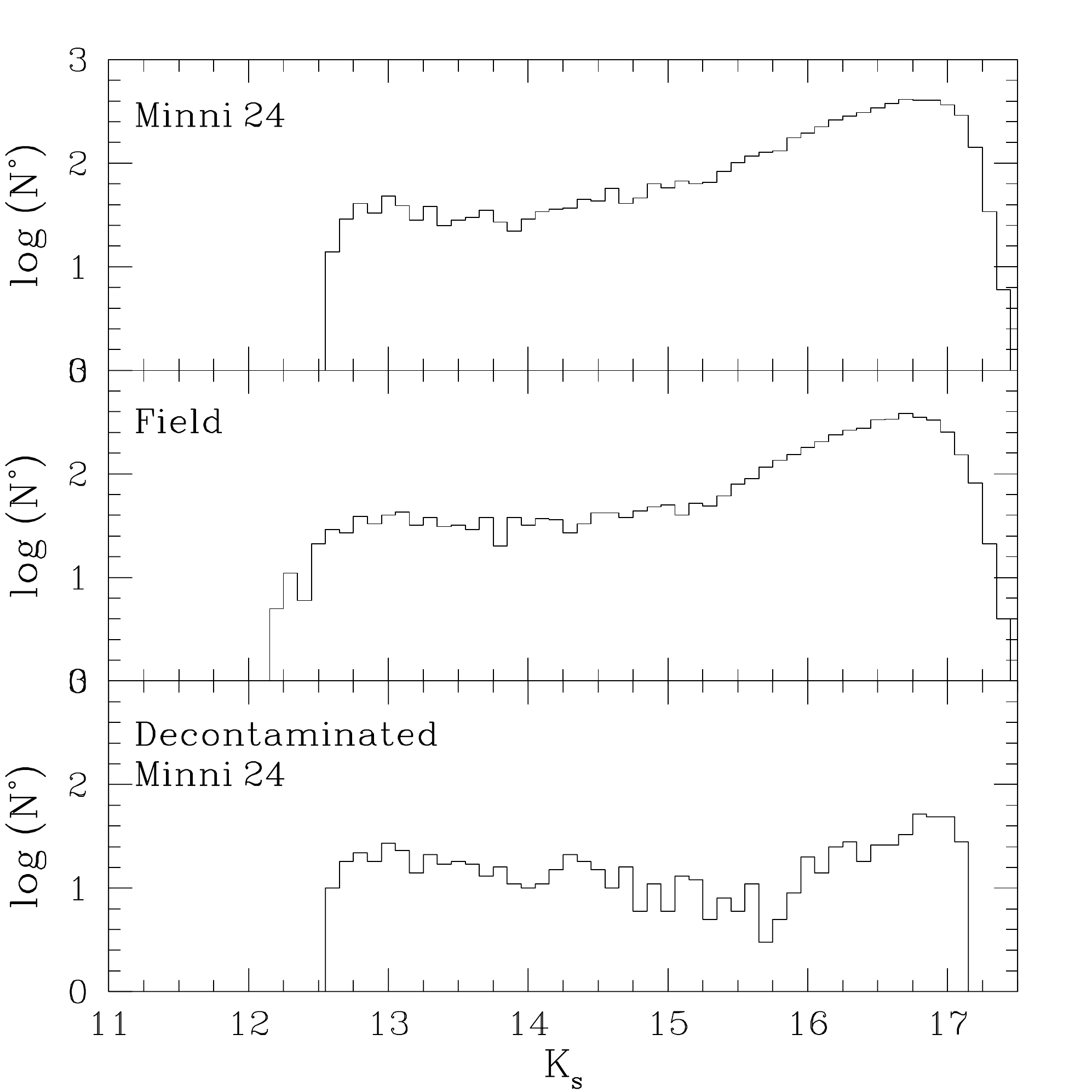}
\includegraphics[scale=.17]{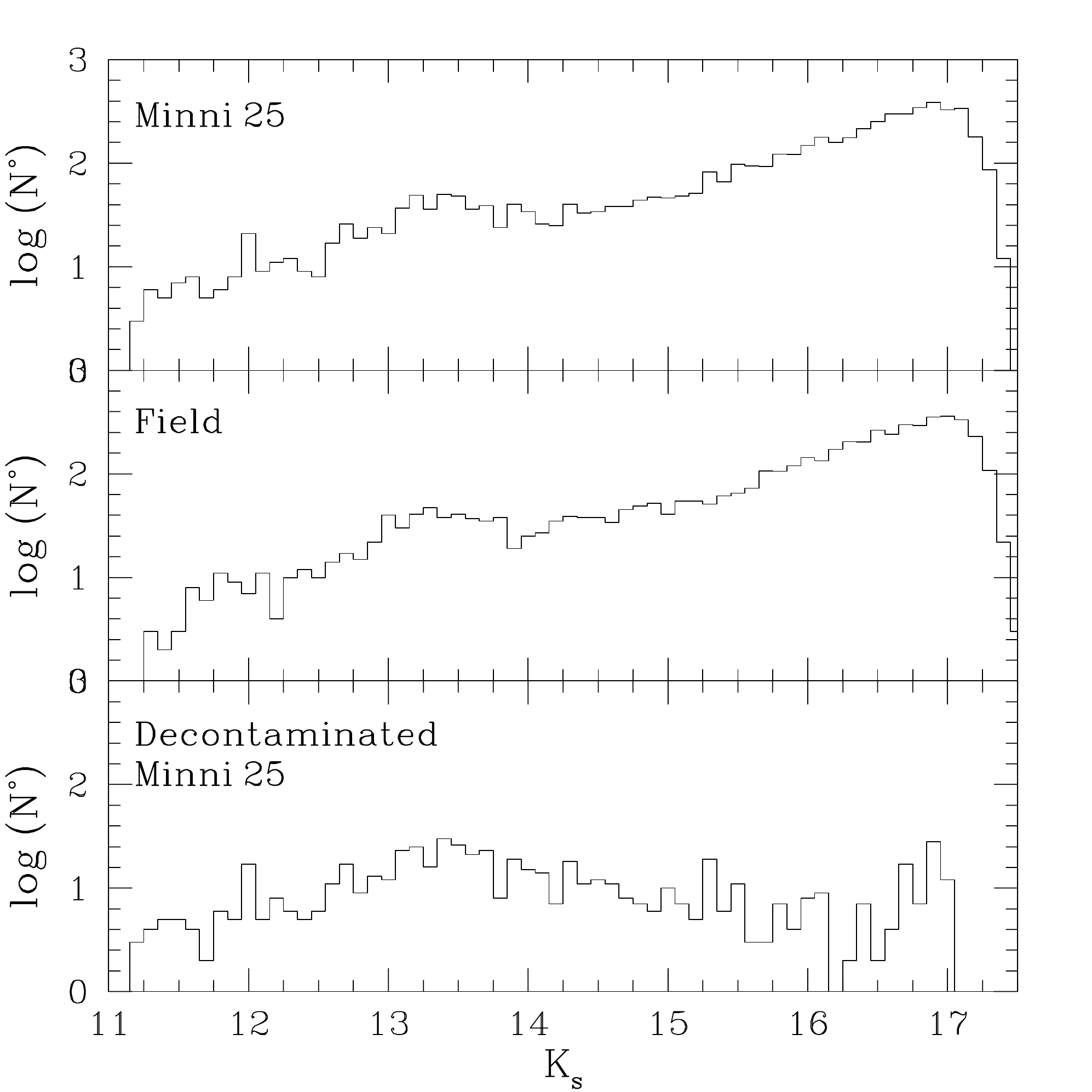}
\includegraphics[scale=.17]{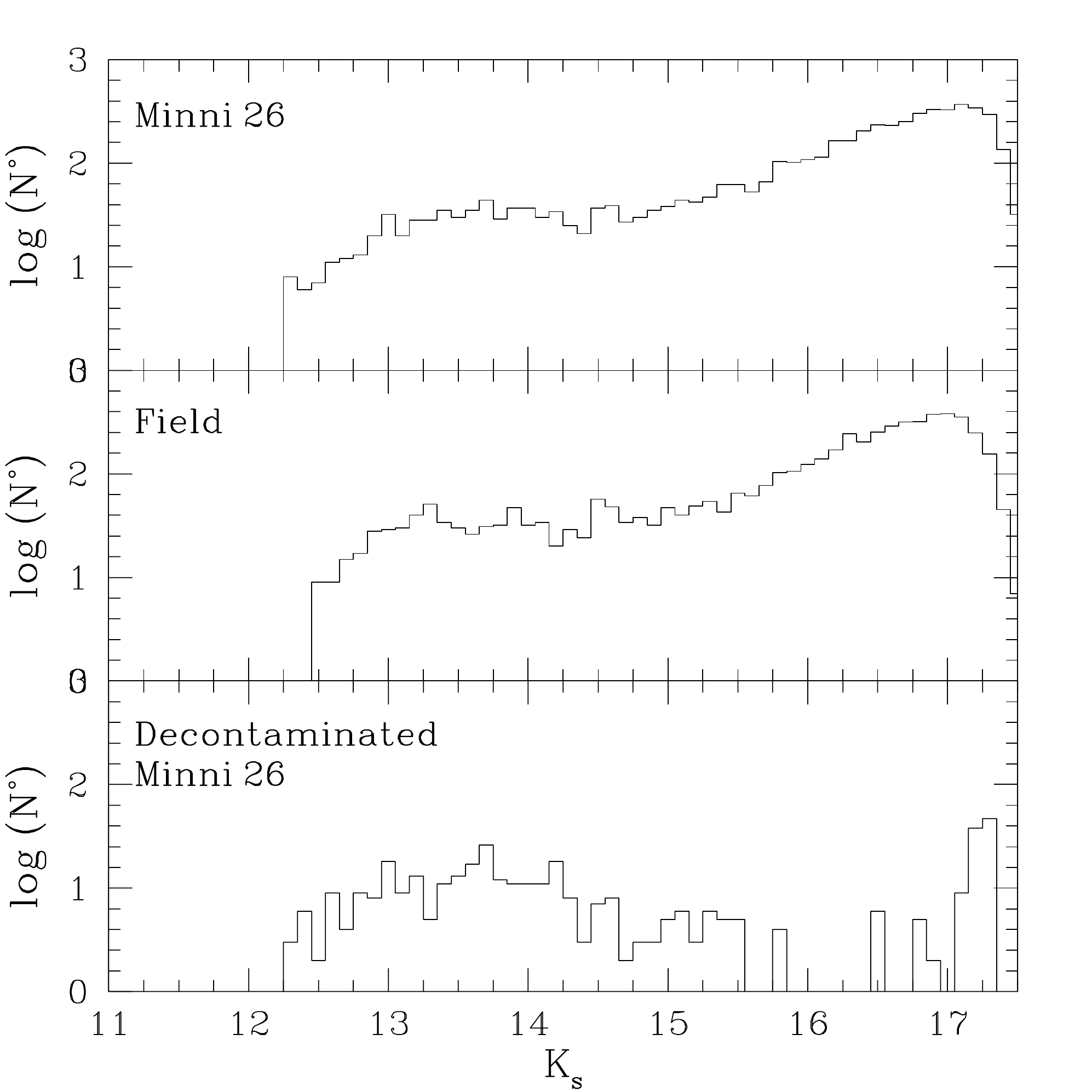}
\includegraphics[scale=.17]{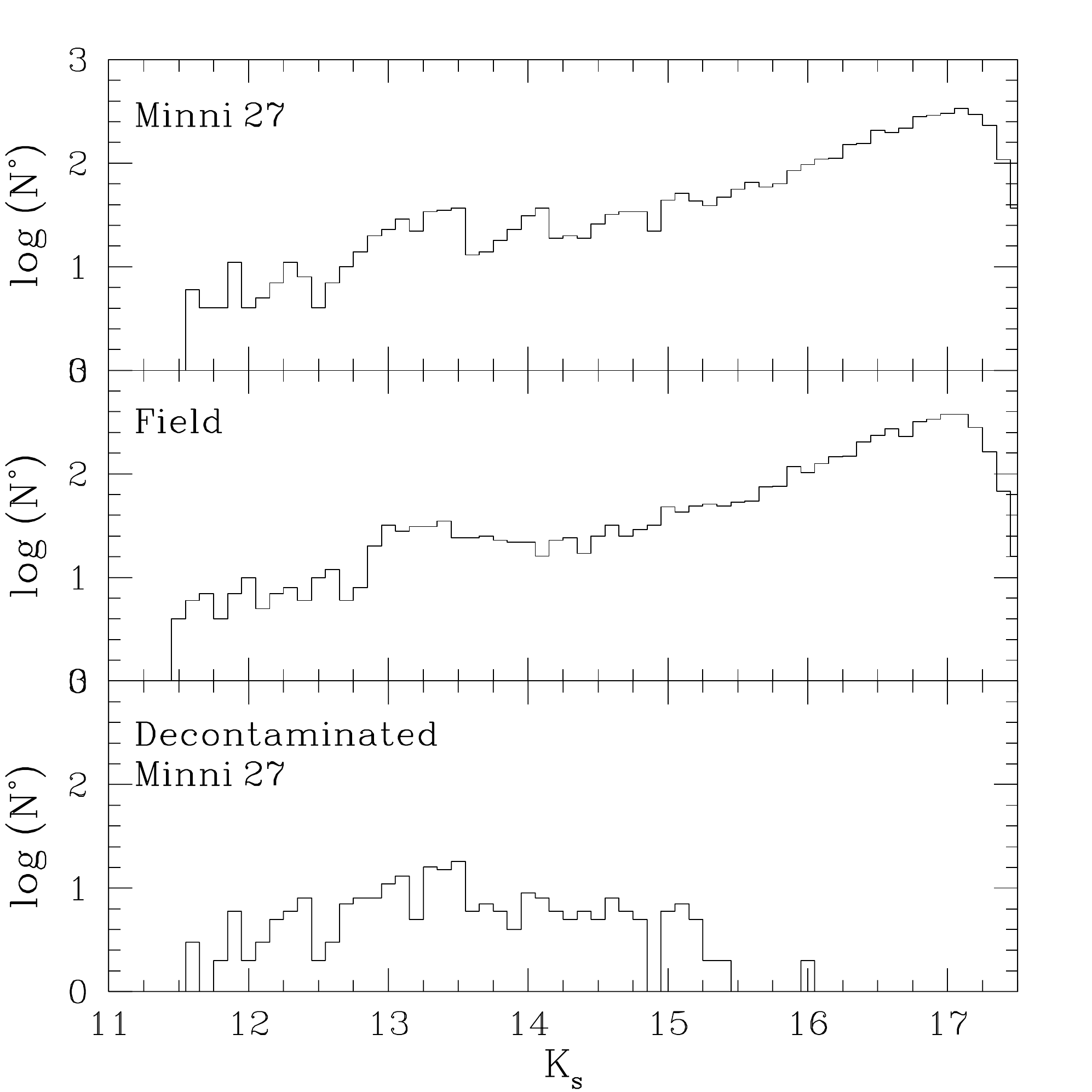}
\includegraphics[scale=.17]{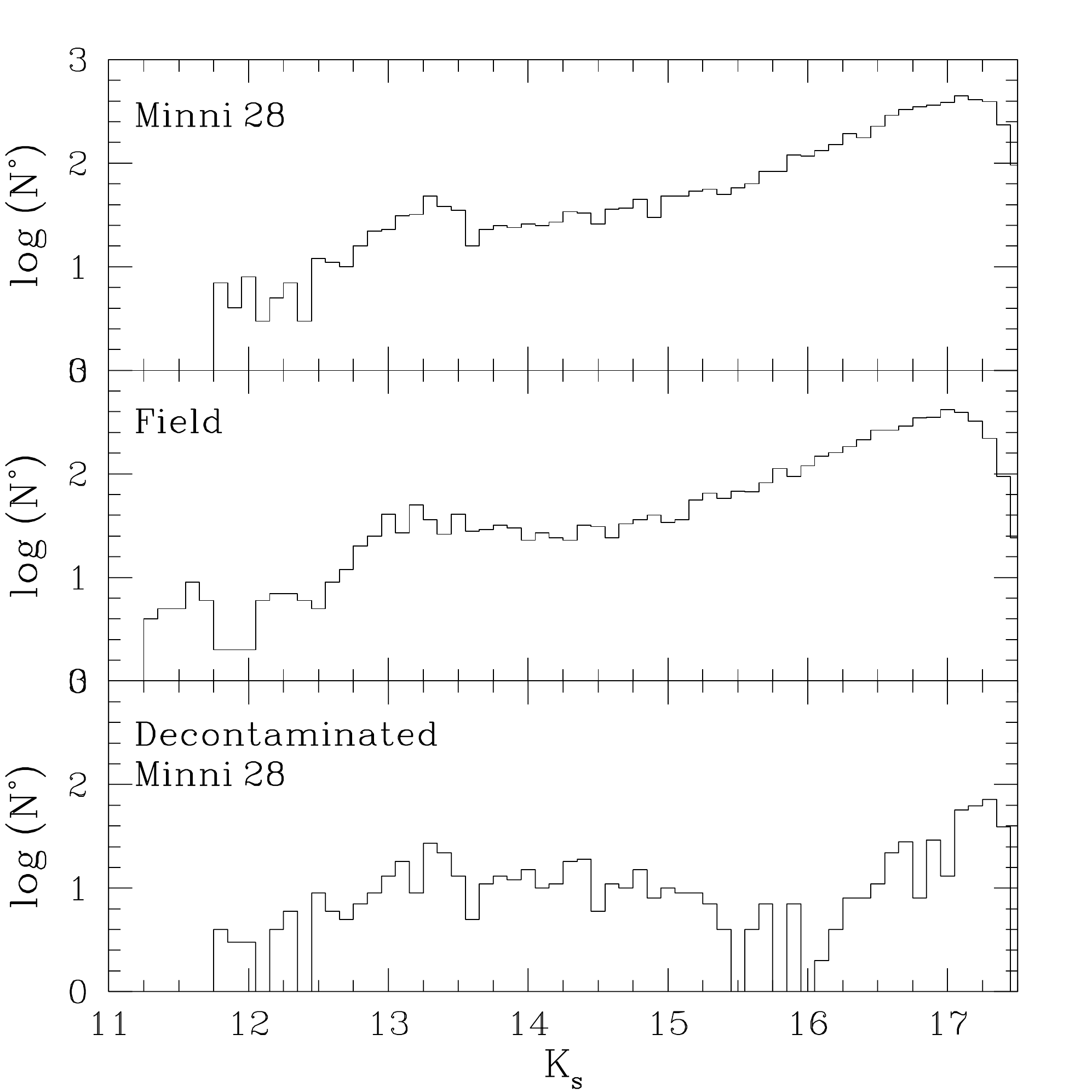}
\includegraphics[scale=.17]{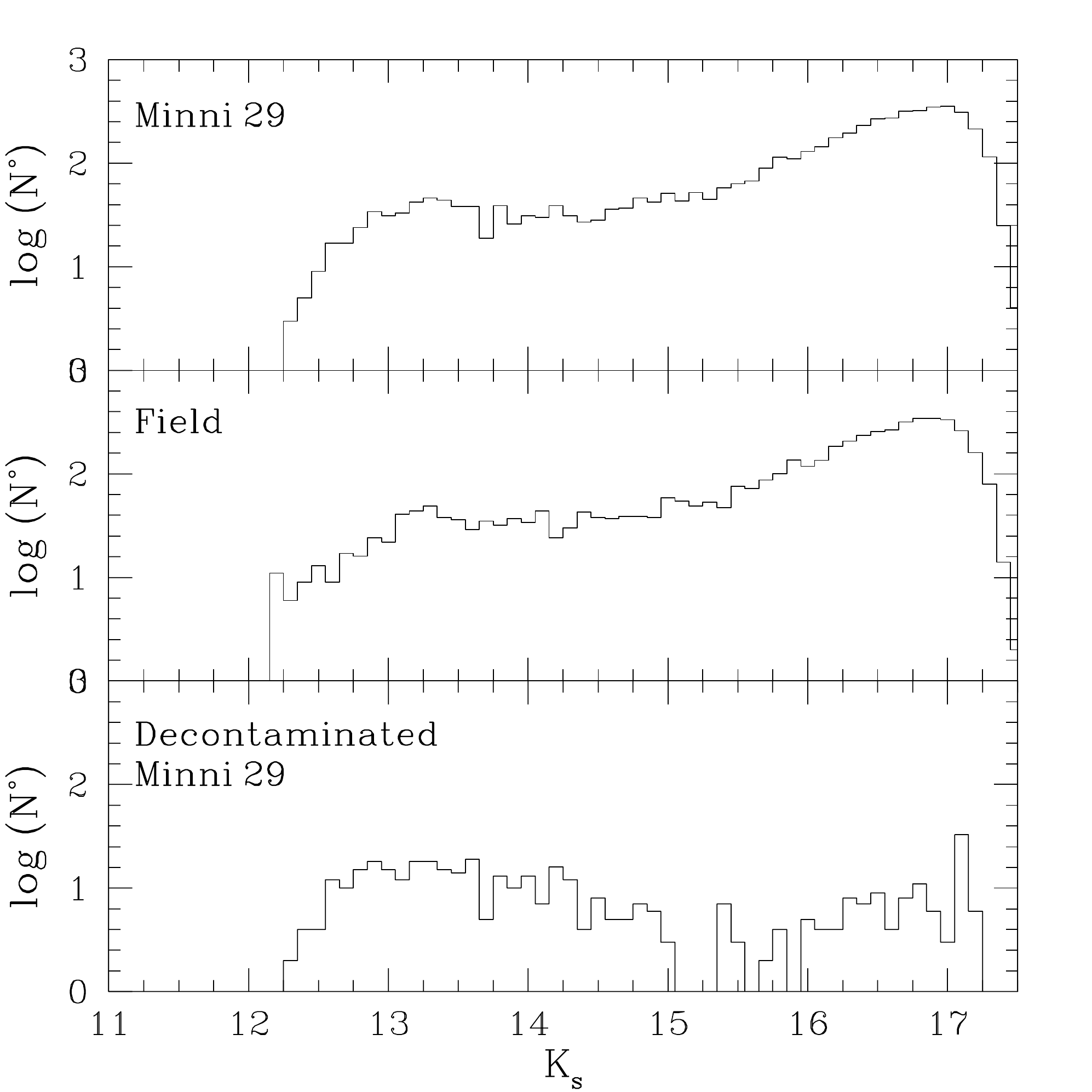}
\includegraphics[scale=.17]{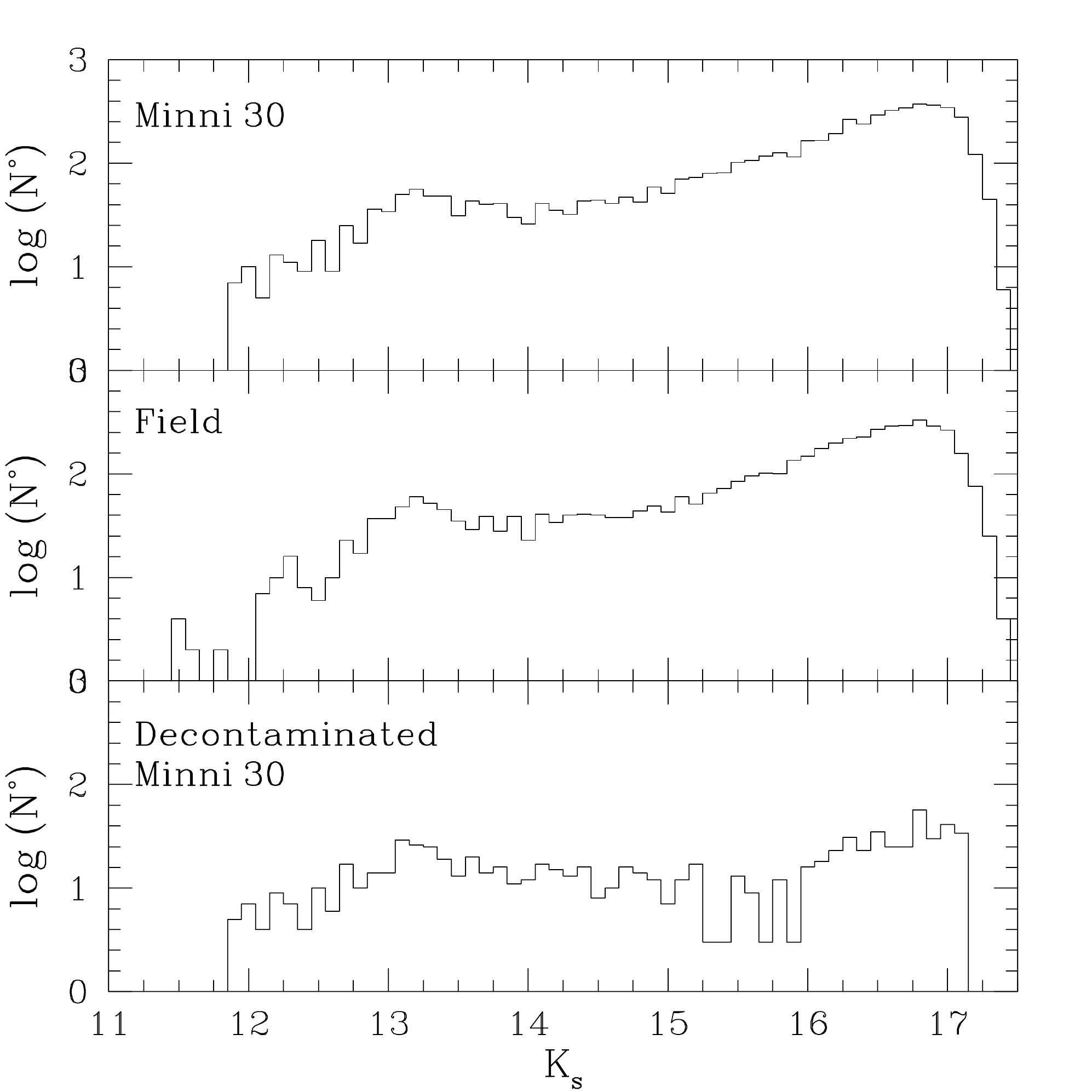}
\includegraphics[scale=.17]{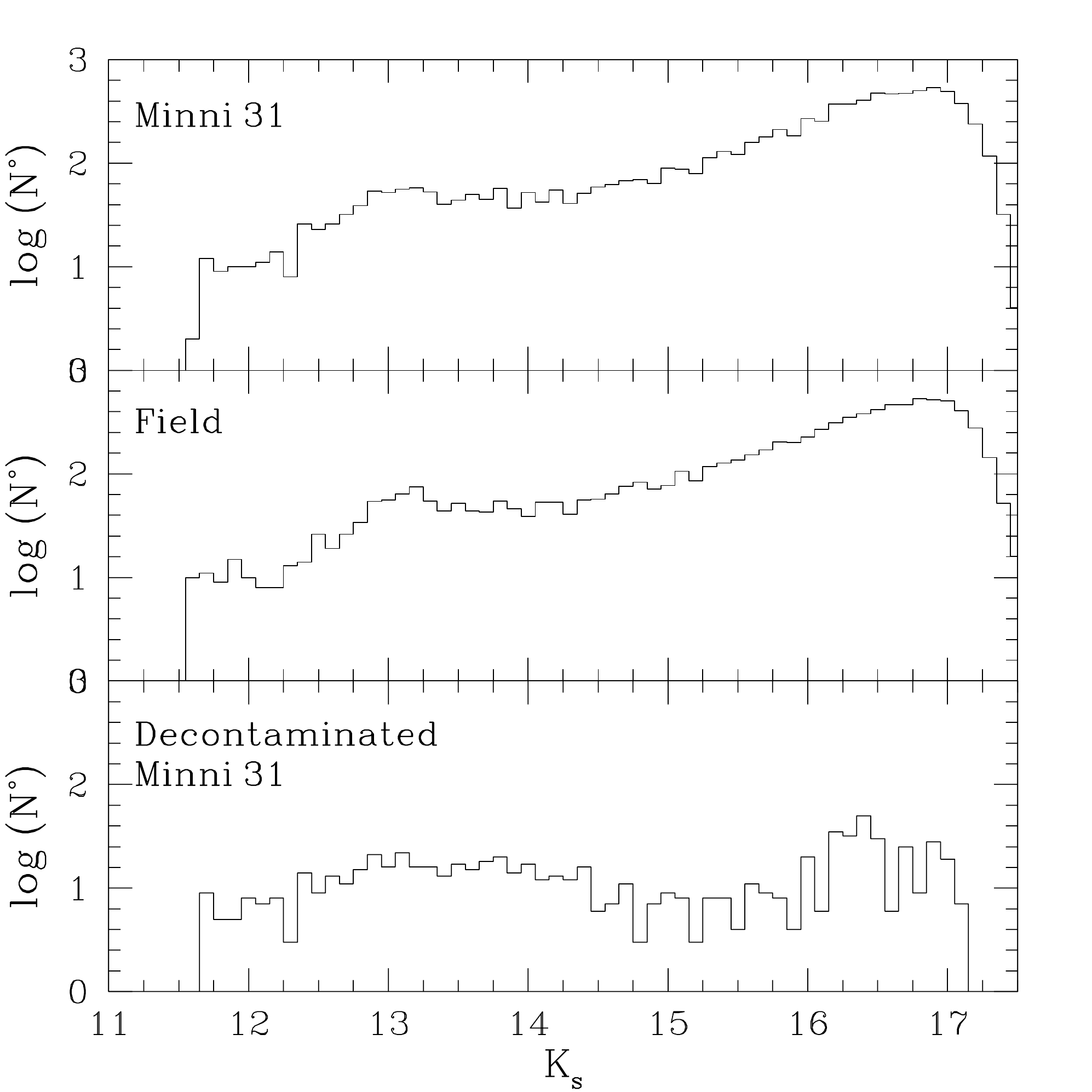}
\includegraphics[scale=.17]{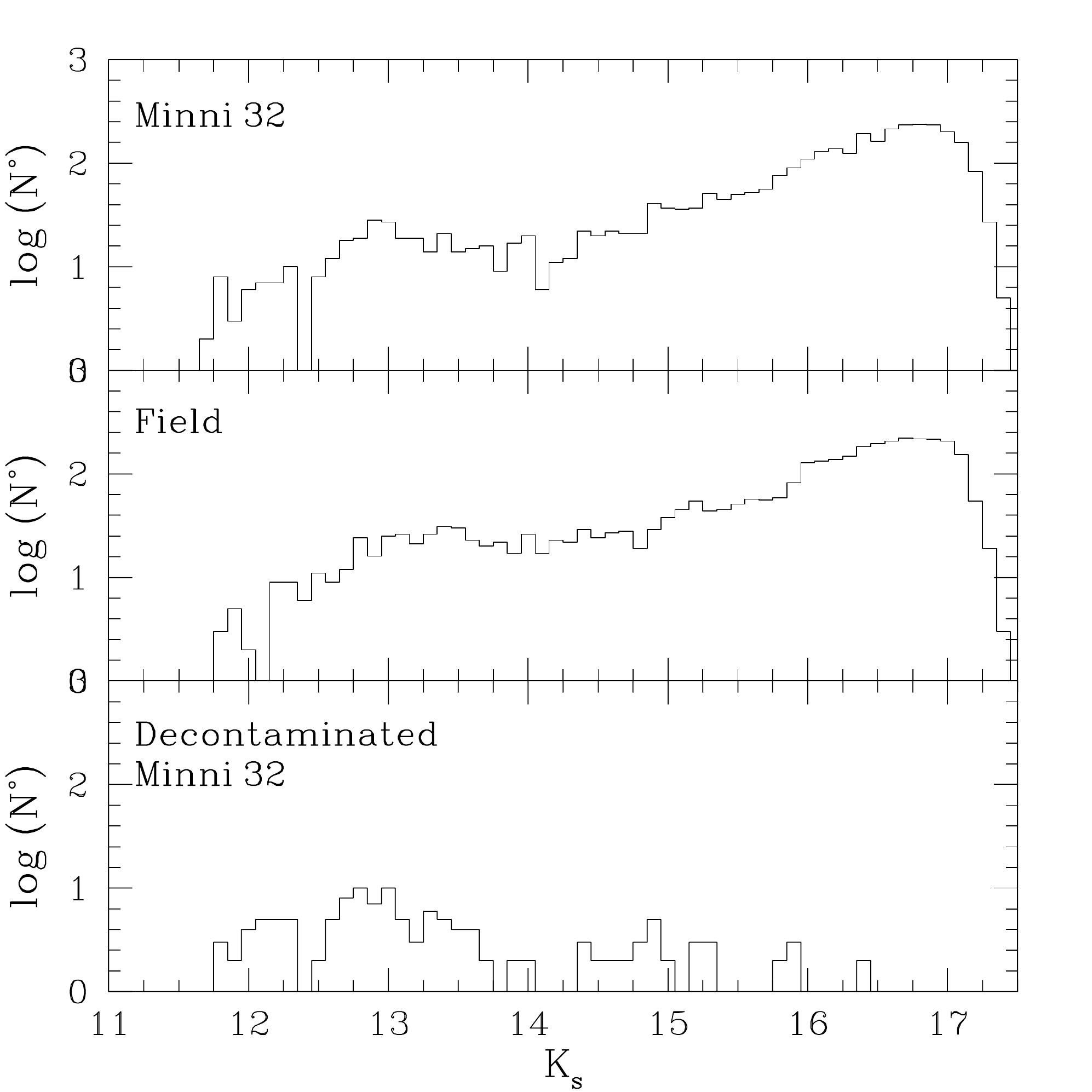}
\includegraphics[scale=.17]{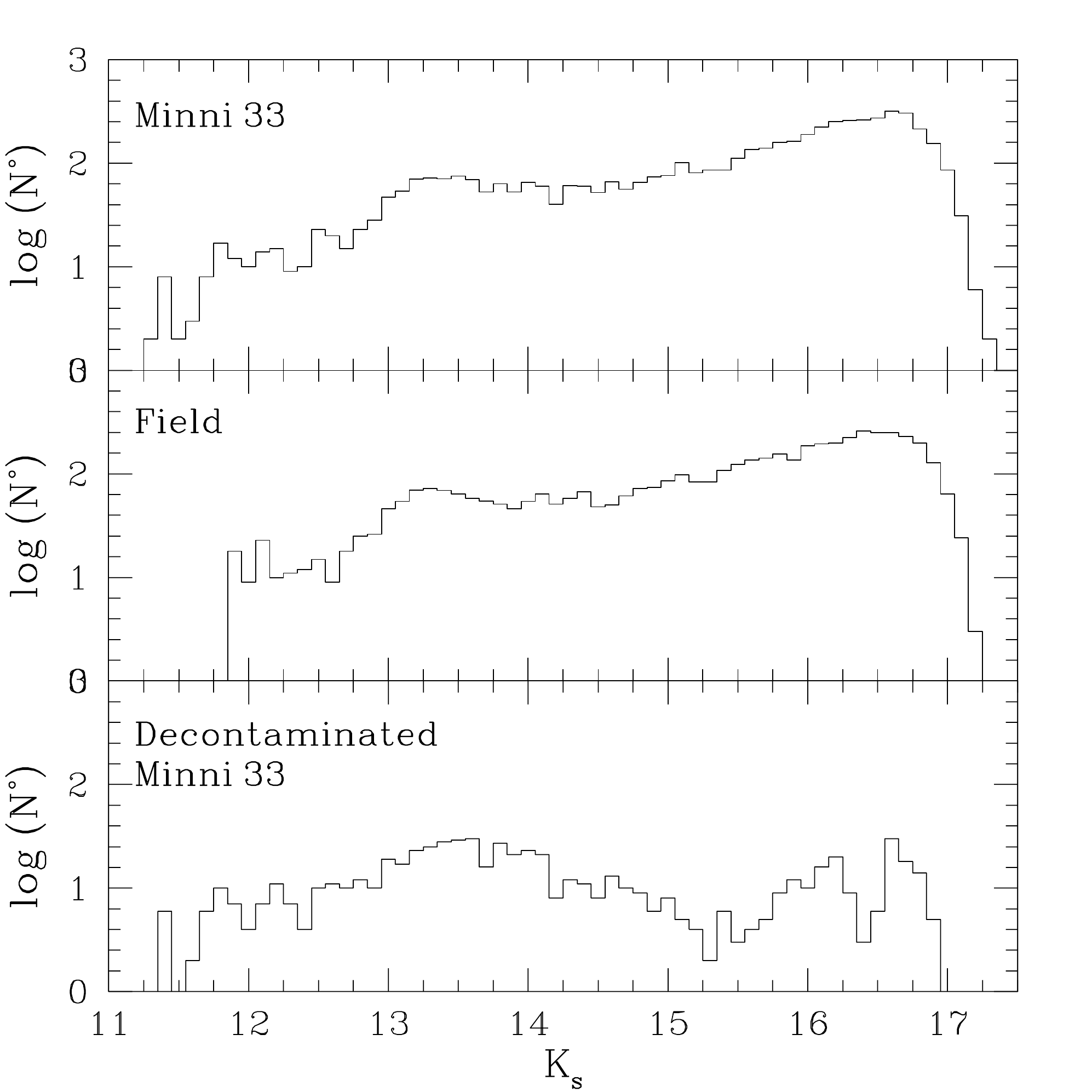}
\includegraphics[scale=.17]{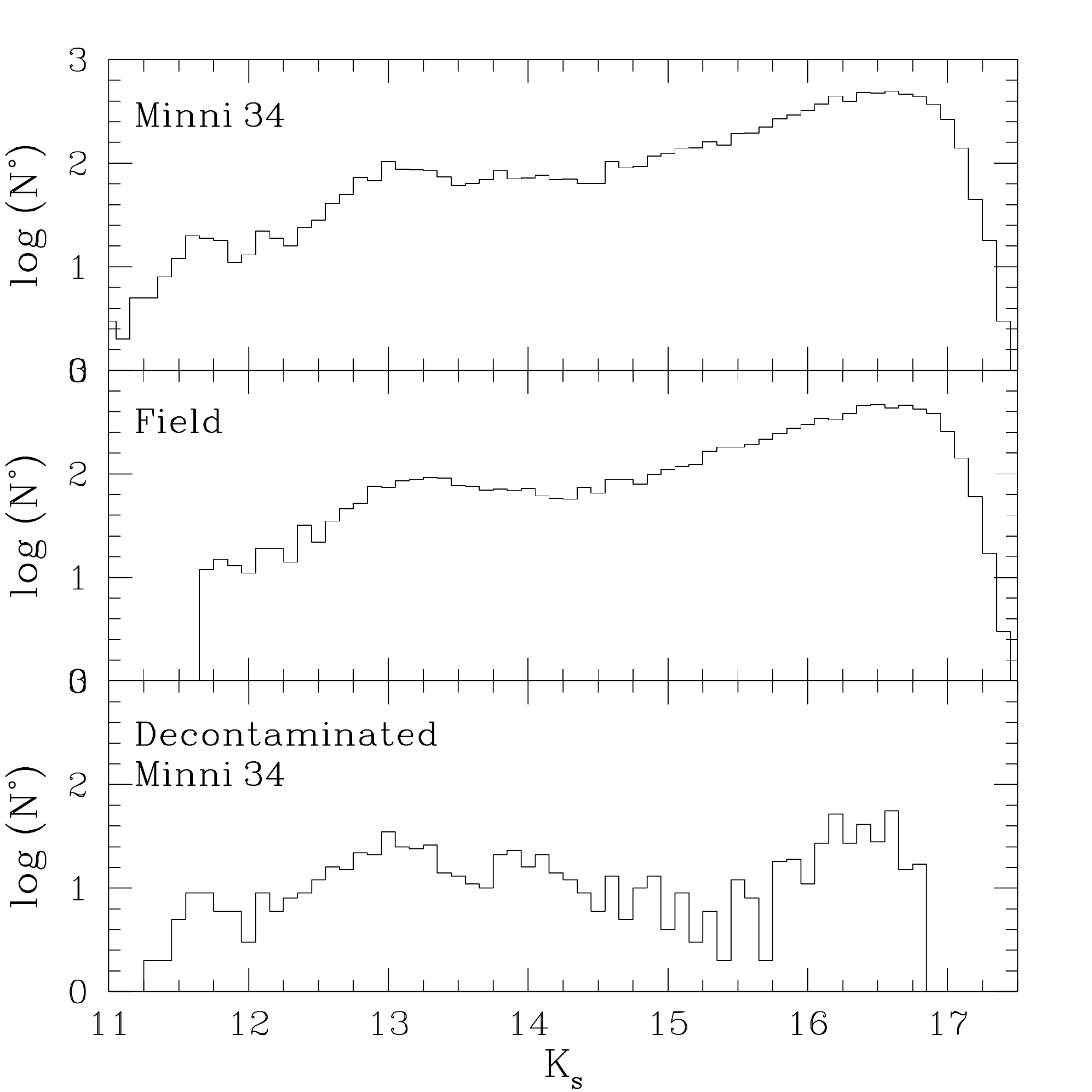}
\includegraphics[scale=.17]{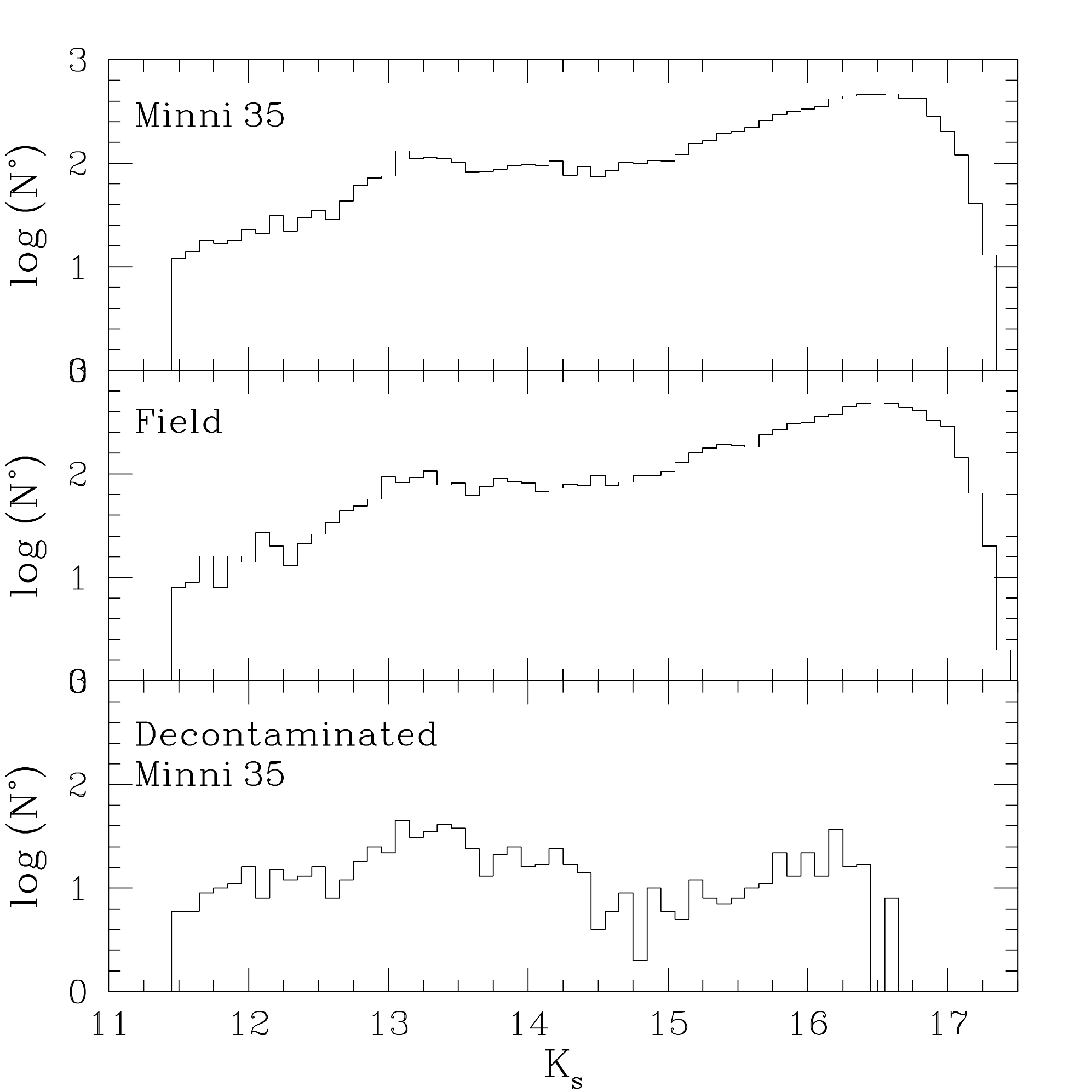}
\includegraphics[scale=.17]{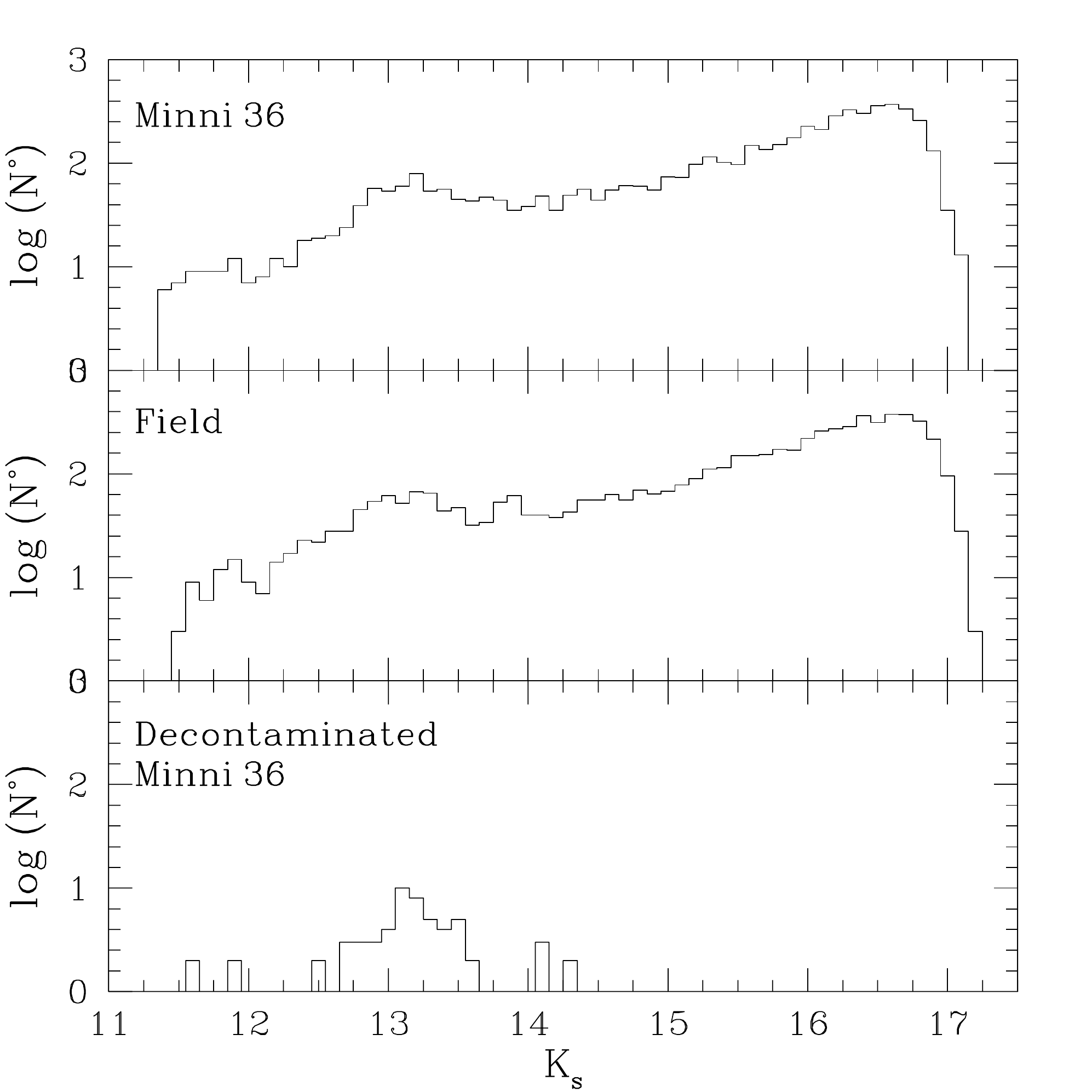}
\includegraphics[scale=.17]{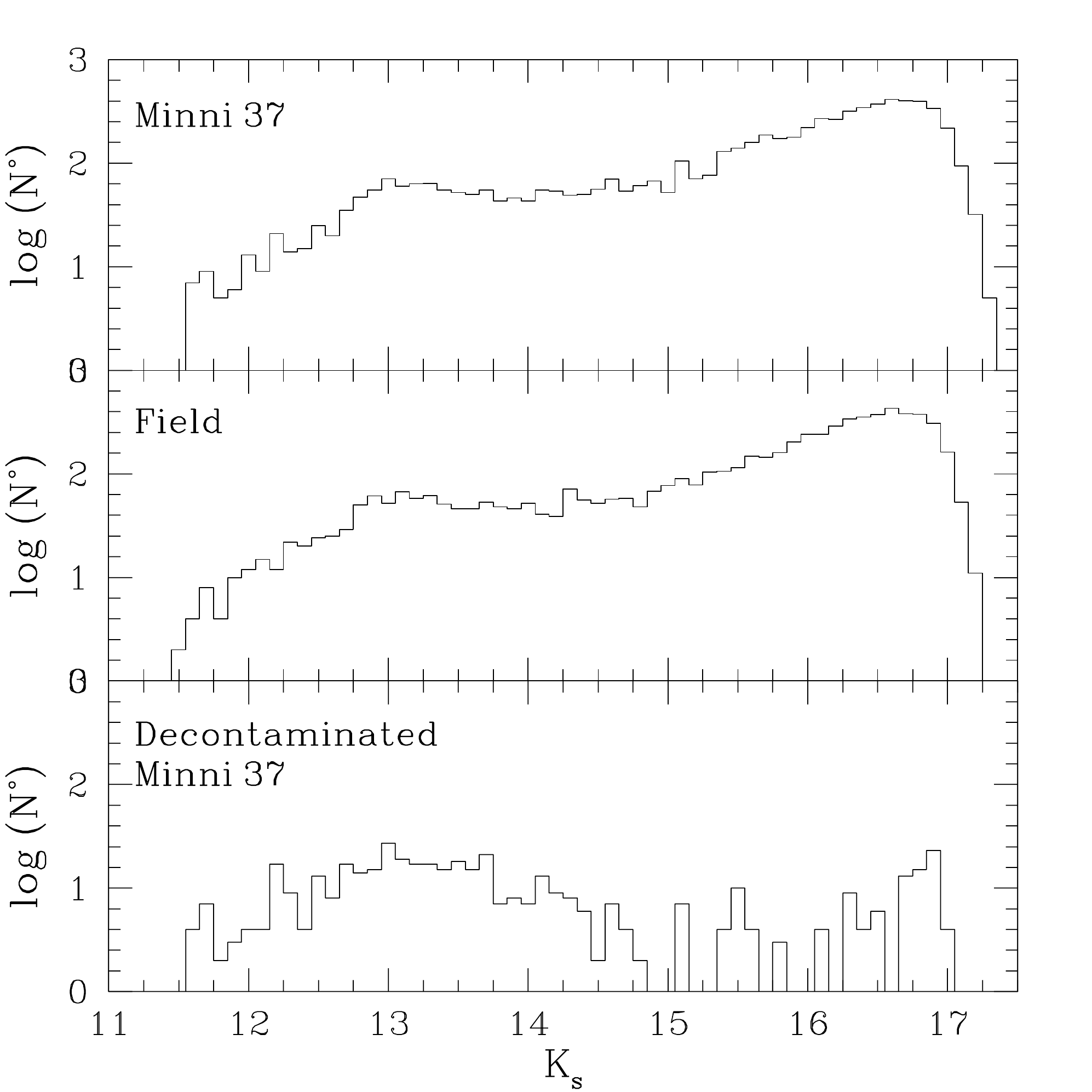}
\includegraphics[scale=.17]{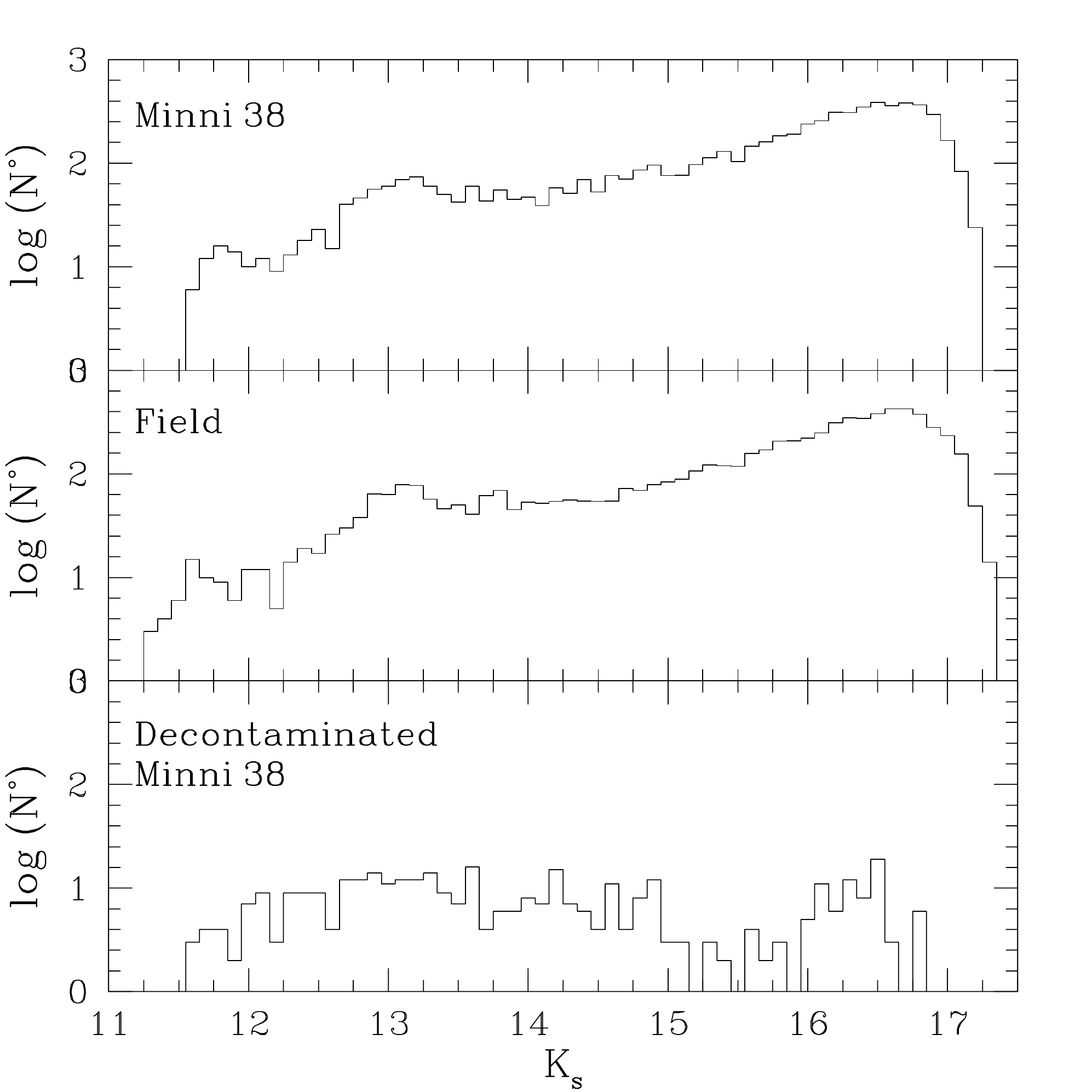}
\includegraphics[scale=.17]{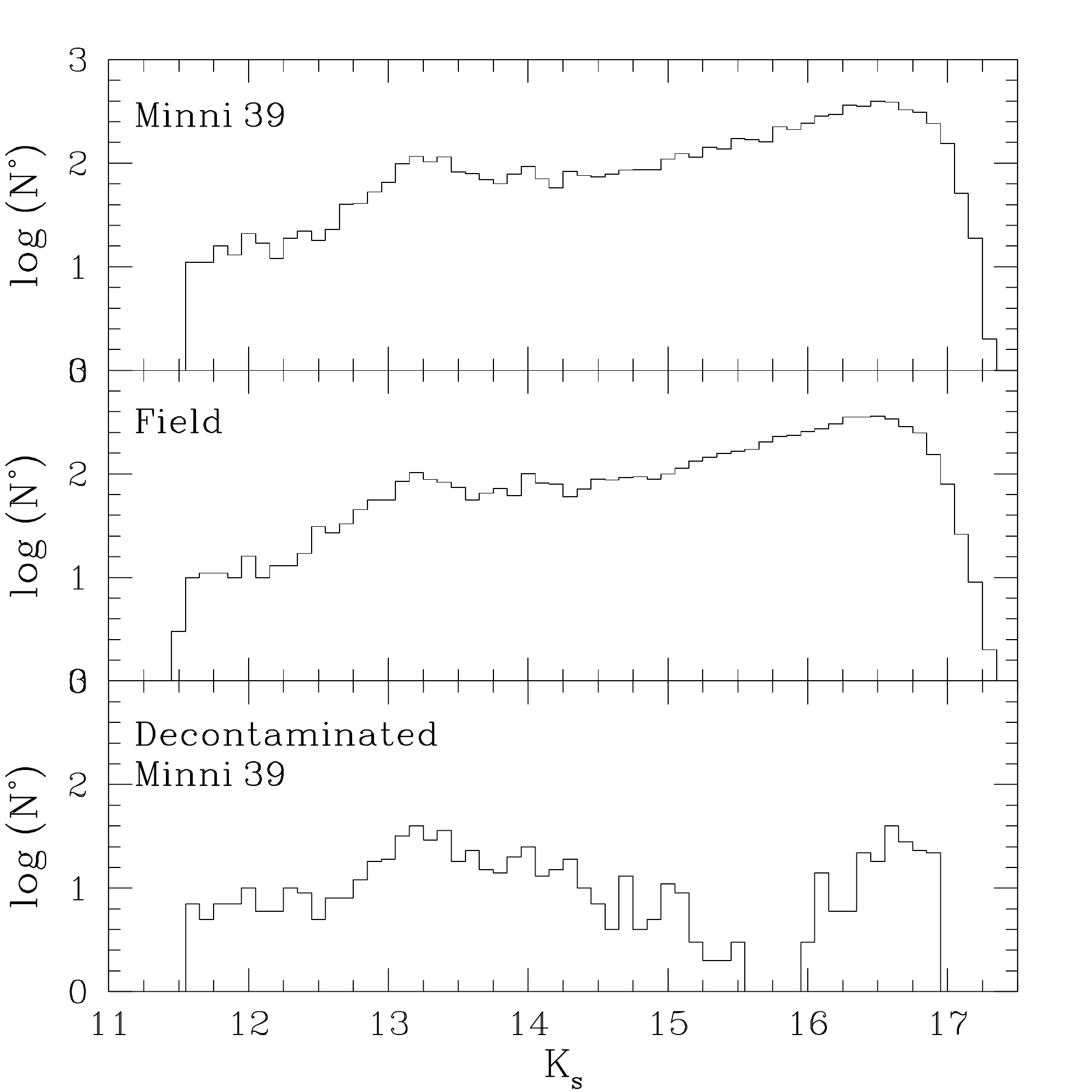}
\includegraphics[scale=.17]{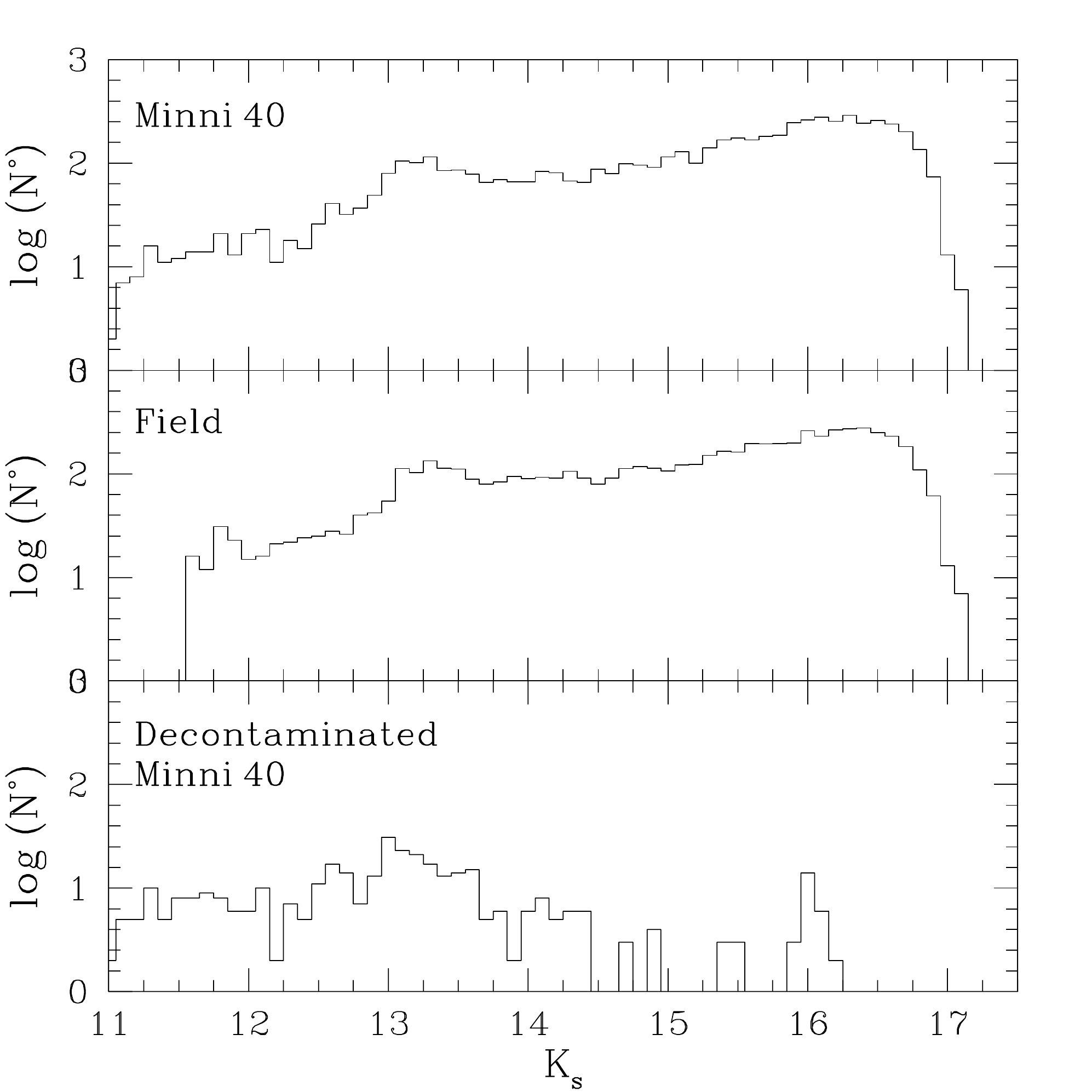}
\includegraphics[scale=.17]{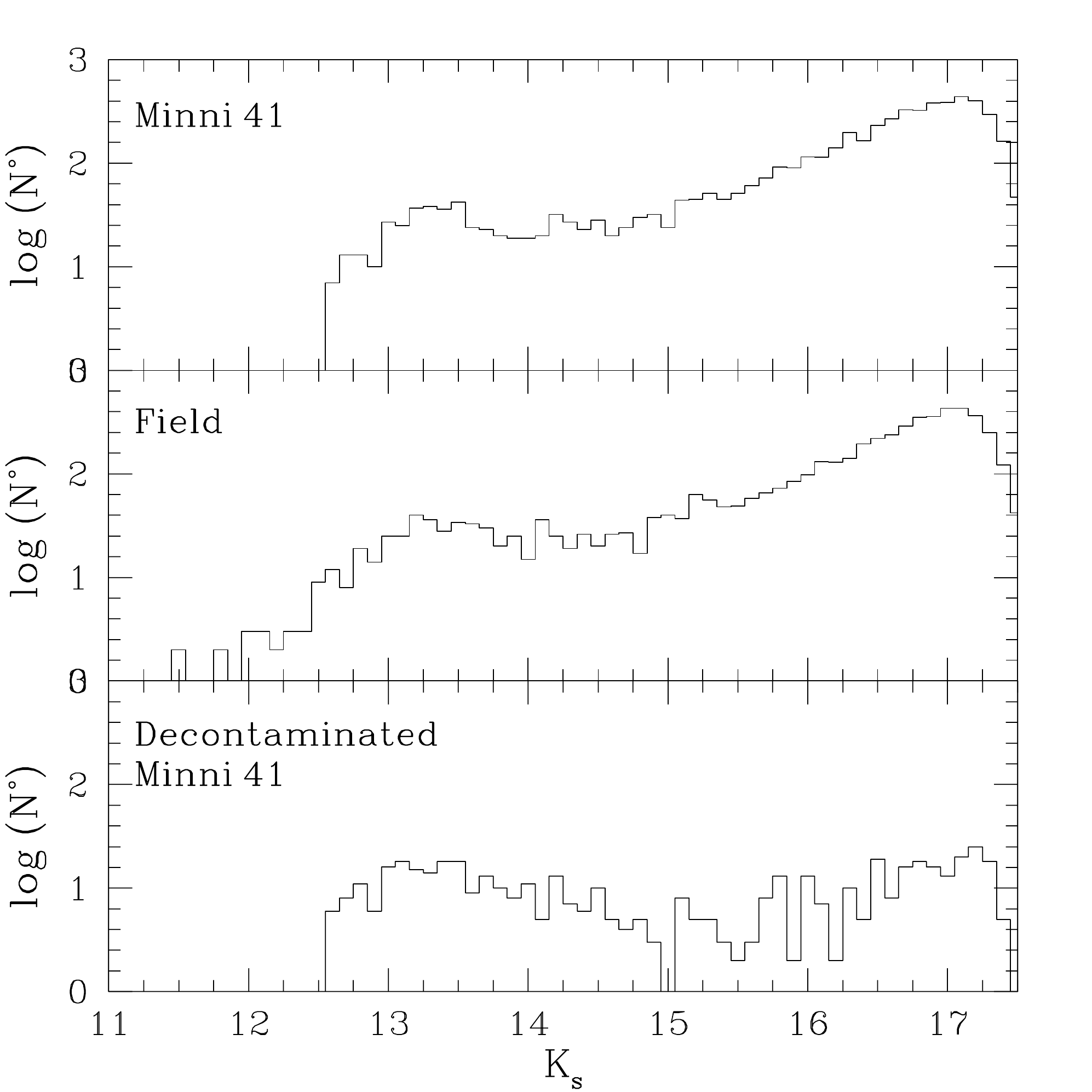}
\includegraphics[scale=.17]{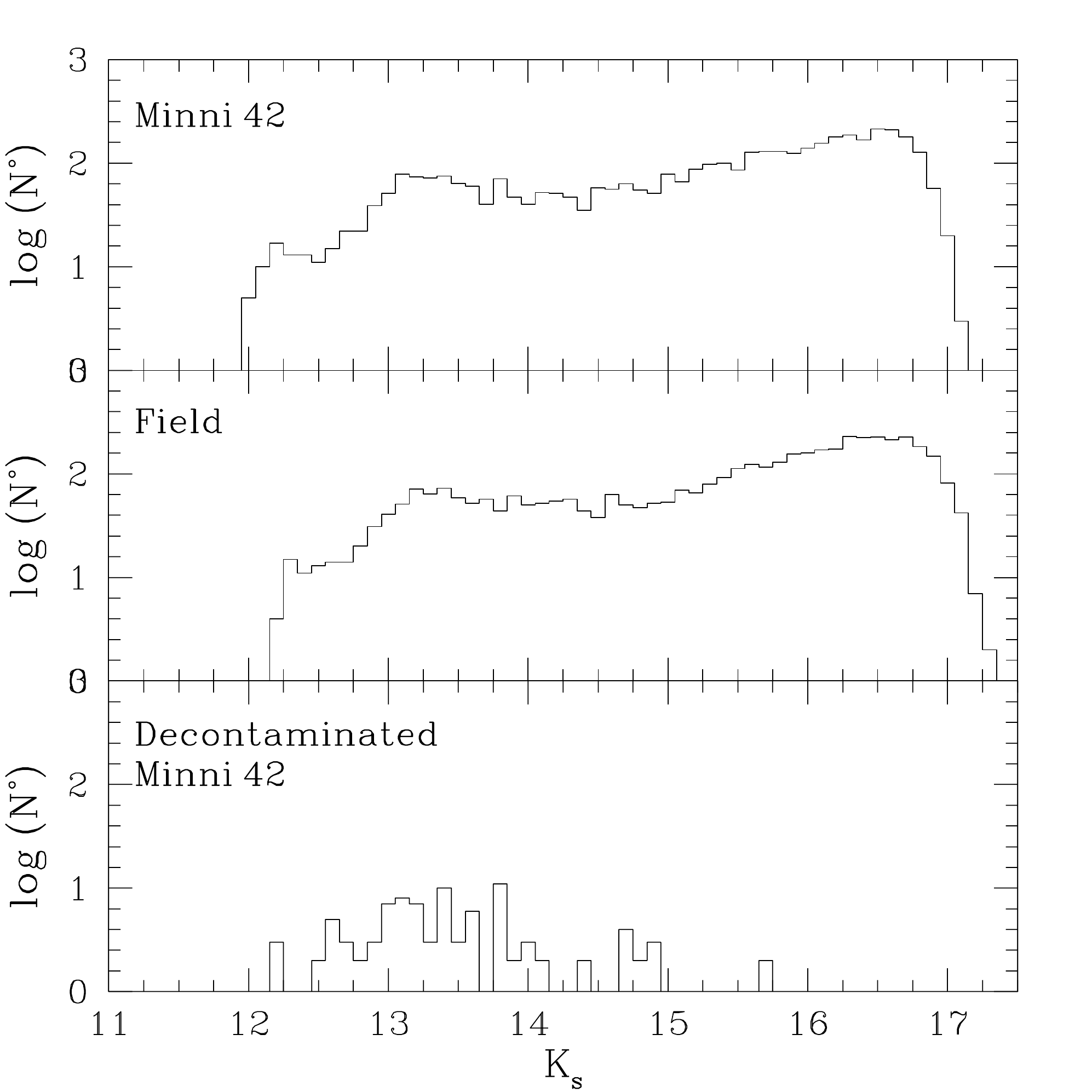}
\includegraphics[scale=.17]{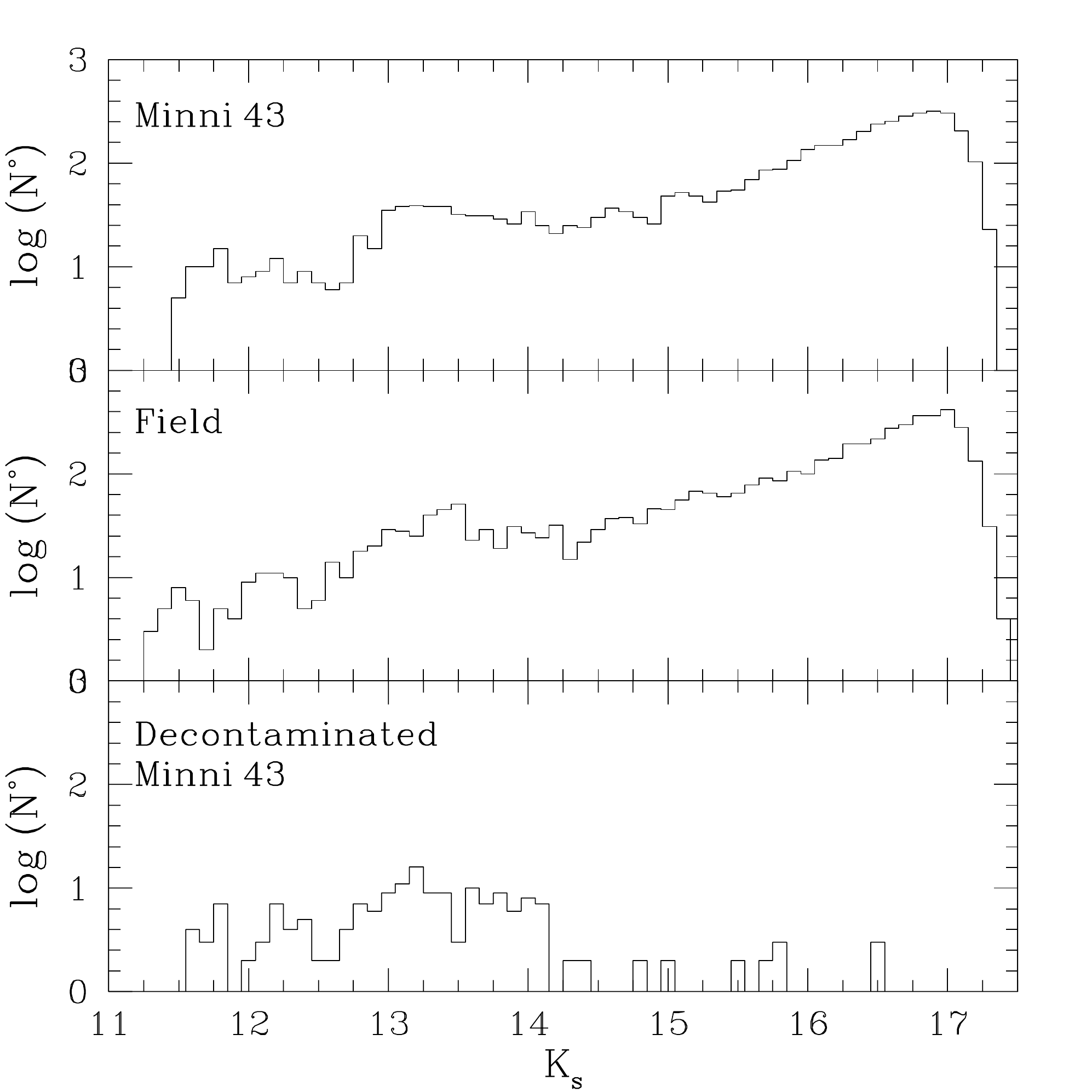}
\includegraphics[scale=.17]{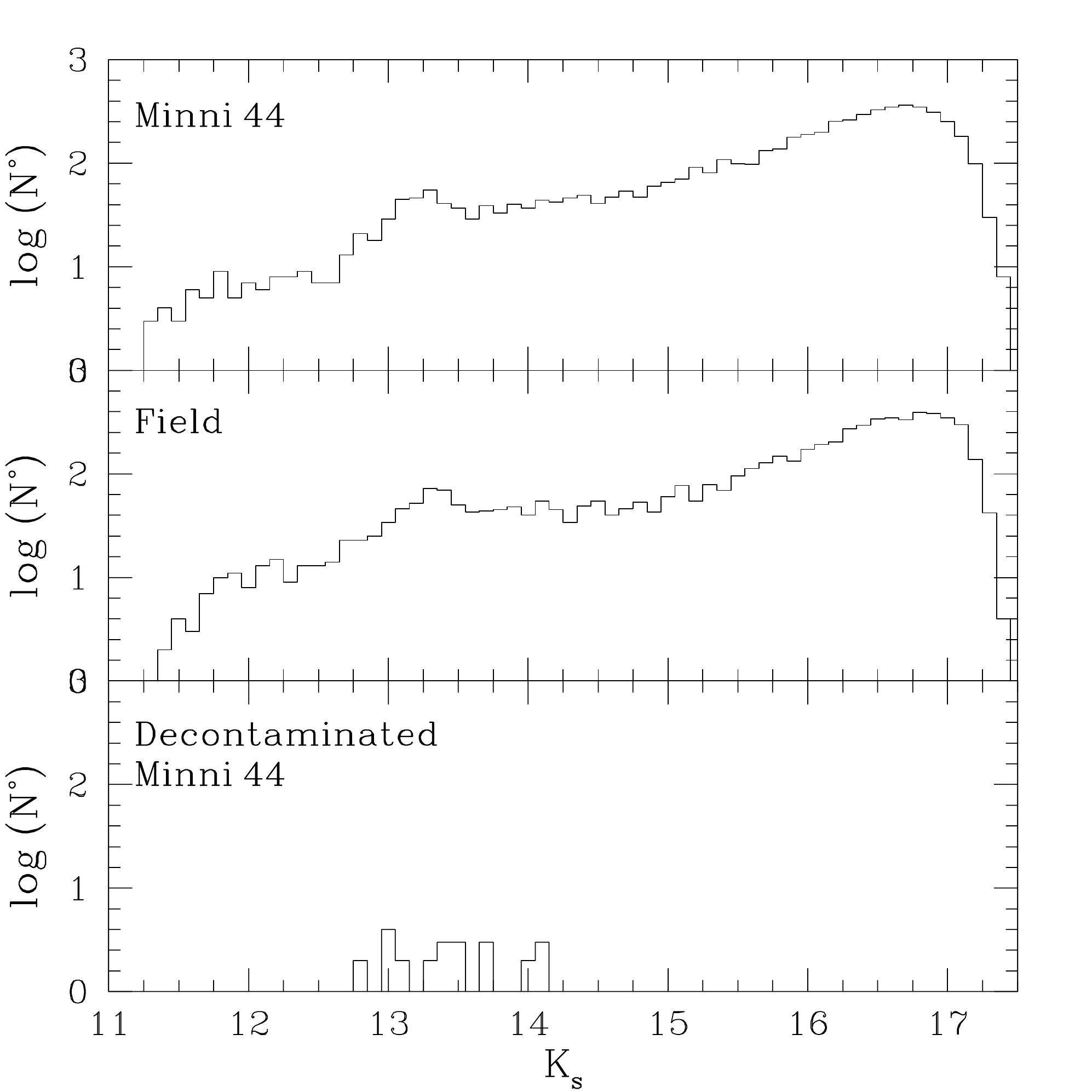}
\caption{Luminosity functions (LFs) showing for each cluster (from top to bottom): the cluster area, the background field, and the decontaminated cluster, respectively.
\label{fig4}  }
\end{figure*}

Figure \ref{fig4} shows the luminosity functions (LFs) for the cluster area, field, and decontaminated cluster, respectively.
We use the decontaminated cluster LFs in order to measure the cluster heliocentric distances, after applying a colour cut to select RGs, thus avoiding foreground main sequence stars of the Galactic disc. The standard deviation of the decontaminated LF distribution is $1.3$ on average. The colour cuts for the individual clusters vary according to their intestellar extinctions, being typically $J-K_s \sim 0.8 - 1.2$ mag. We note that some clusters (e.g. Minni\,36, Minni\,44) have substracted LFs that are consistent with about zero within the errors. \\

Columns 8 and 9 of Table \ref{tab1} list our final assessment of the GC candidates taking into account both decontamination methods, where the bad candidates are labelled as NO and discarded. On the other hand, the good candidates are labelled as YES, and in the cases where there is some doubt (e.g., a wide RGB, low number of stars in the LF, etc.) the candidates are labelled as YES? or NO?. These doubtful cases should be subjected to further study.\\

\section{Variable Stars: RR Lyrae and Miras} 

We performed a search for RR\,Lyrae and Mira stars located within 5 arcmin of the GC candidate centres using coordinates and periods from the Optical Gravitational Lensing Experiment \citep[OGLE;][]{soszynski11,soszynski13}. OGLE provided us with observed $V$ and $I$ magnitudes, oscillation periods and Fourier phases. We follow the procedure described by \citet{pietru15} to obtain the RR Lyrae distances. Briefly, we obtained the absolute $I$ and $V$-band magnitudes ($M_{I}$, $M_{V}$) by applying the period-luminosity-metallicity relation from \cite{catelan04}. Metallicities injected in that equation were determined using the period and Fourier phases derived from the OGLE light curves following \cite{smolec05}.  After that, distances can be straight forwardly obtained through the classical distance modulus relation $I - M_I = 5log(d) - 5 + A_I$,  where $A_I = 0.7465E(V-I)+1.37E(J-Ks)$ and $E(J-K_s)$ was obtained from \citet{gonzalez12} extinction maps, for each RR Lyrae star. For Mira variable stars, we have performed a procedure similar to that used for the RR\,Lyrae stars, without estimating metallicities. The resulting stars had their coordinates cross-matched with the 2MASS point source catalogue \citep{skrutskie06}. In this way, we obtained $K_s$-band magnitudes for the Mira stars. We note that these magnitudes have not been phase corrected so an intrinsic dispersion is to be expected. In order to derive the heliocentric distances of our GC candidates, we first calculated their absolute magnitudes using the relation given by \cite{whitelock08}: $M_{K_s} = -3.51 (log P - 2.38) -7.25$, where $M_{K_s}$ is the absolute $K_s$-band magnitude, $P$ the pulsation period in days, and $-7.25$ the zero-point of the Period-Luminosity ($K_s$) relation. Then, the classic distance modulus equation was used to find the heliocentric distances. \\

\begin{figure*}
\includegraphics[scale=.17]{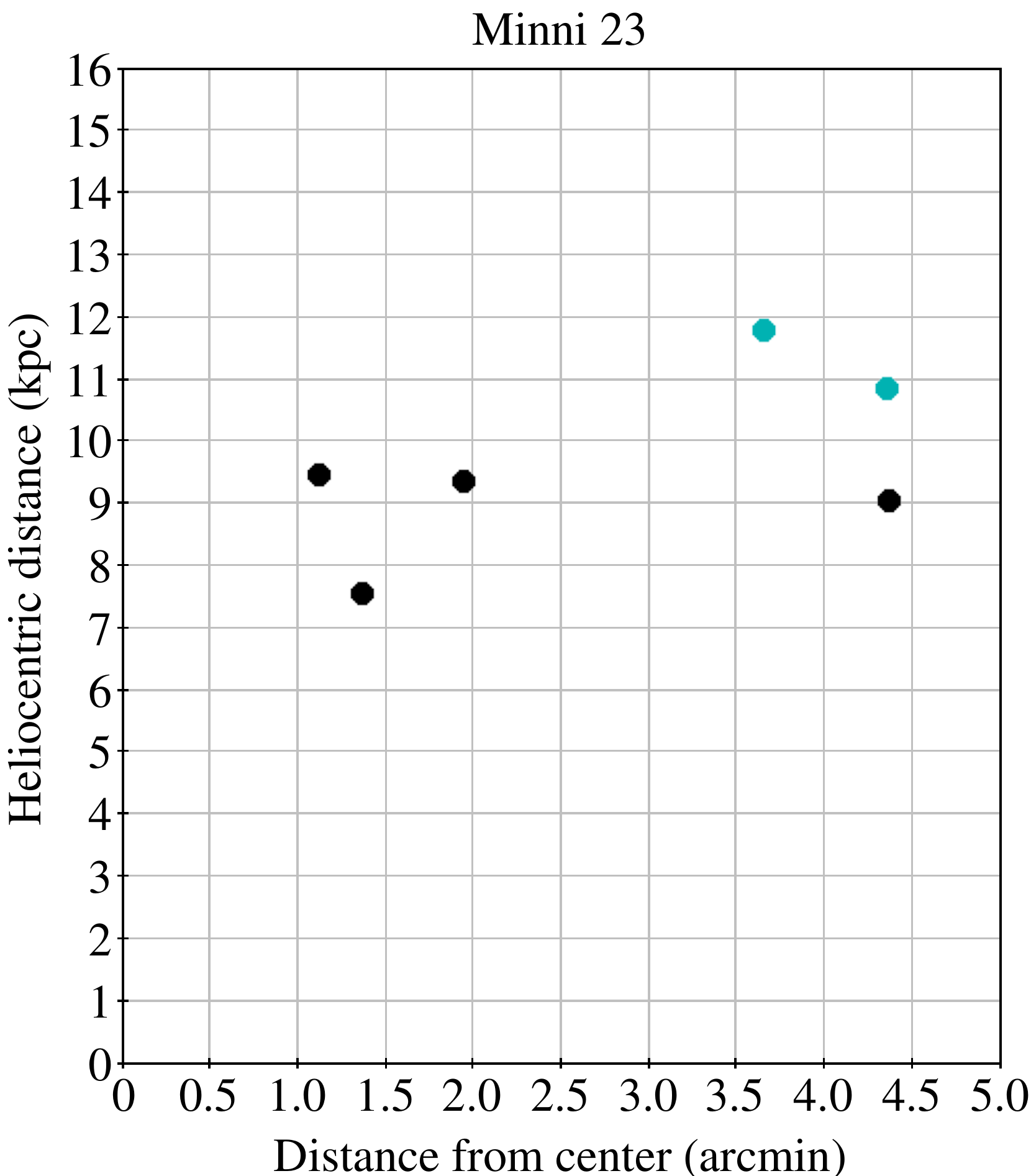}
\includegraphics[scale=.17]{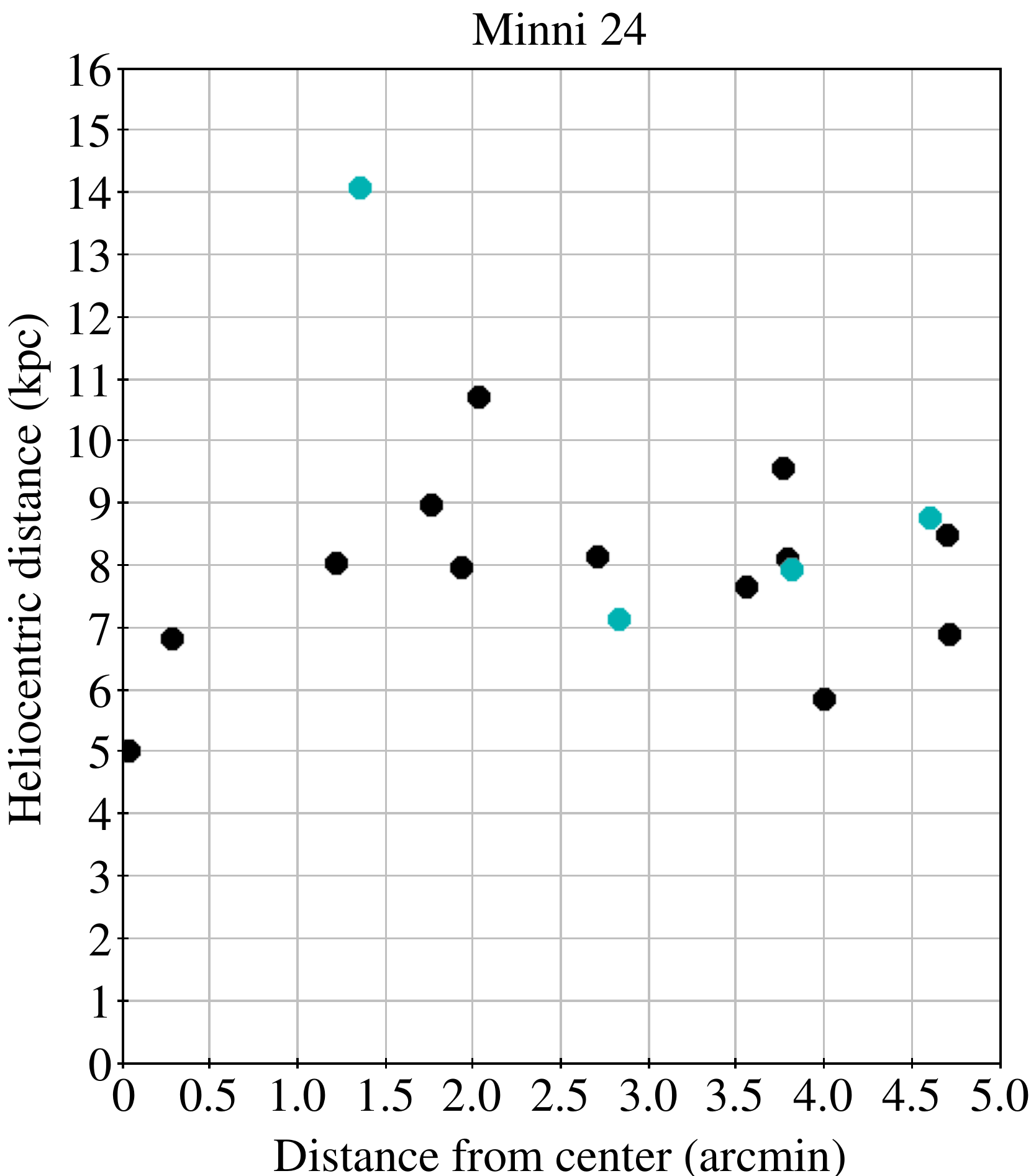}
\includegraphics[scale=.17]{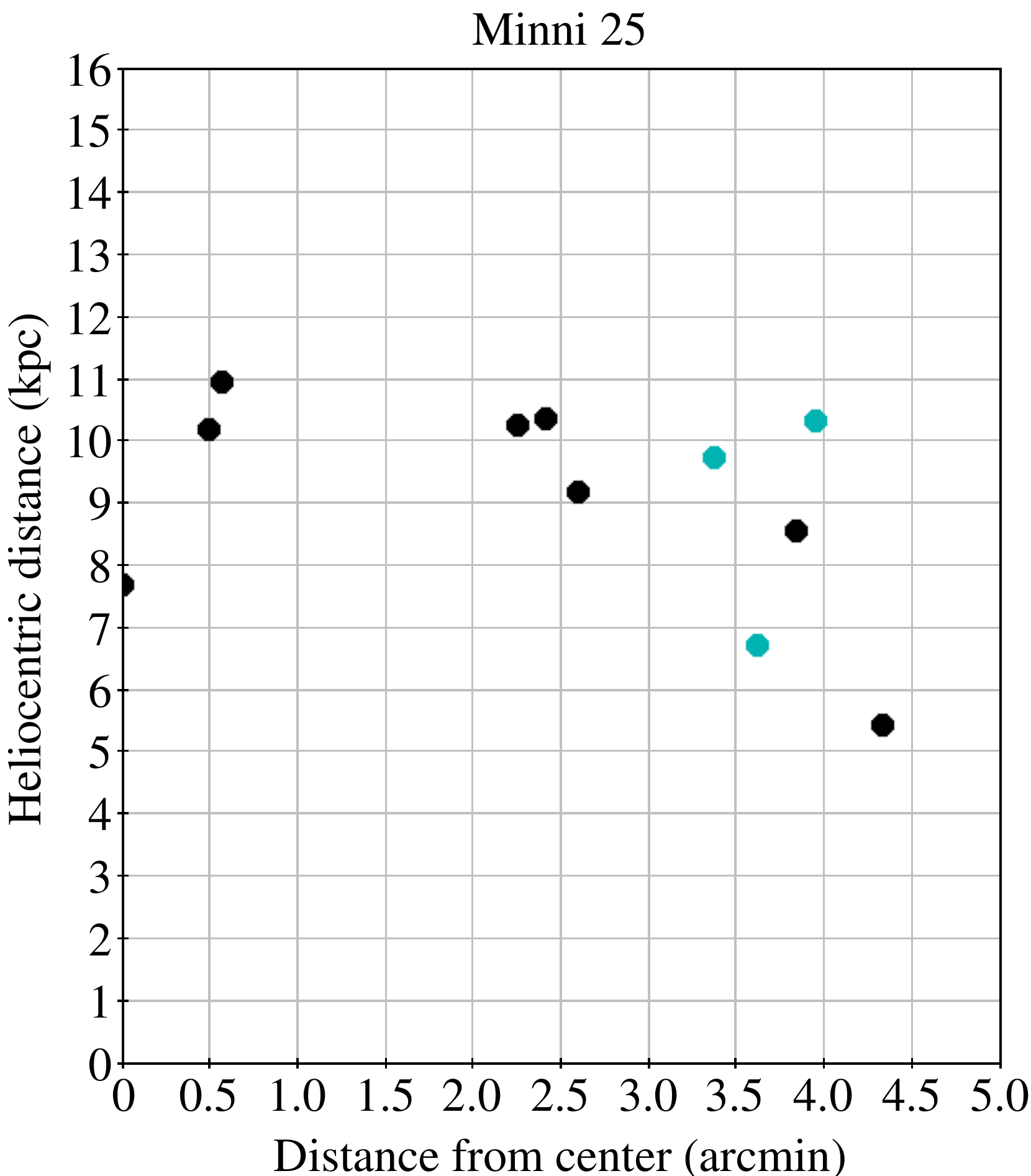}
\includegraphics[scale=.17]{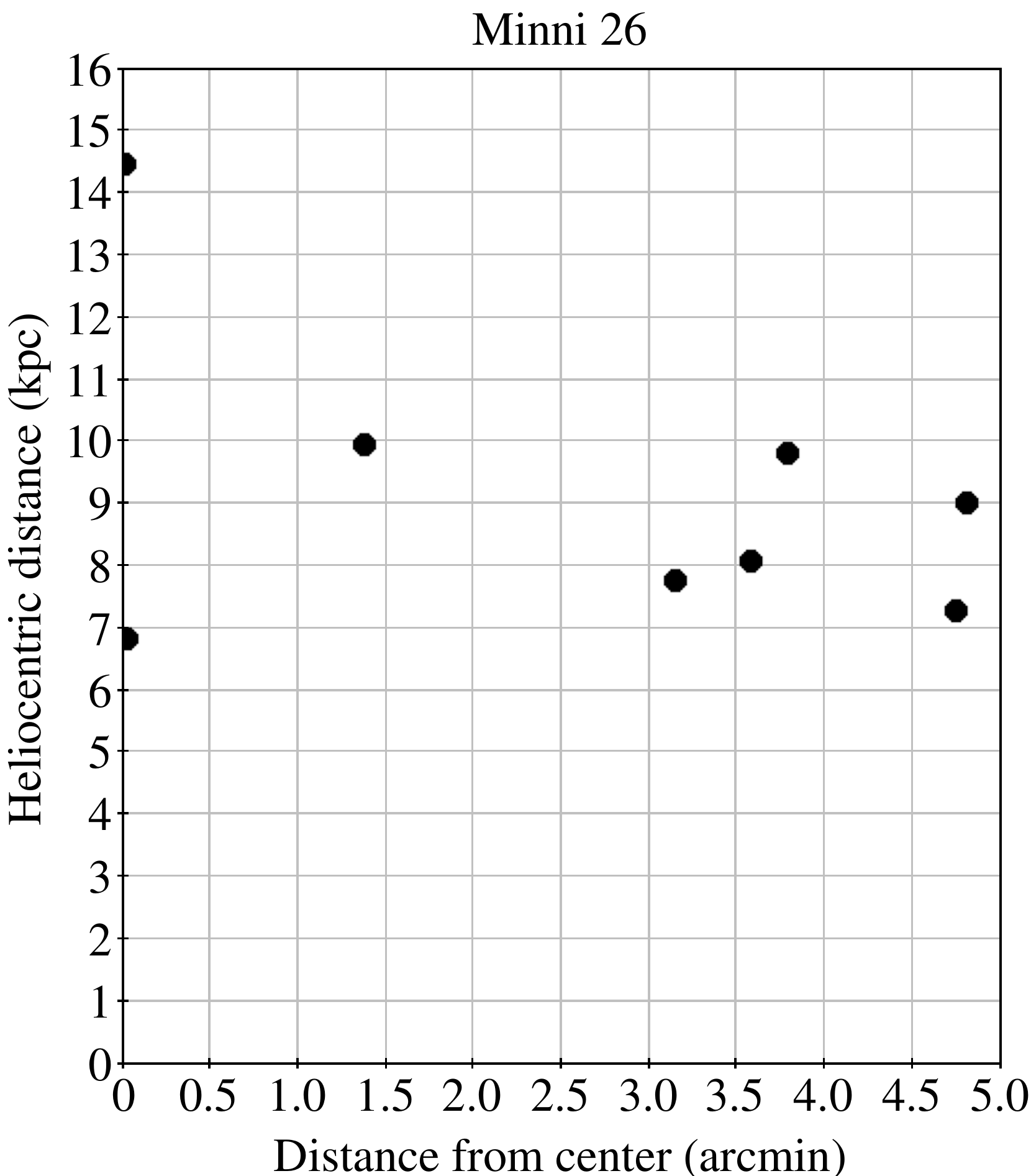}
\includegraphics[scale=.17]{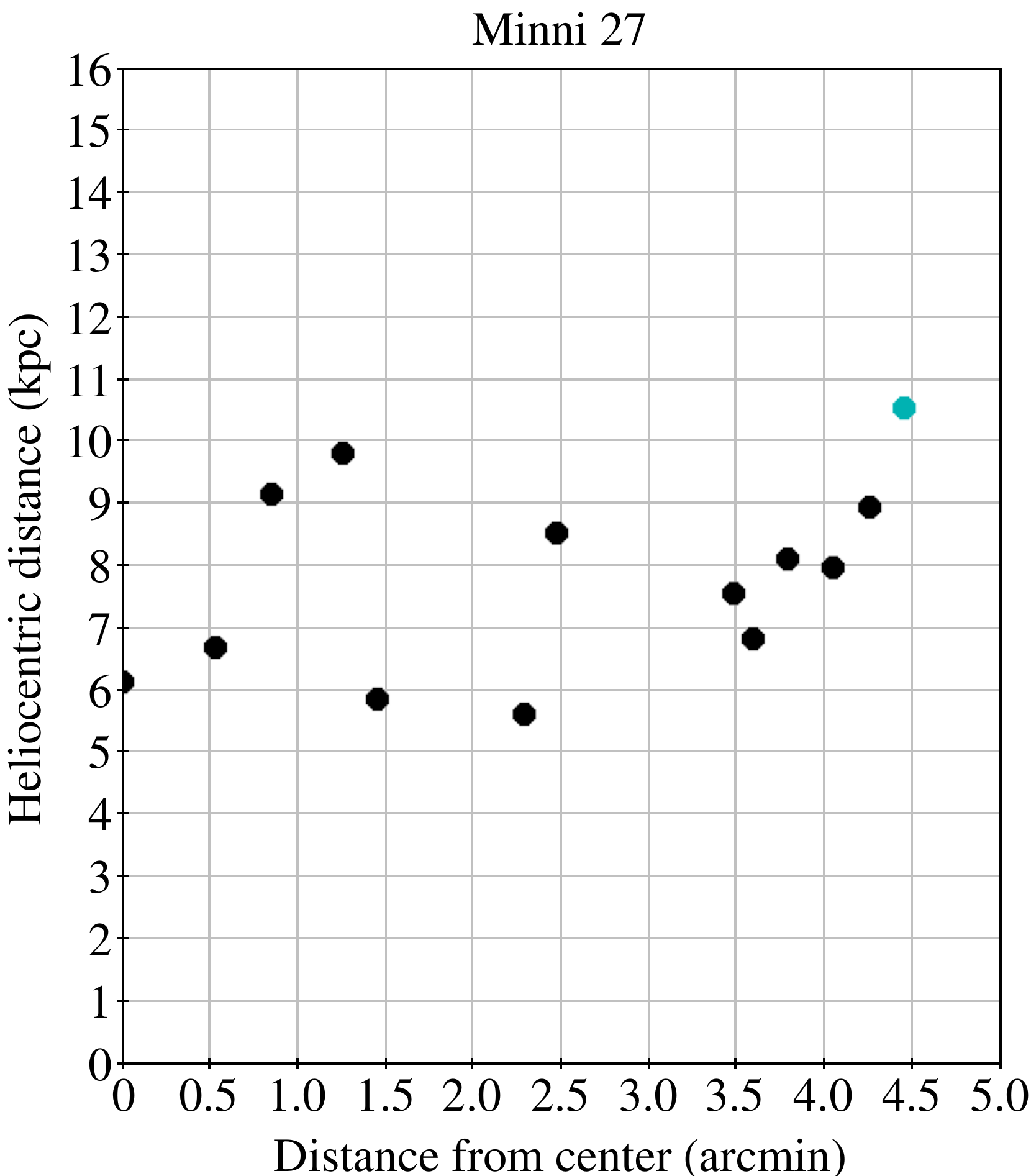}
\includegraphics[scale=.17]{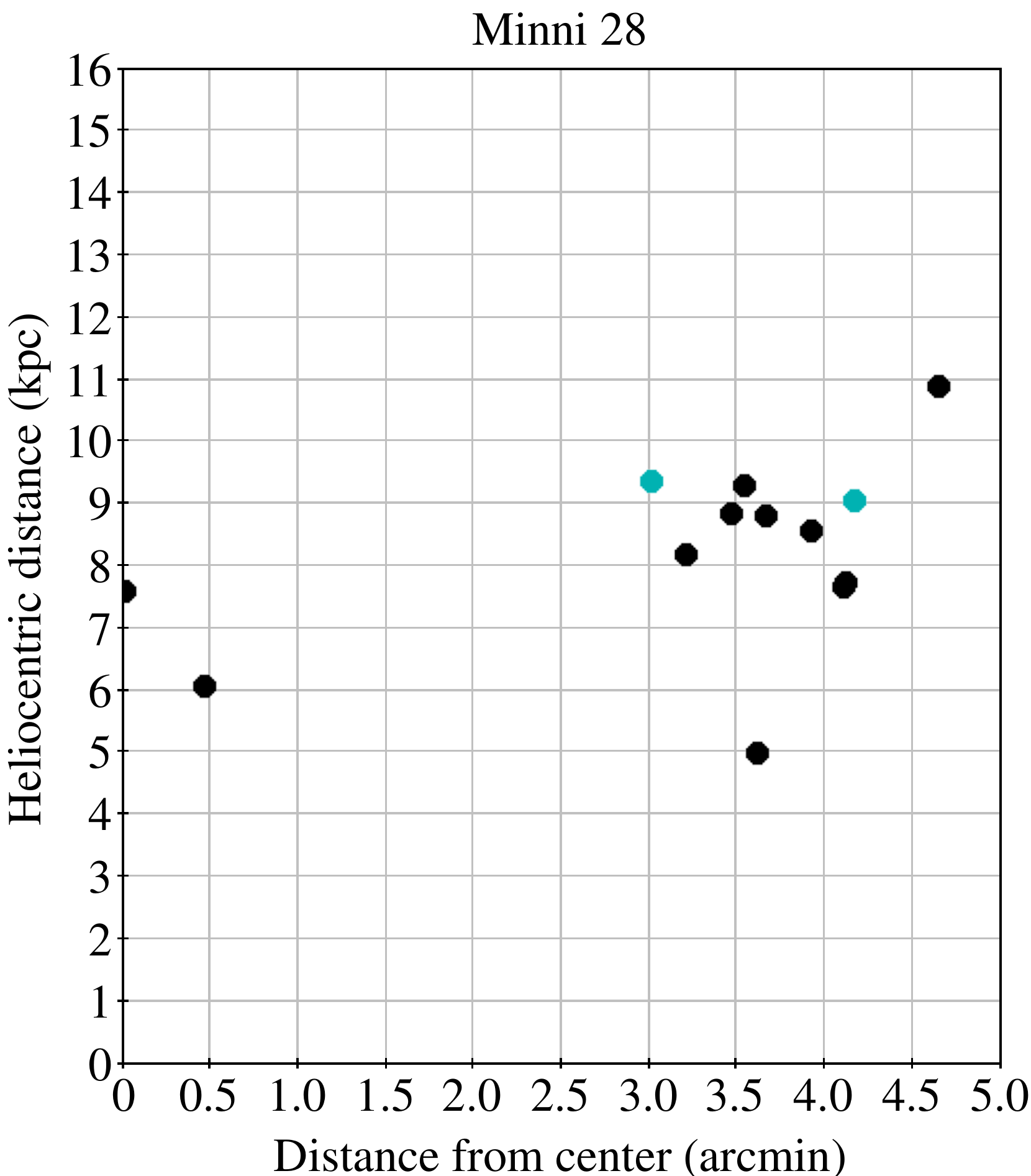}
\includegraphics[scale=.17]{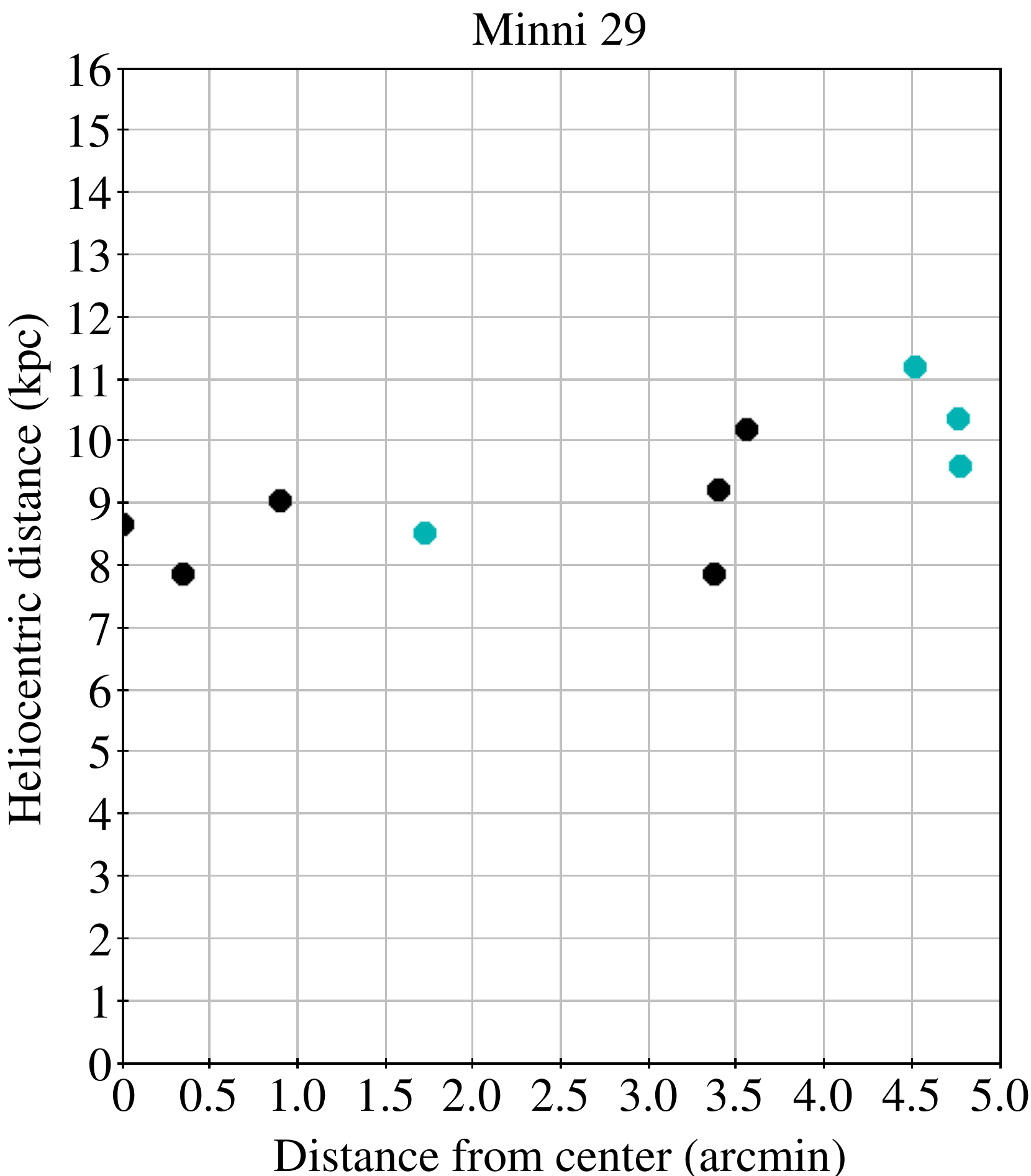}
\includegraphics[scale=.17]{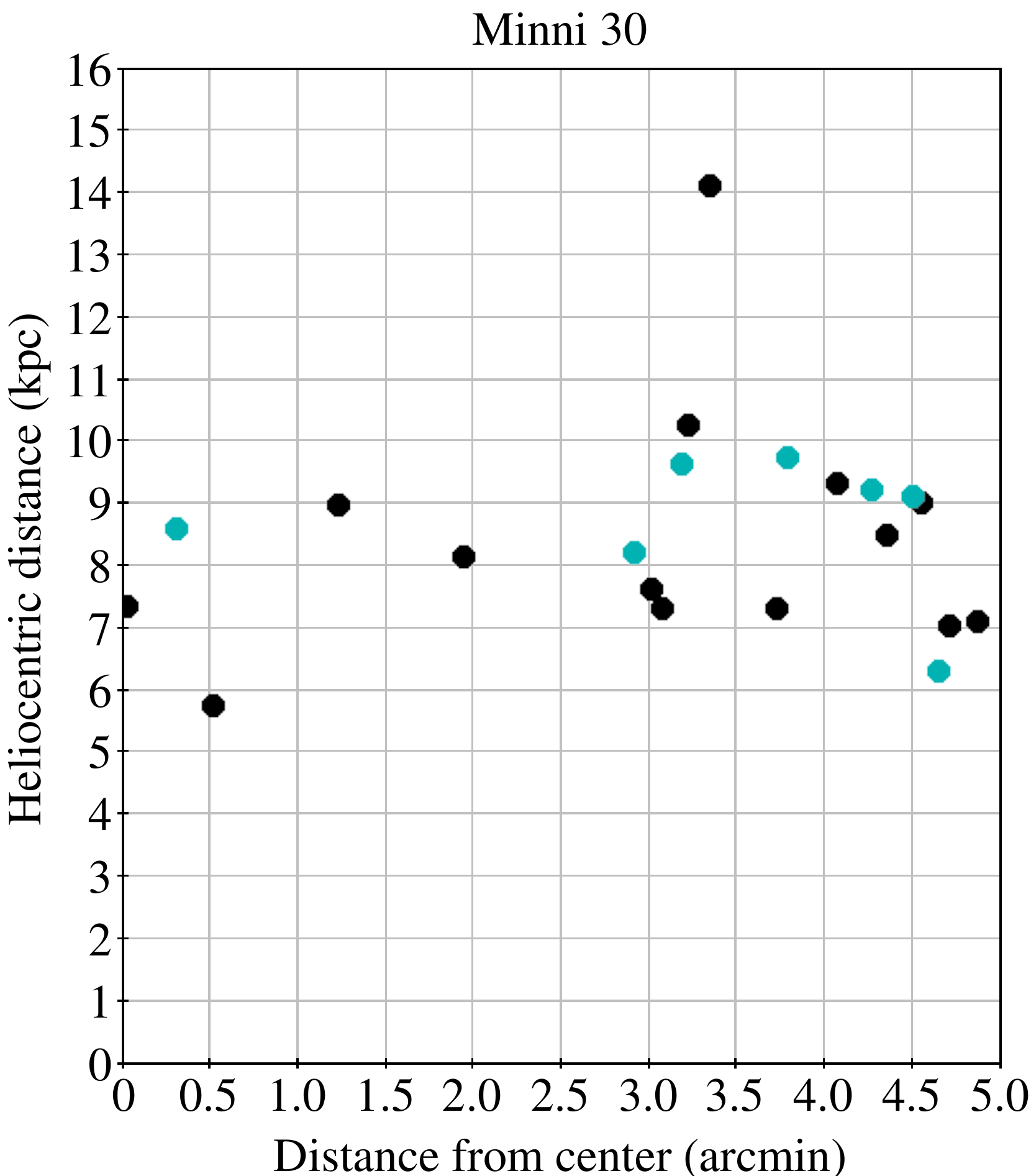}
\includegraphics[scale=.17]{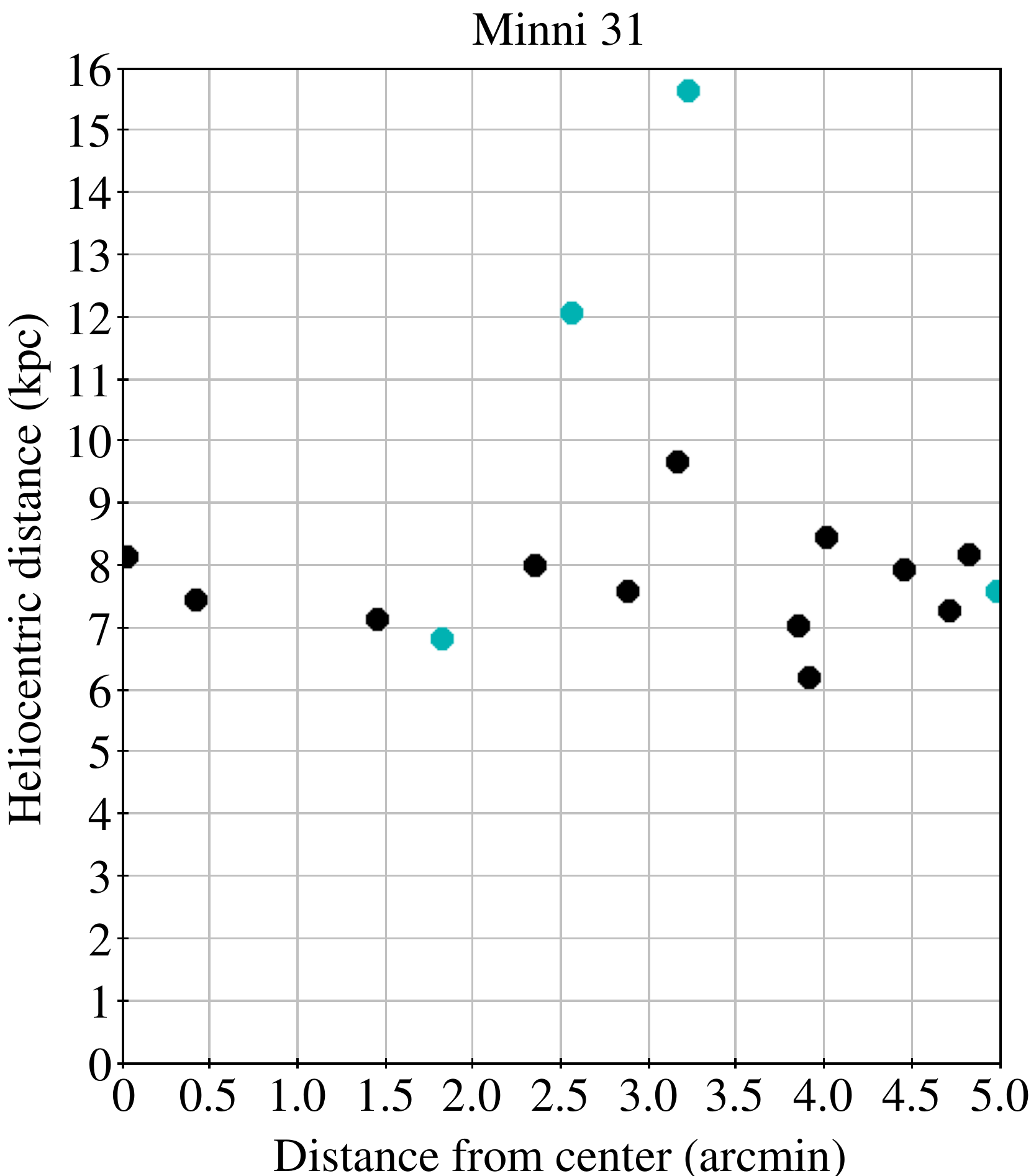}
\includegraphics[scale=.17]{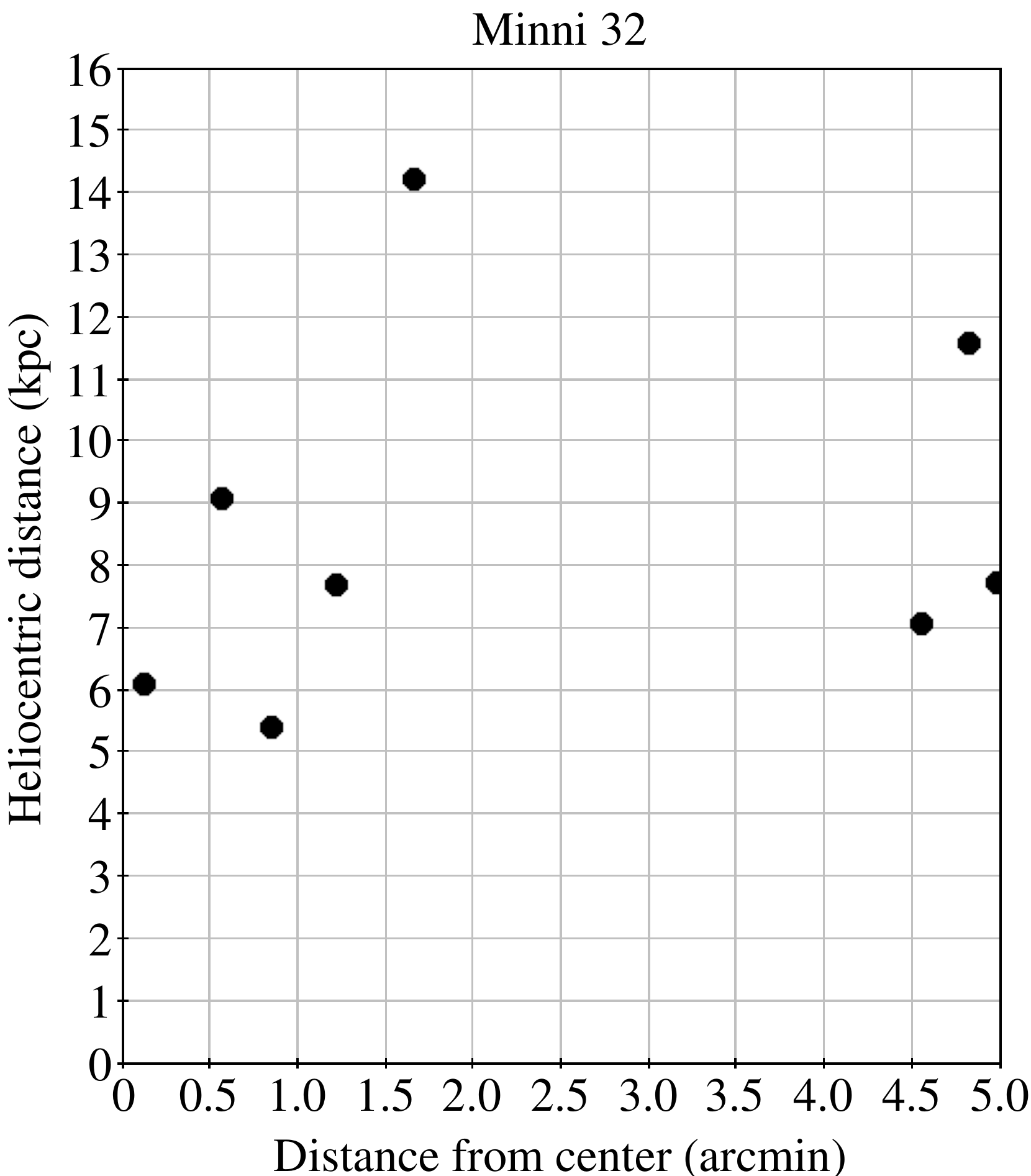}
\includegraphics[scale=.17]{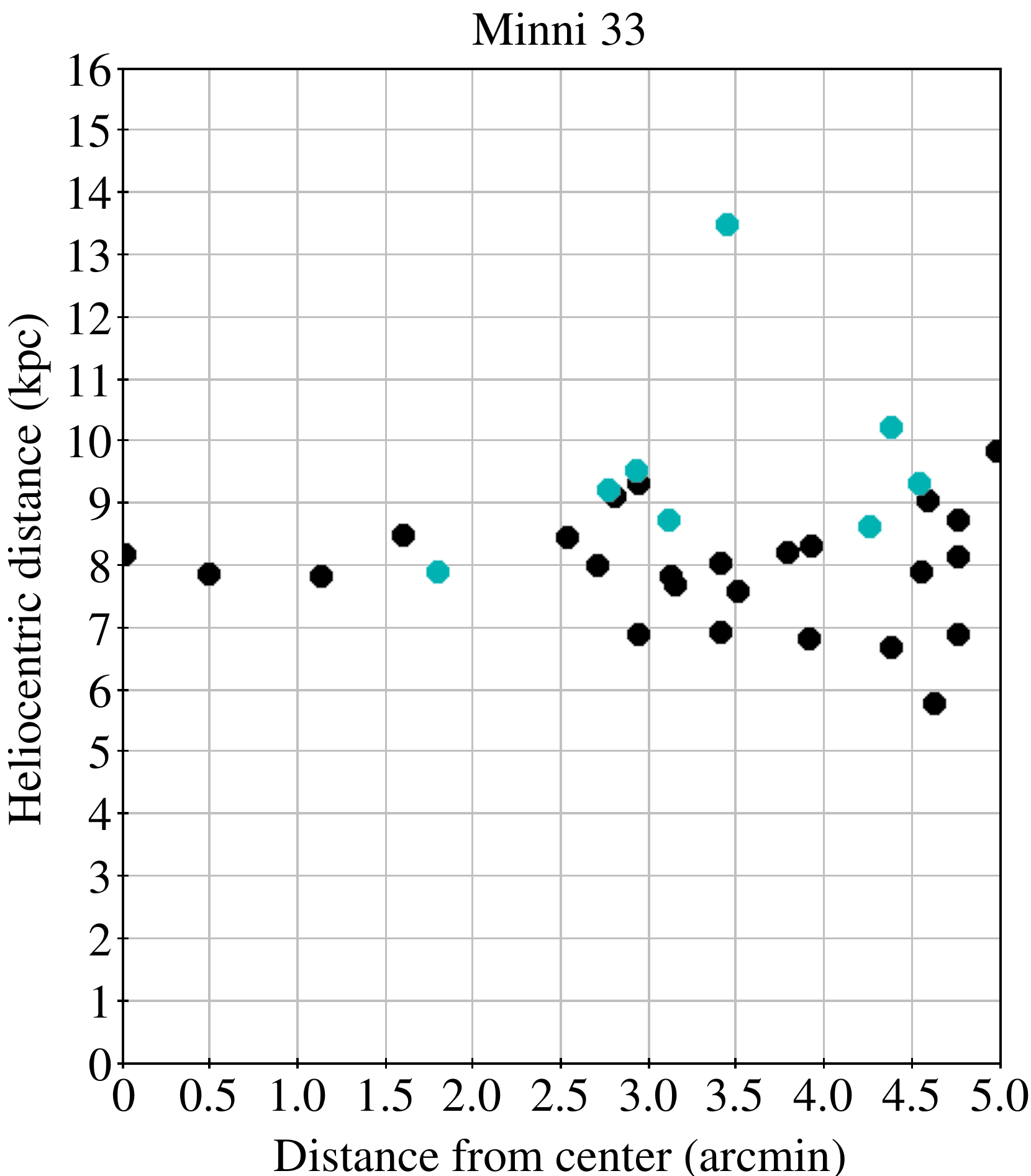}
\includegraphics[scale=.17]{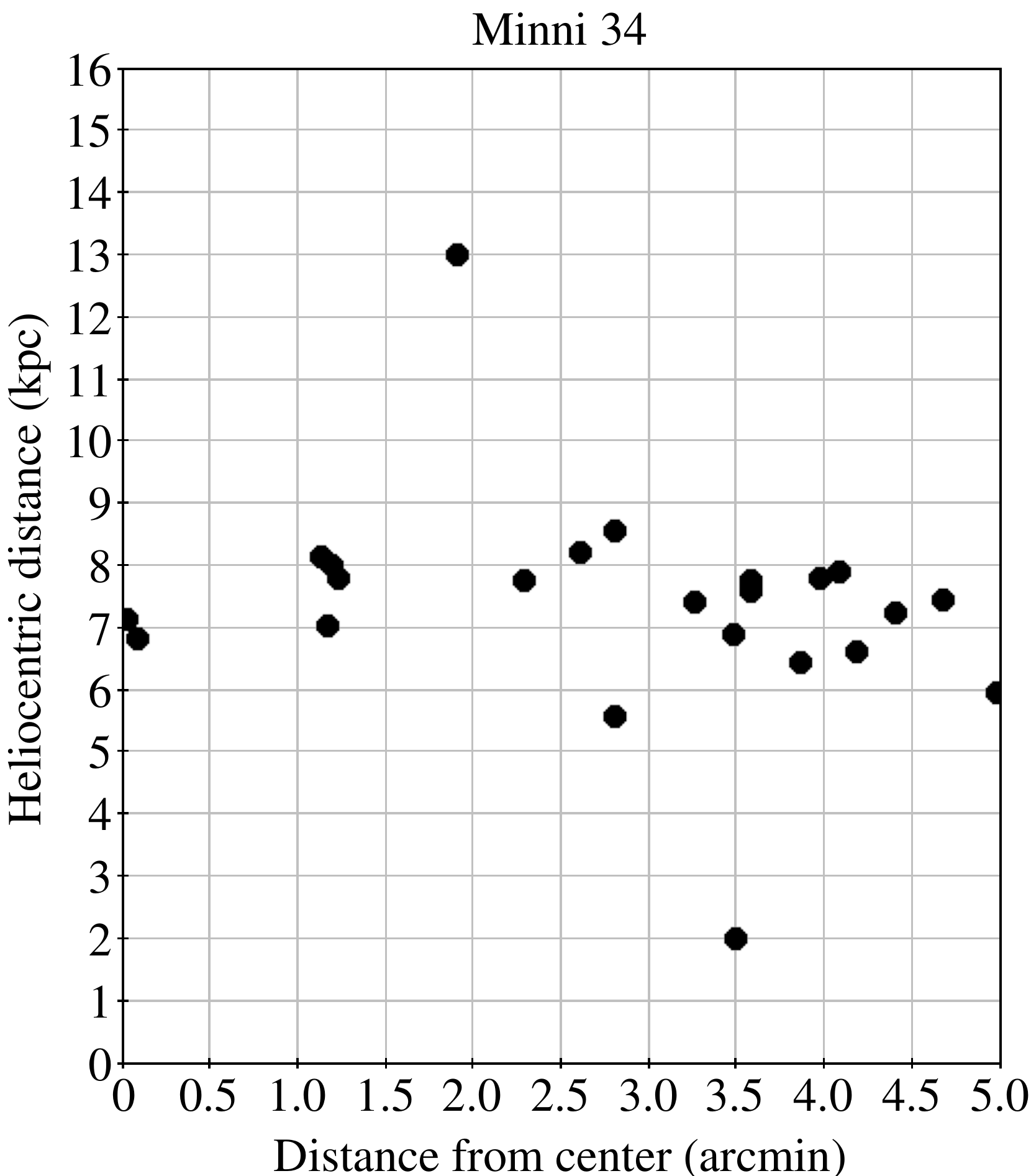}
\includegraphics[scale=.17]{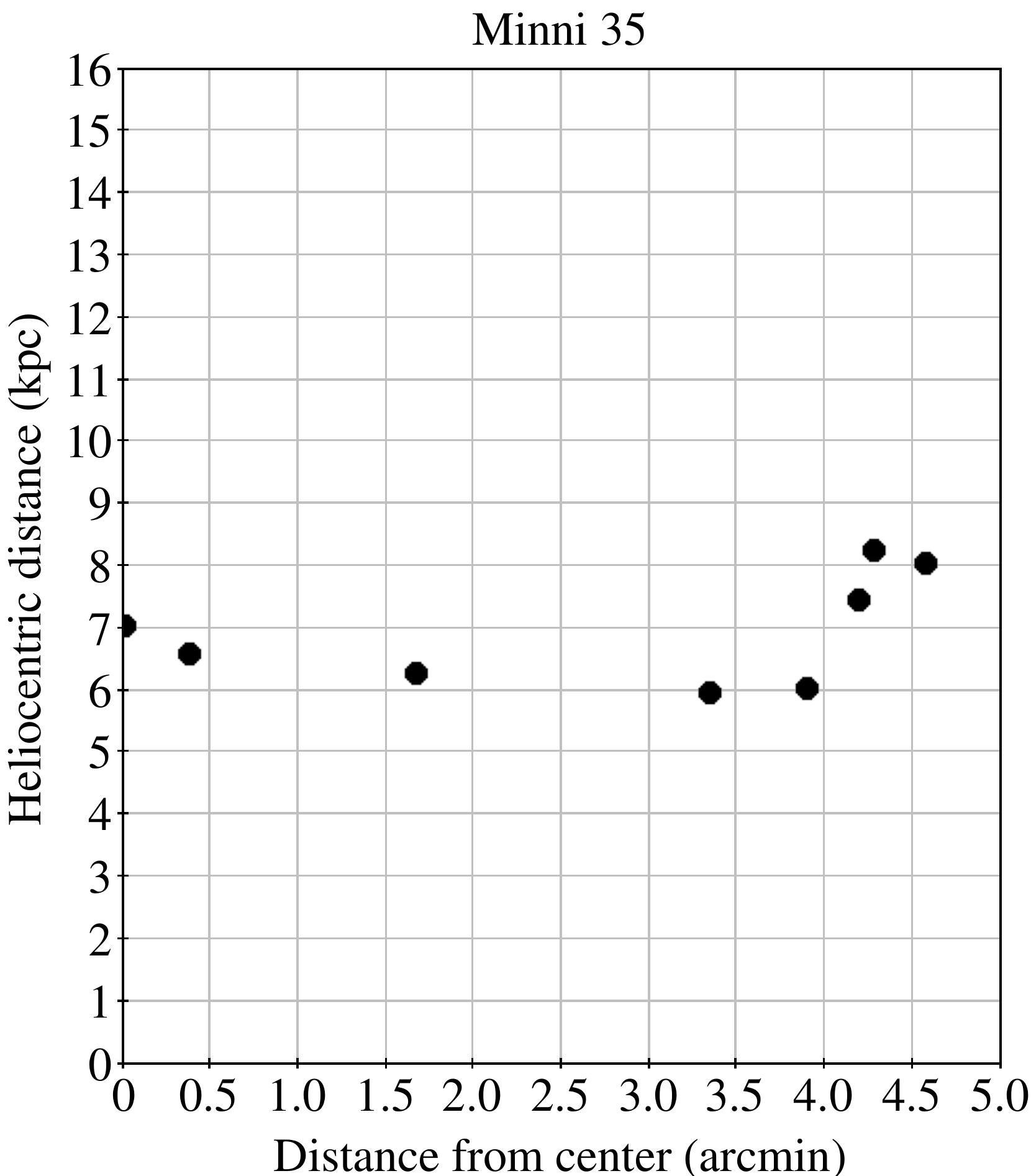}
\includegraphics[scale=.17]{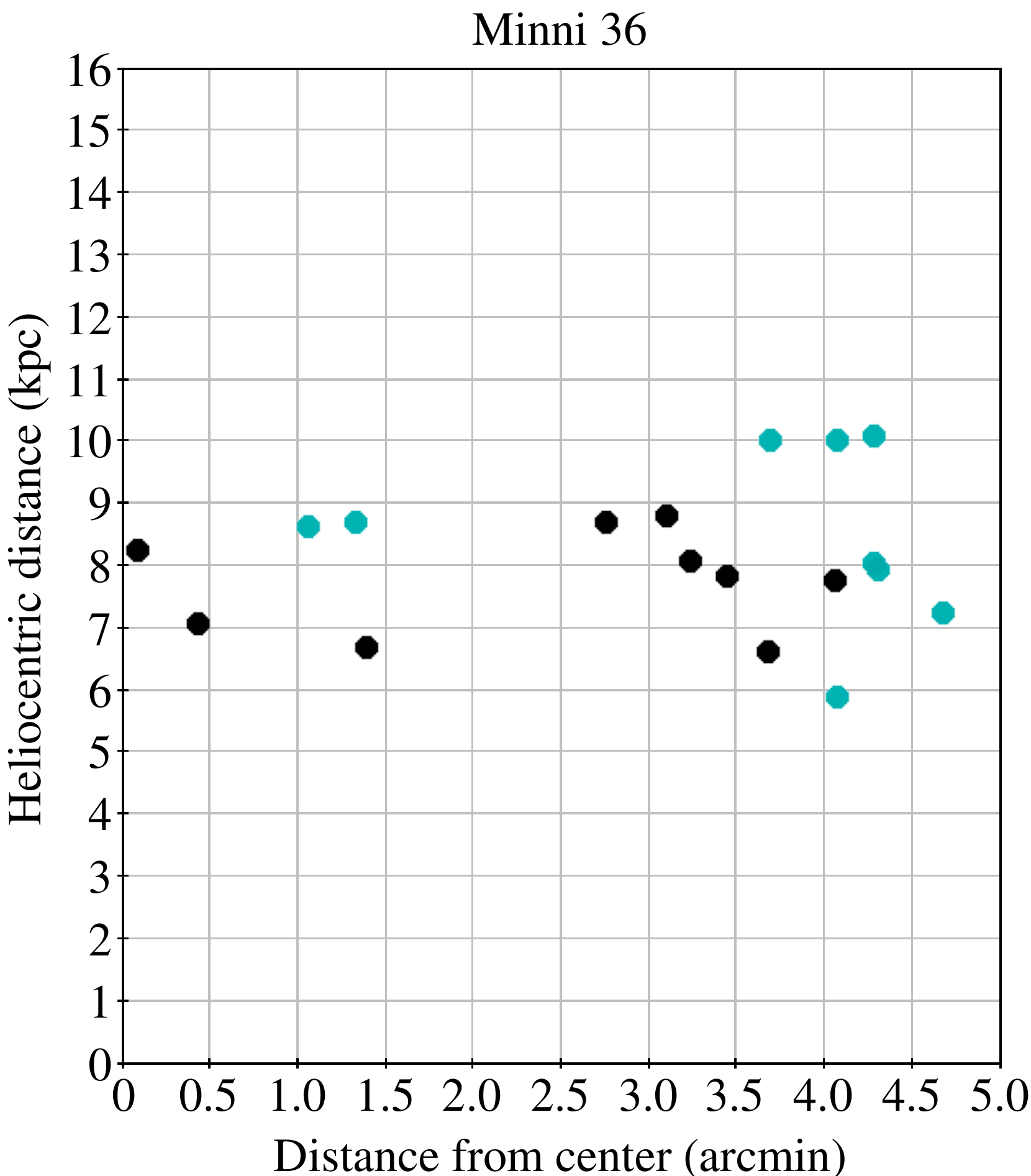}
\includegraphics[scale=.17]{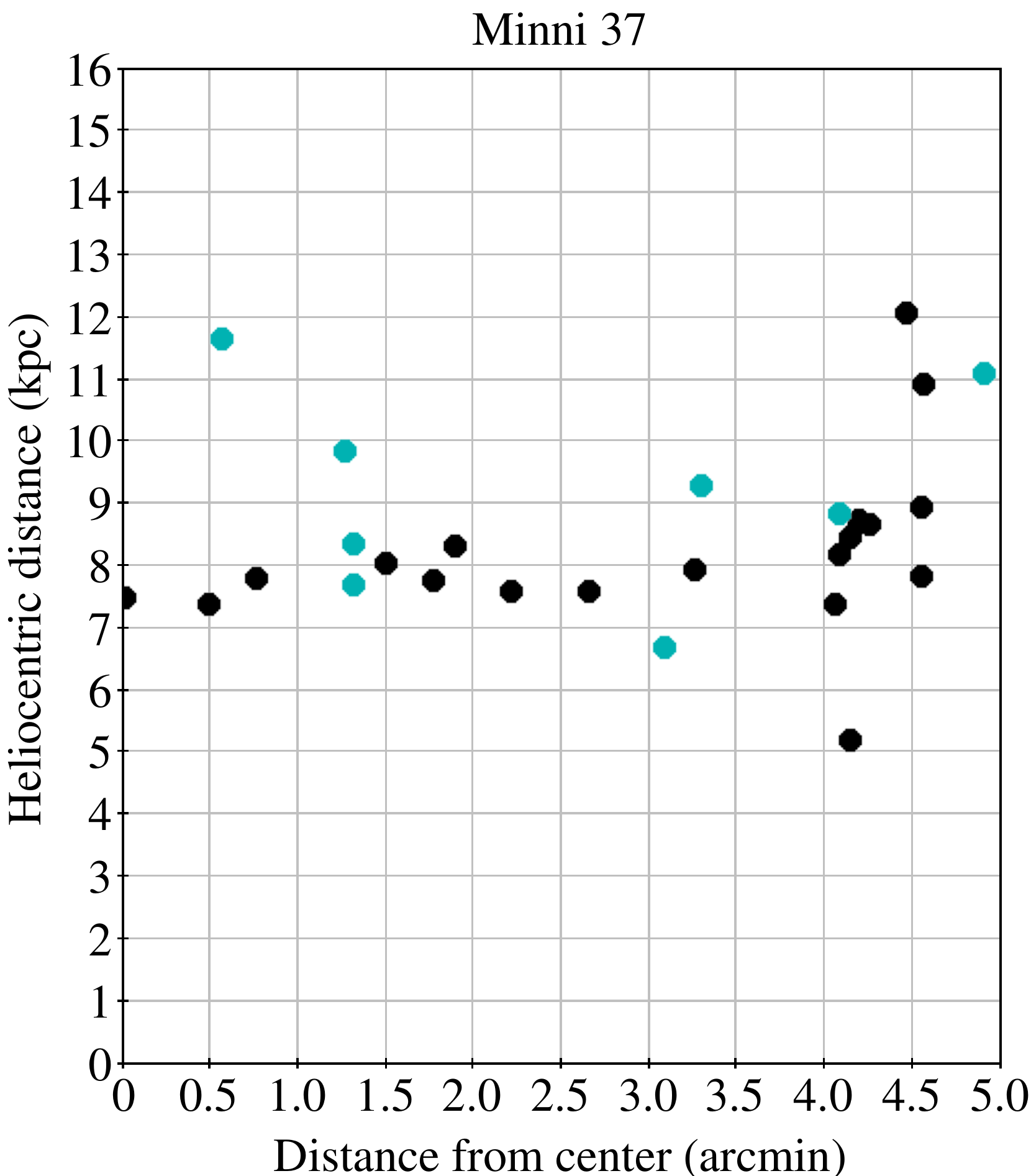}
\includegraphics[scale=.17]{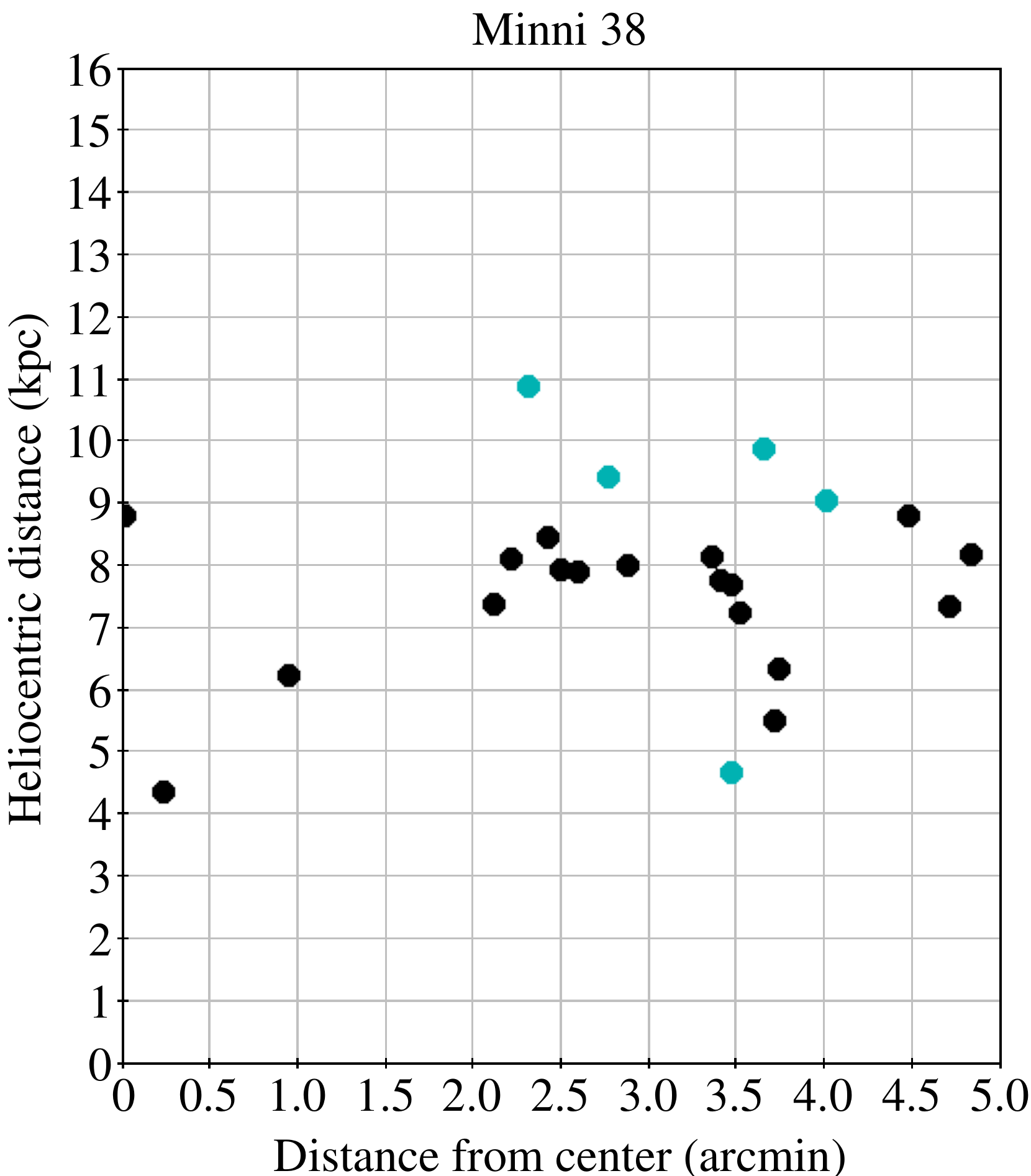}
\includegraphics[scale=.17]{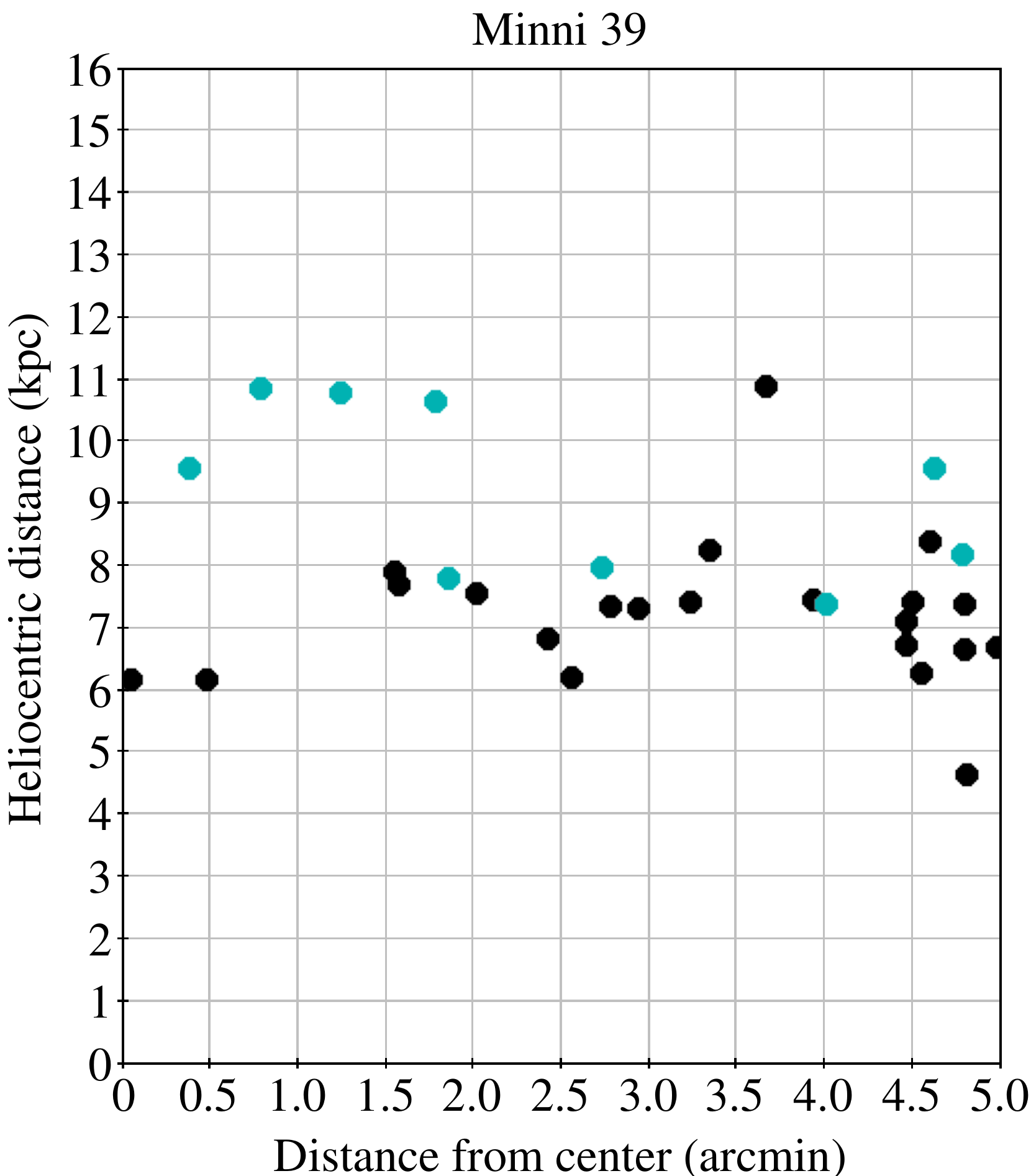}
\includegraphics[scale=.17]{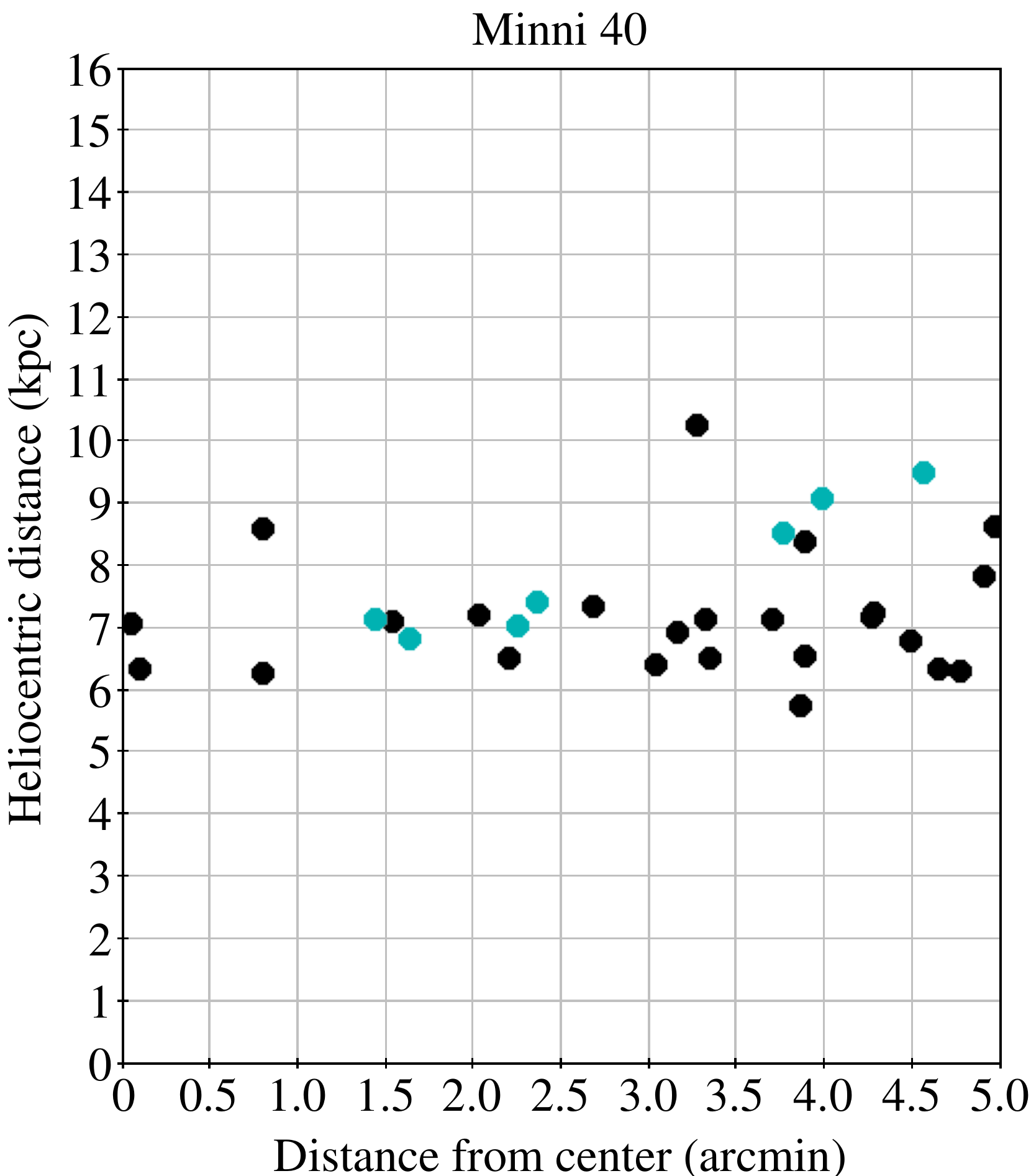}
\includegraphics[scale=.17]{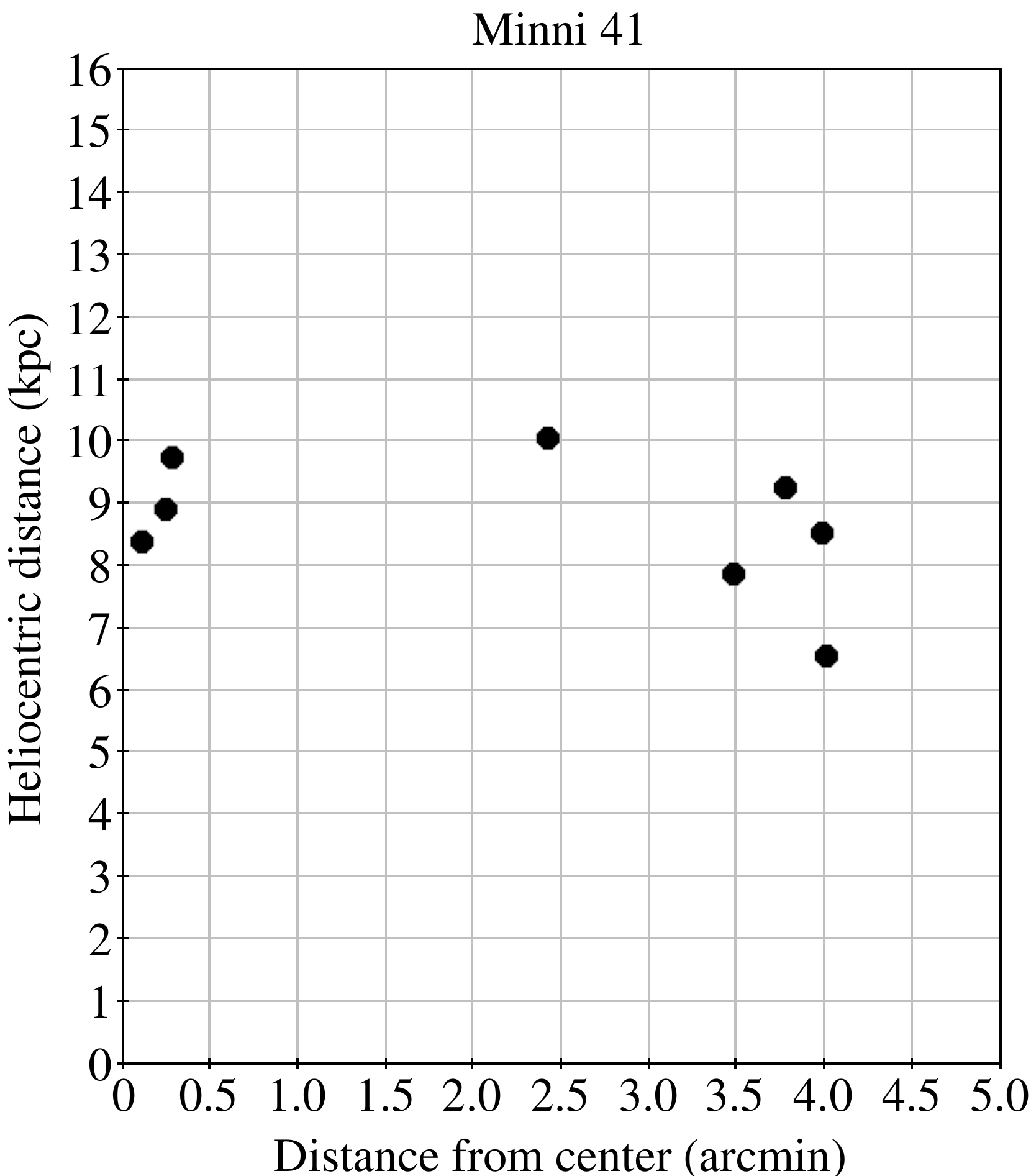}
\includegraphics[scale=.17]{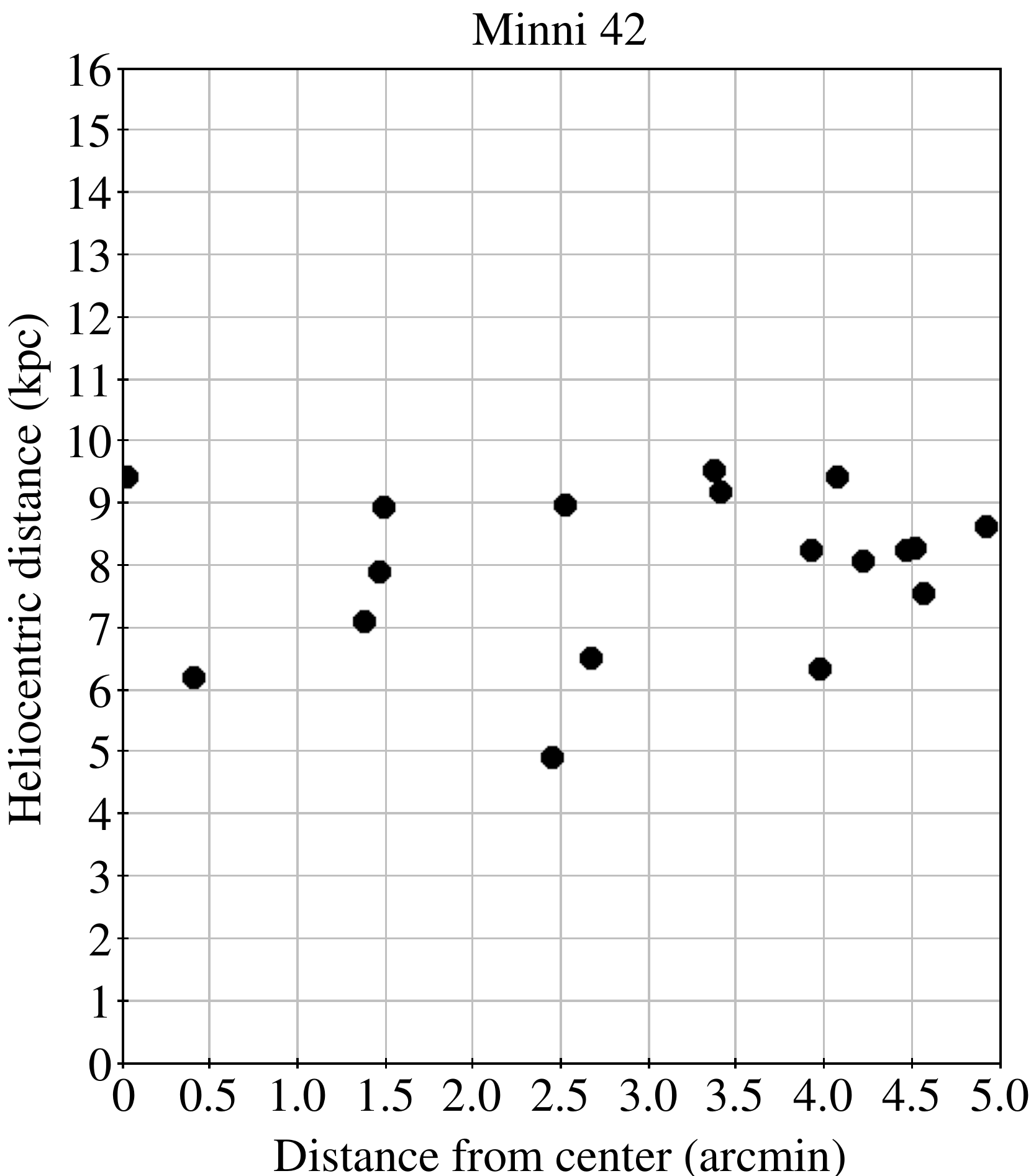}
\includegraphics[scale=.17]{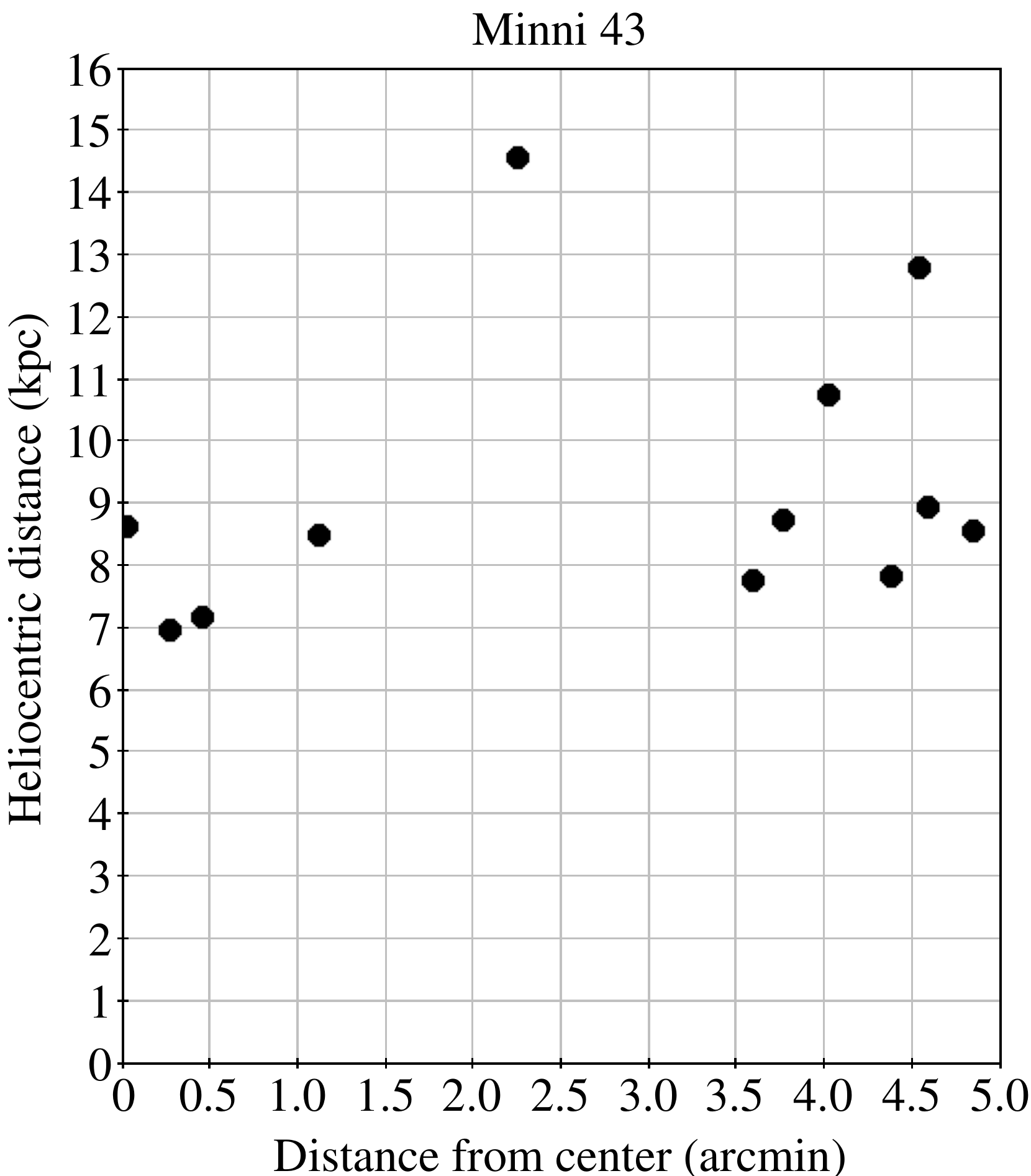}
\includegraphics[scale=.17]{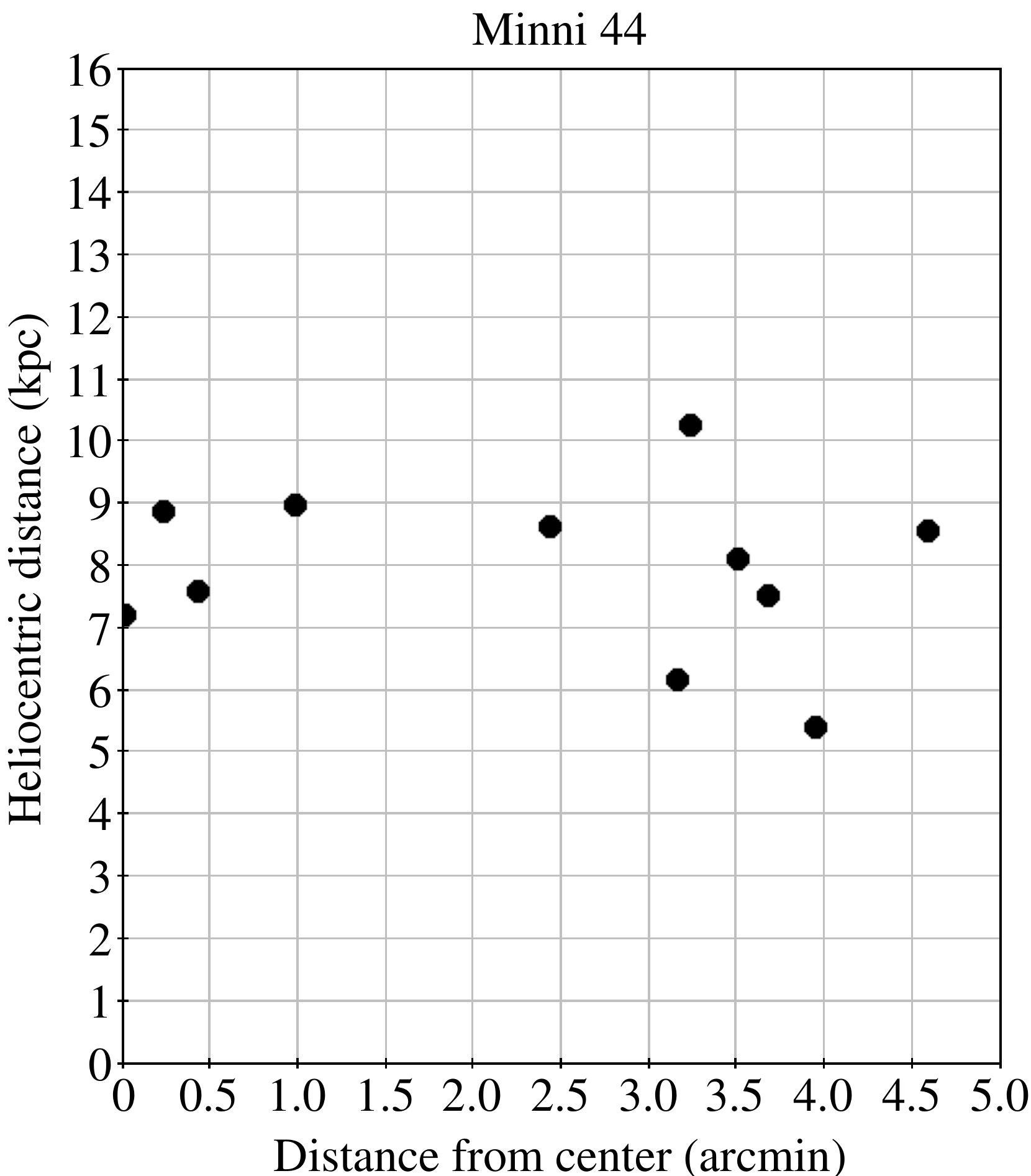}
\caption{Distances measured using individual RR Lyrae (black circles) and Mira (light-blue circles) variable stars in the fields of the GC candidates. On average, assuming that they are all field stars, the RR Lyrae distances appear to be systematically shorter than the Mira distances. This is  because the brightest Miras (with $K_s < 11$ mag)  are saturated in the VVV photometry.
\label{fig5}  }
\end{figure*}

Figure \ref{fig5} shows the results of those distances derived using both kinds of variable stars. Clearly, not all these variable stars are members of the clusters. The distance estimates exhibit several spatial concentrations of two or more stars in some selected cases. The probability to find a chance grouping of 2 or more stars in areas that are used for statistical decontamination is discussed in \cite{minniti18b}.
Among those, two RR\,Lyrae stars that grouped together spatially could be observed in Minni\,24, Minni\,25, Minni\,28, Minni\,30, Minni\,33, Minni\,34, Minni\,37, Minni\,38, Minni\,39, Minni\,40 and Minni\,43. Of these, the majority are deemed to be good star cluster candidates, except for Minni\,25 and Minni\,43 that have been discarded (Table \ref{tab1}). \\

\section{Proper motion analysis}

\begin{figure*}
\includegraphics[scale=.285]{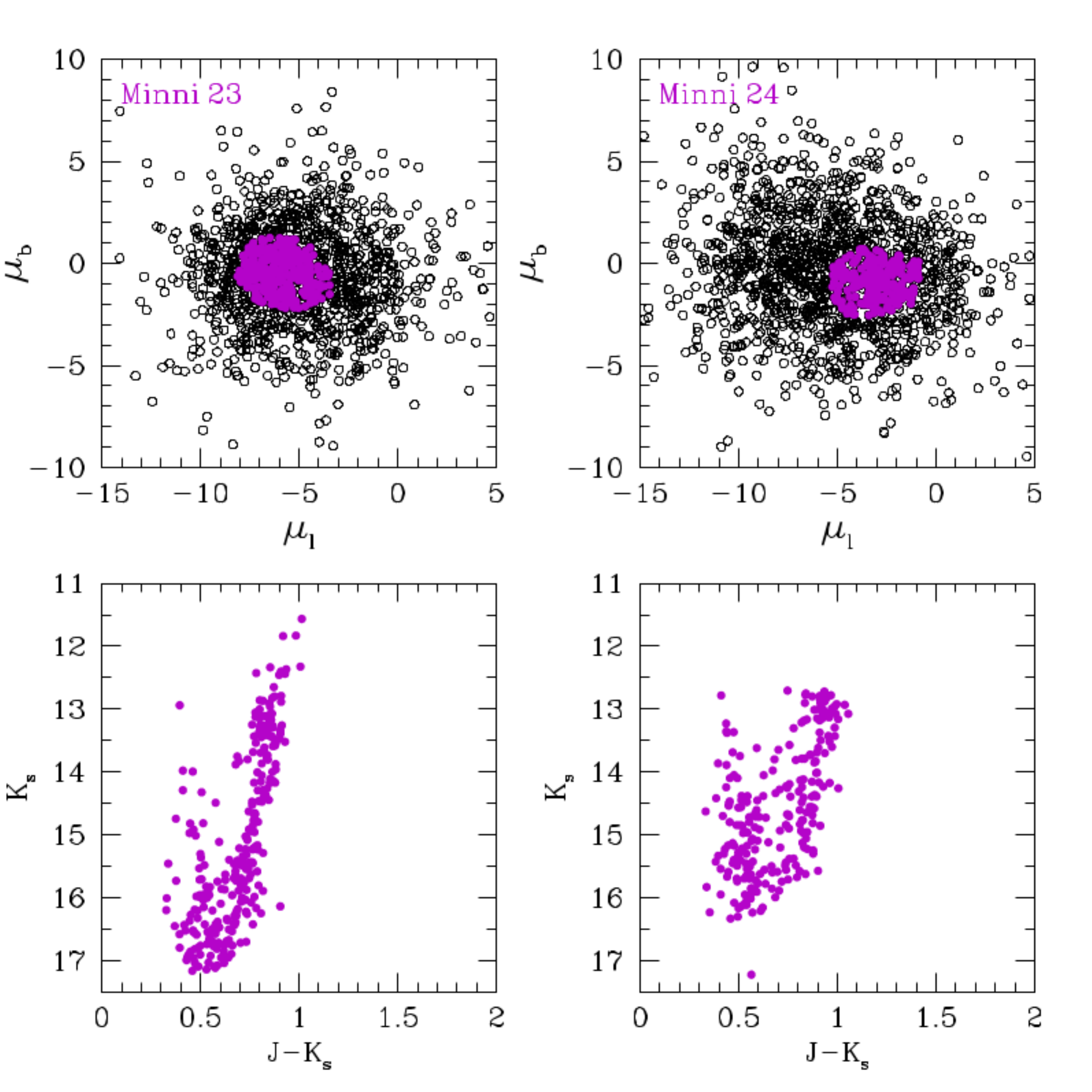}
\includegraphics[scale=.285]{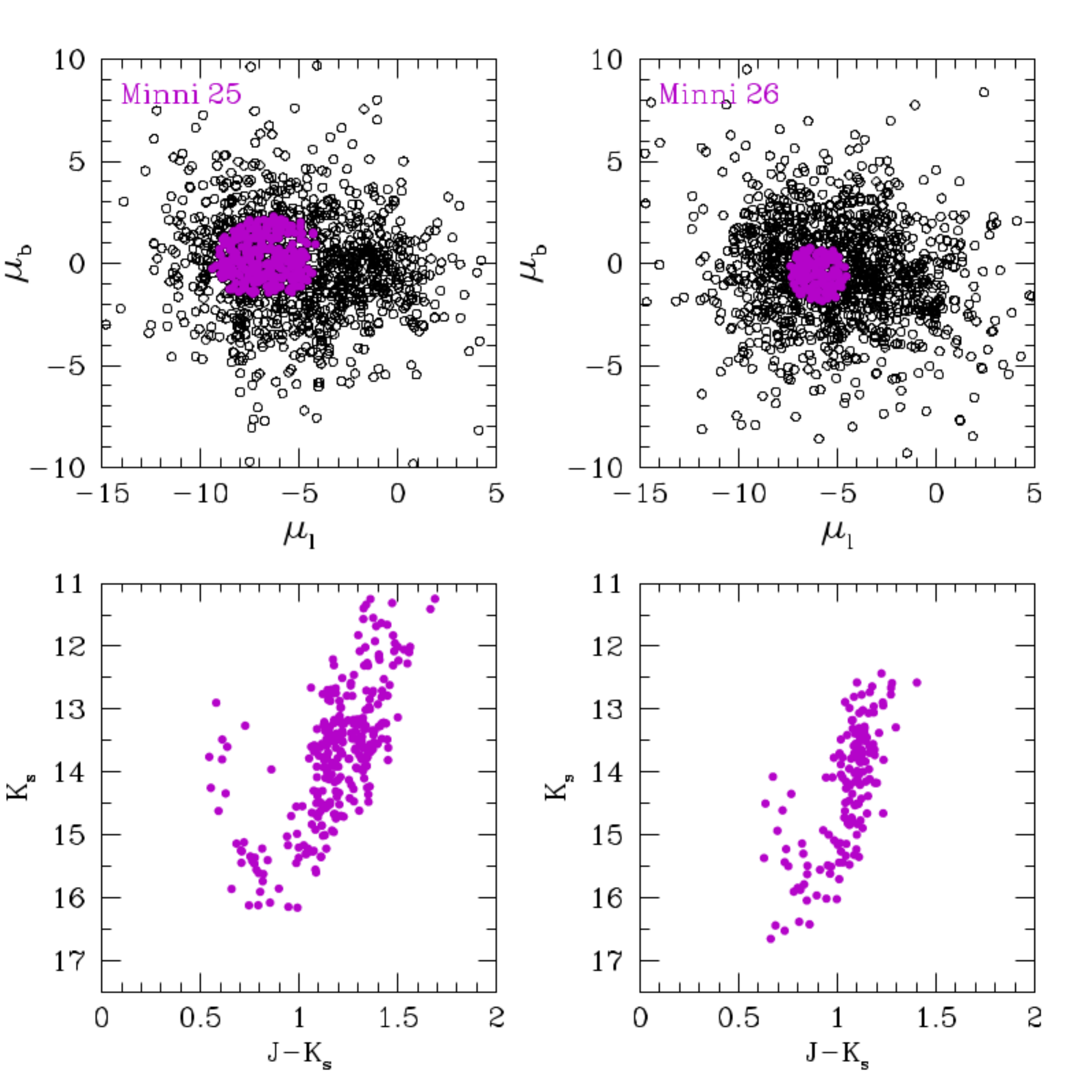}
\includegraphics[scale=.285]{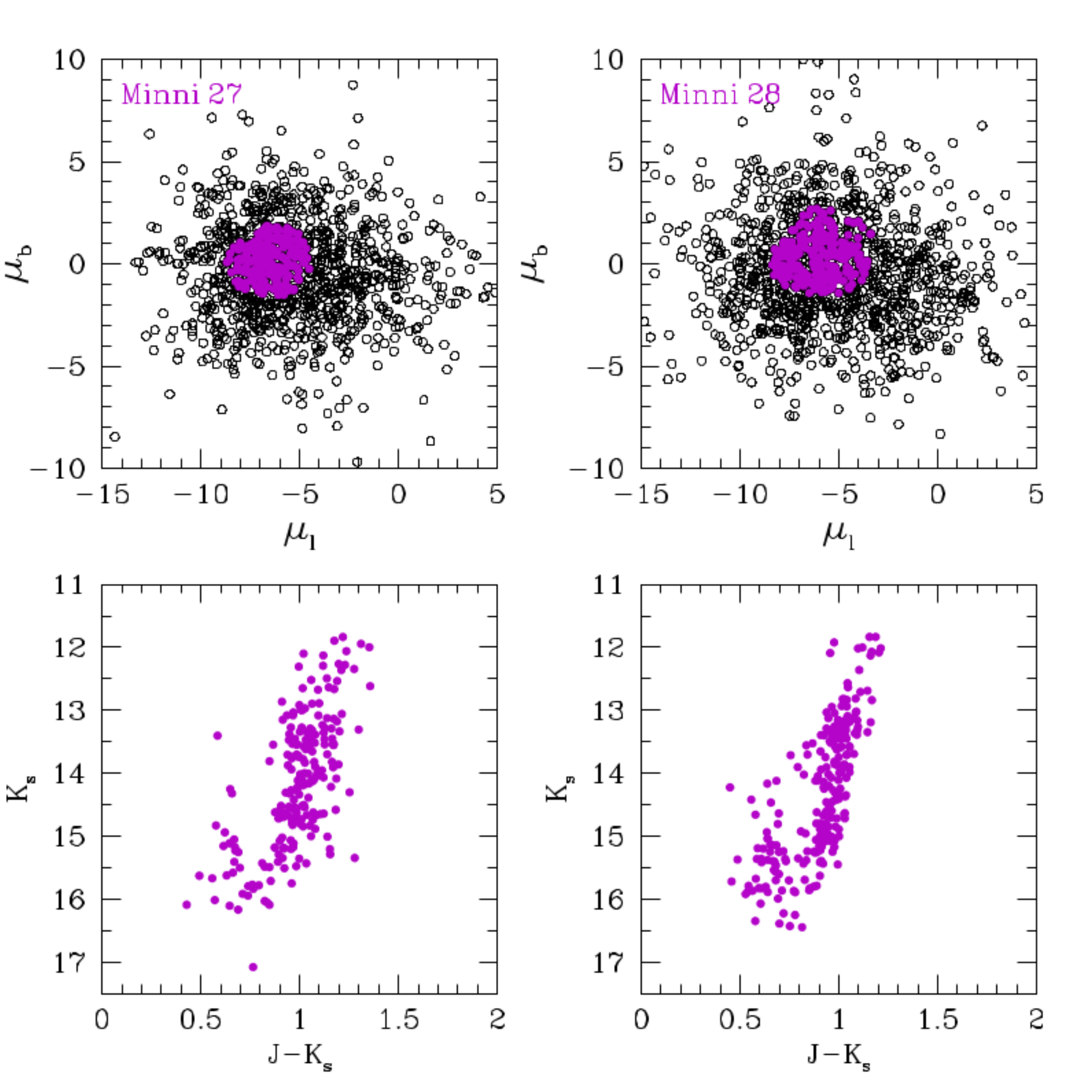}
\includegraphics[scale=.285]{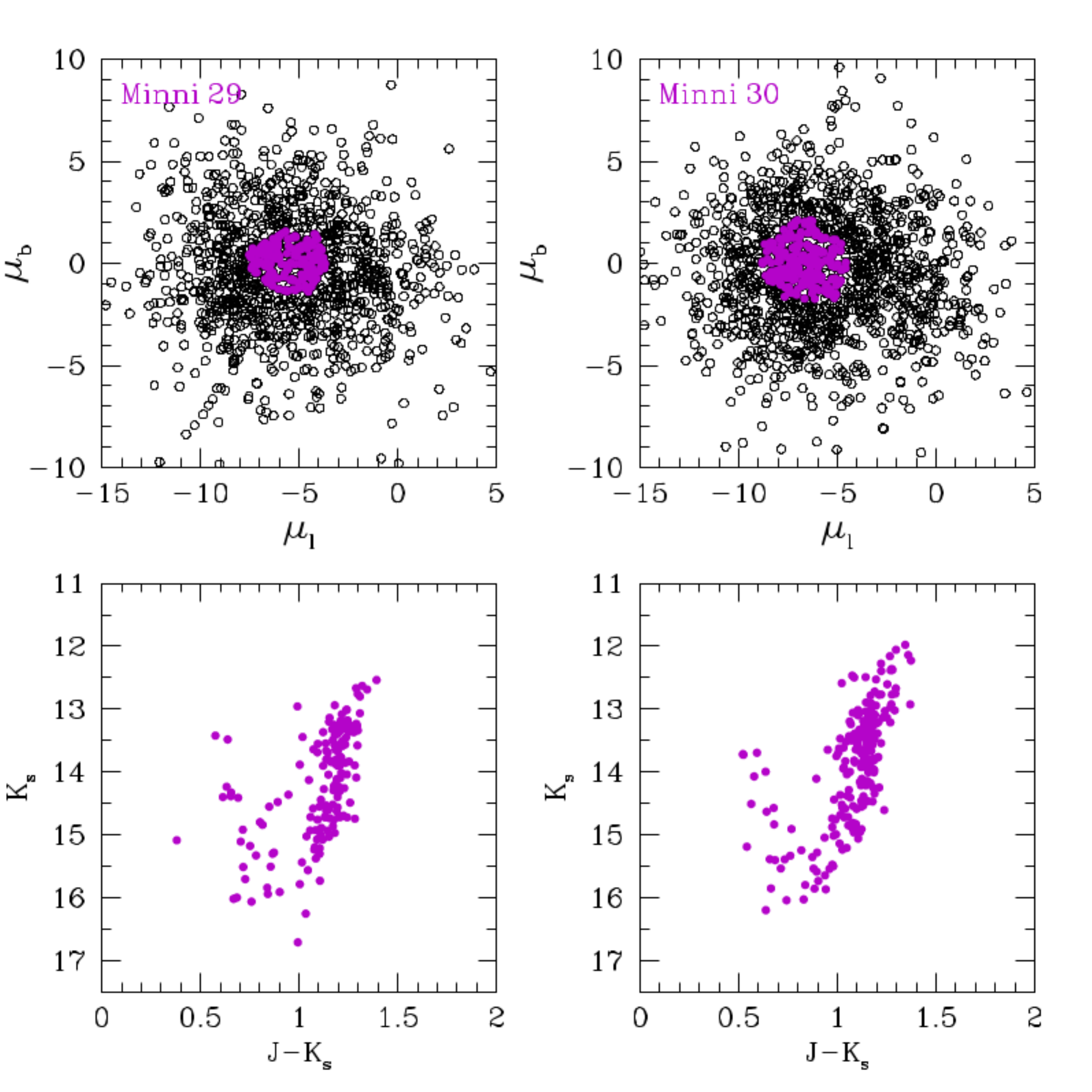}
\includegraphics[scale=.285]{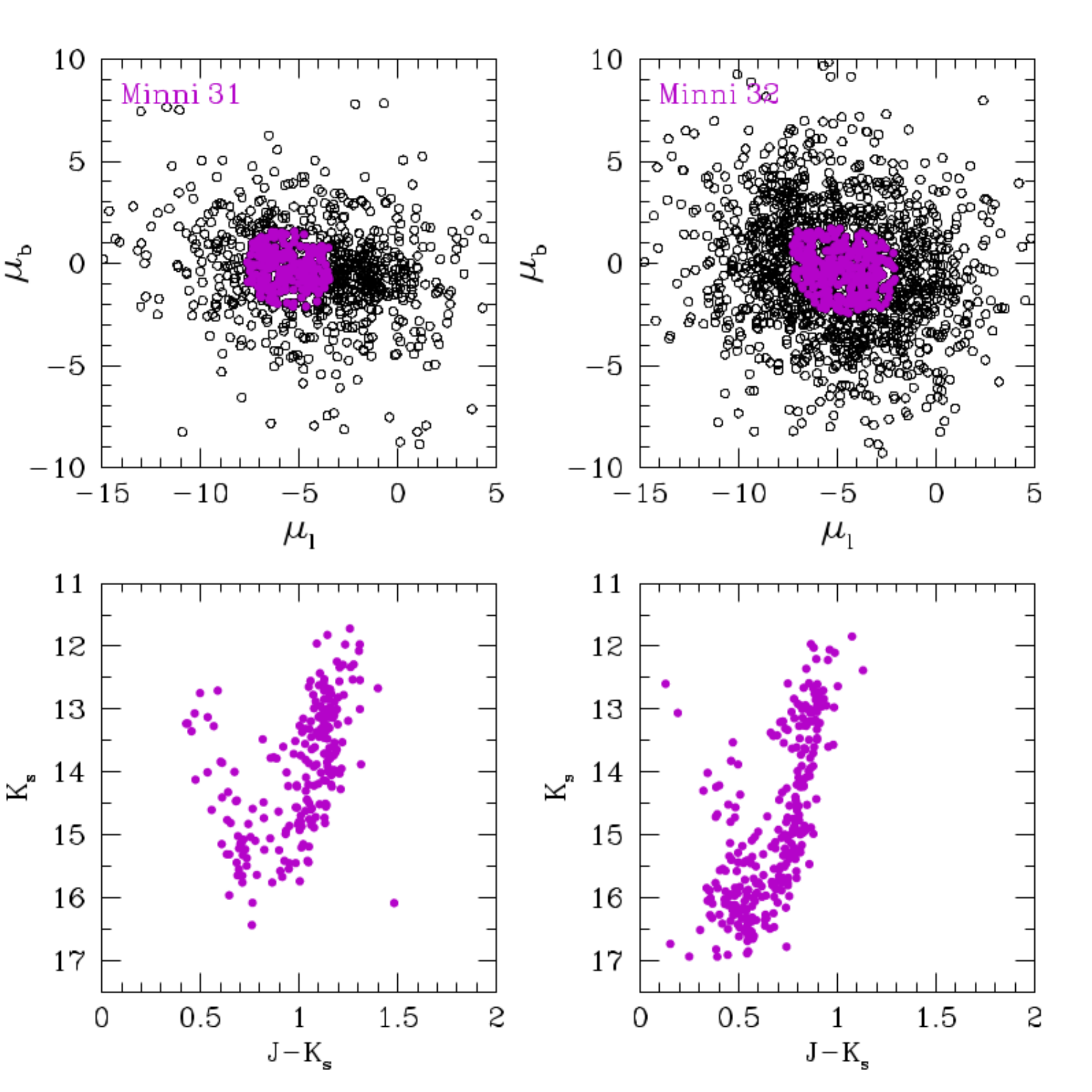}
\includegraphics[scale=.285]{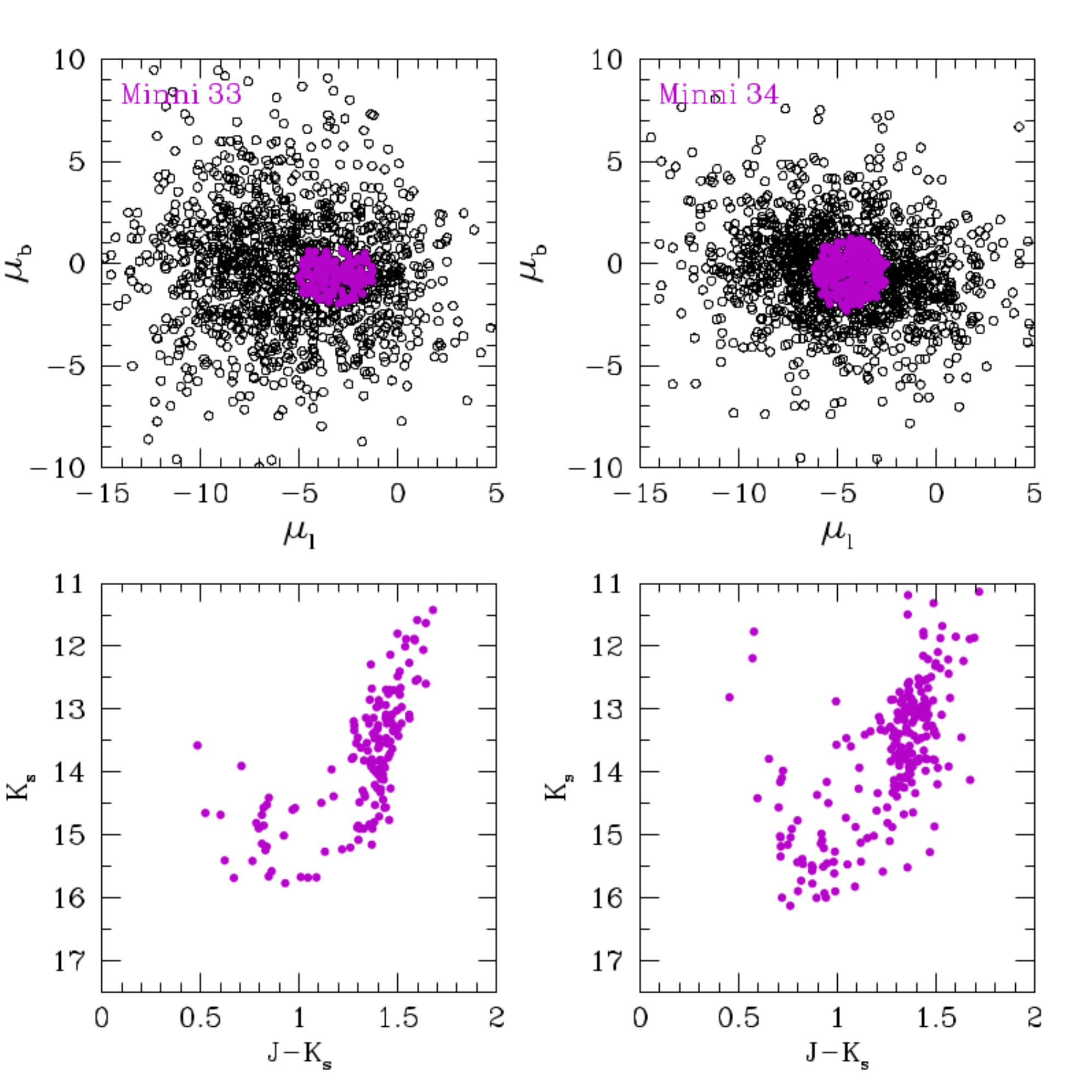}
\includegraphics[scale=.285]{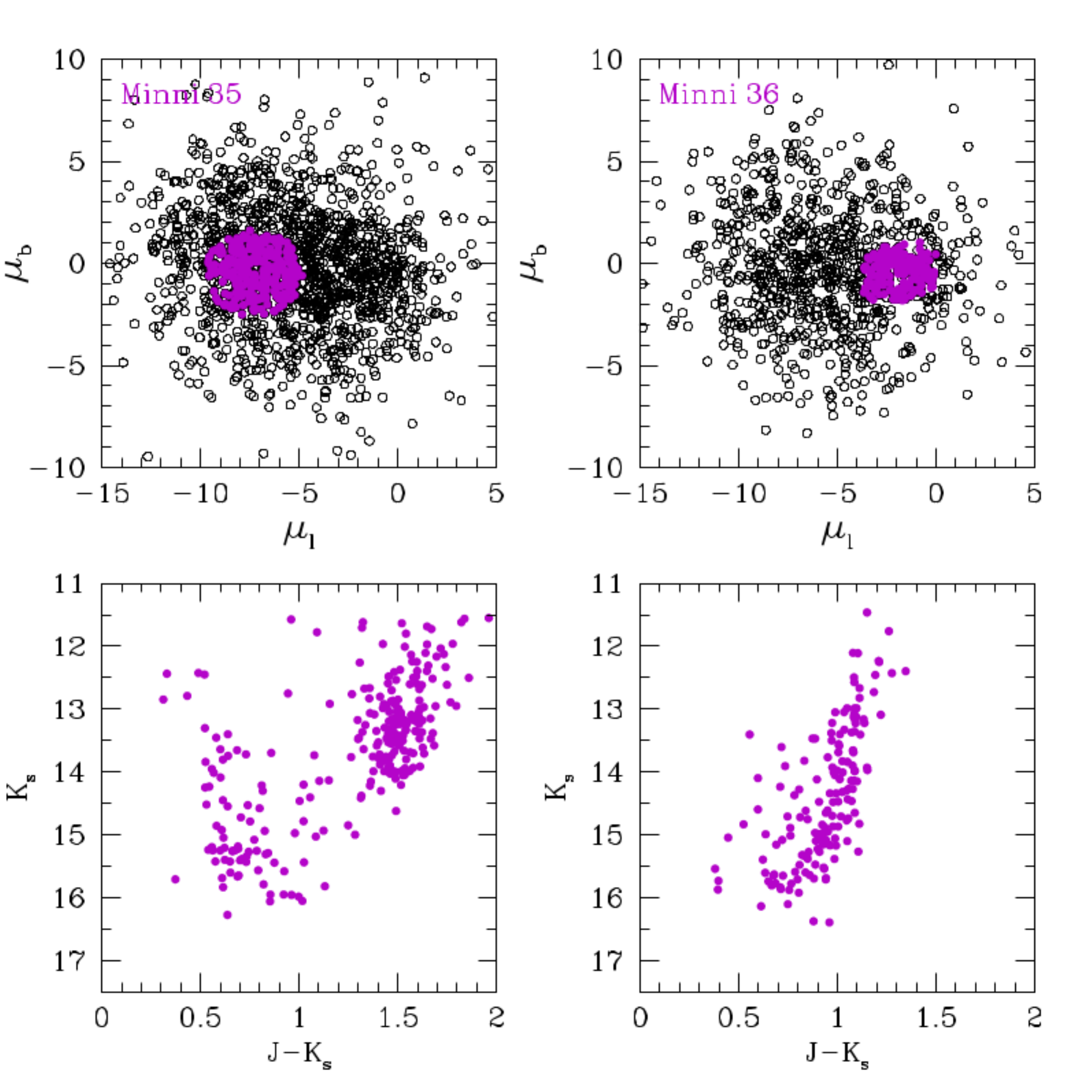}
\includegraphics[scale=.285]{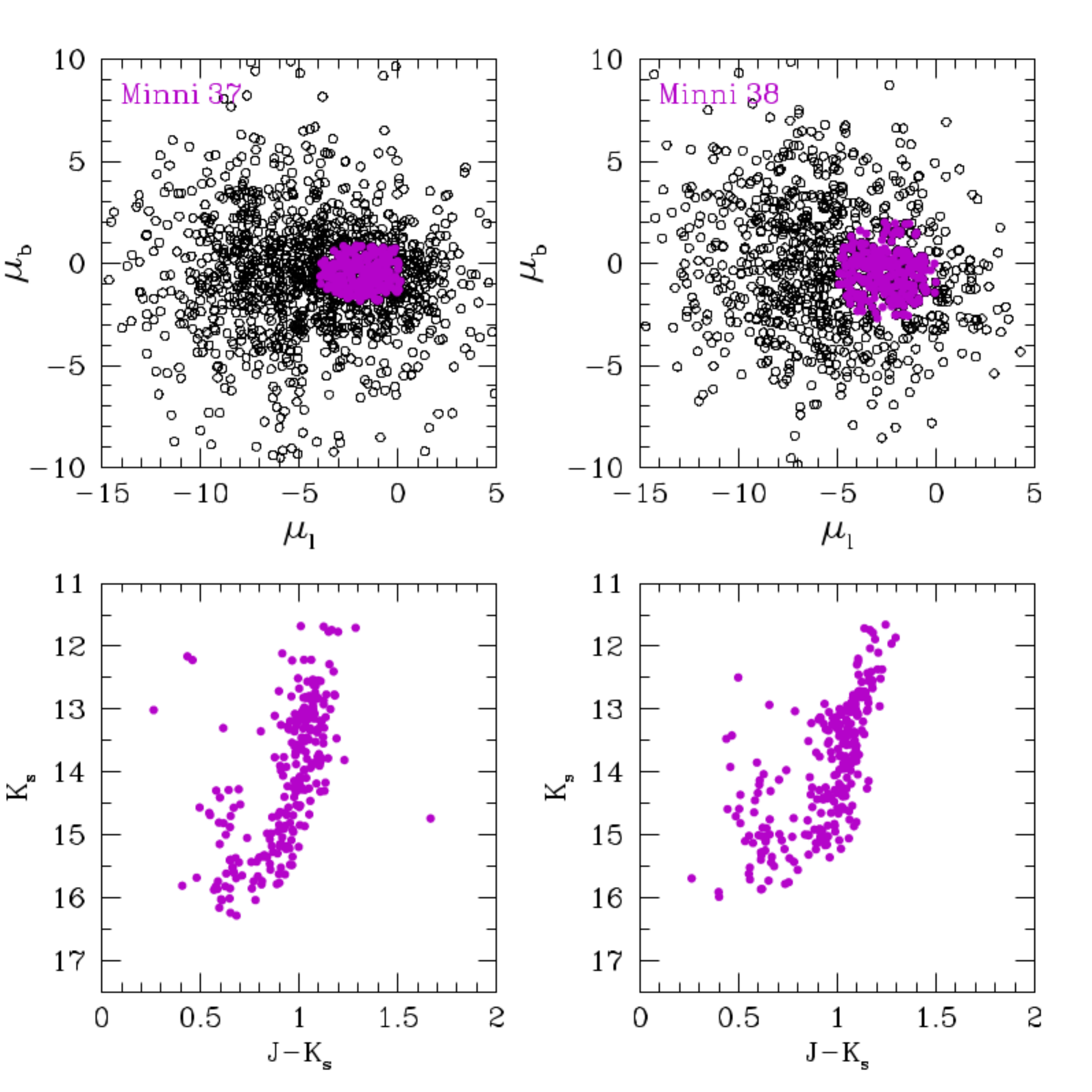}
\includegraphics[scale=.285]{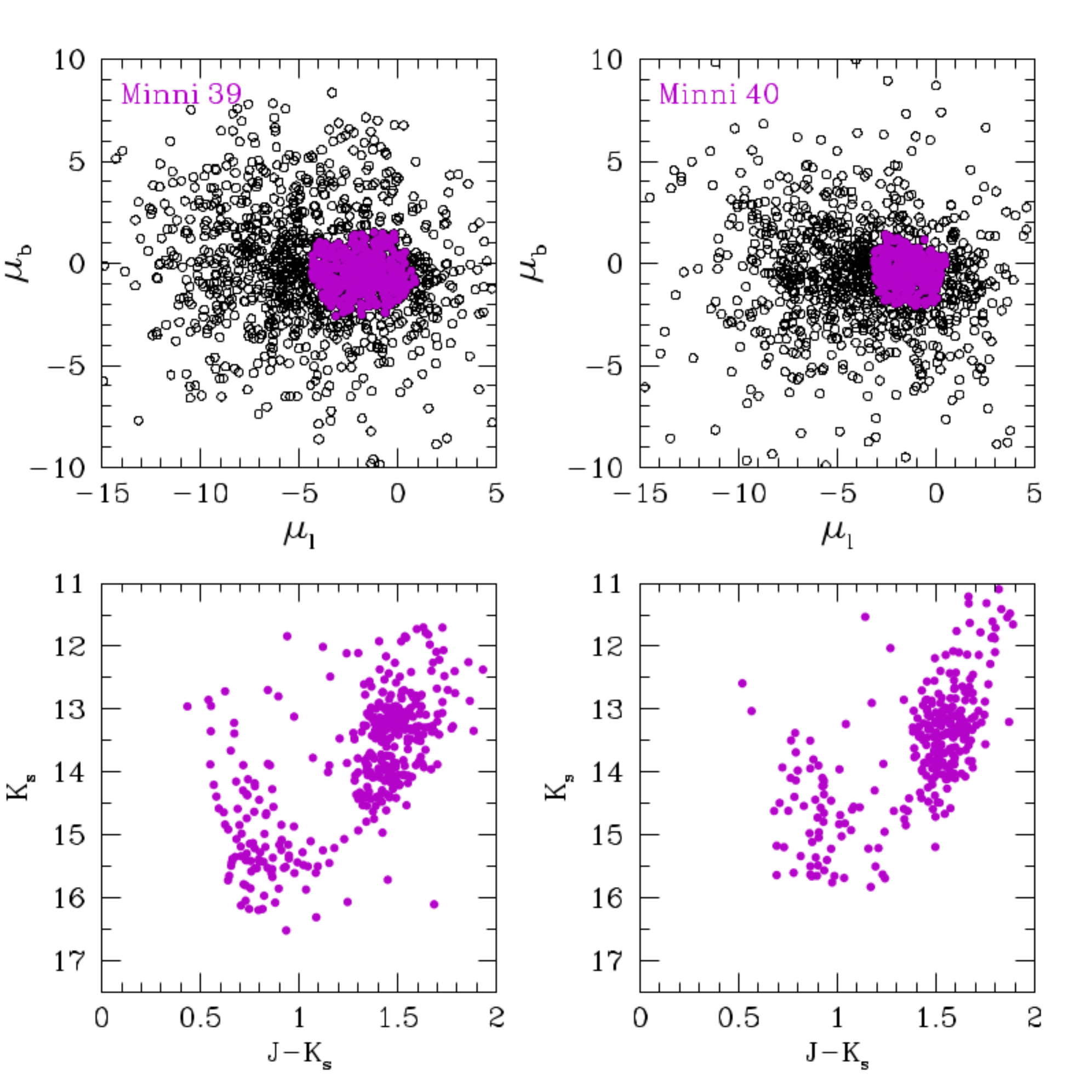}
\includegraphics[scale=.285]{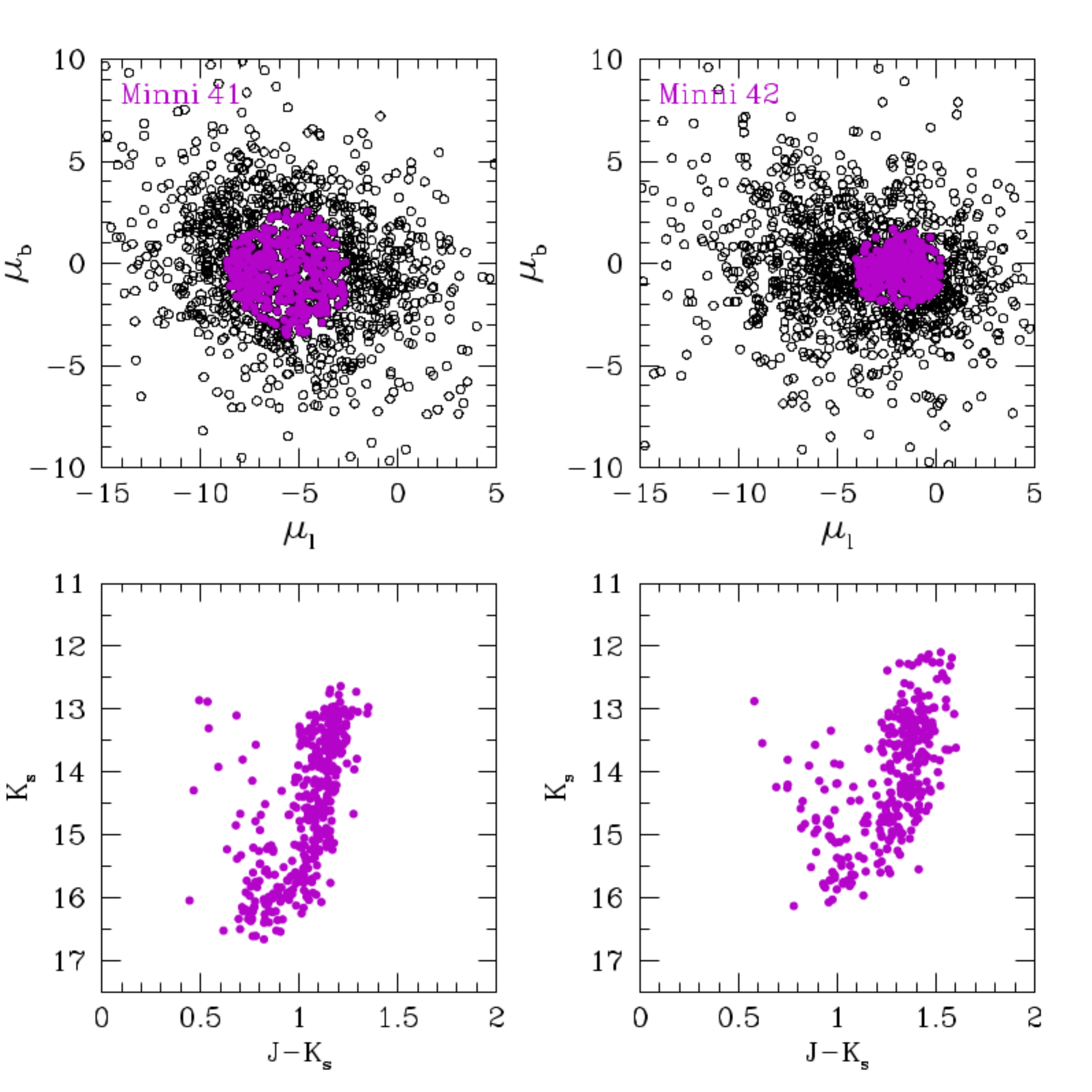}
\includegraphics[scale=.285]{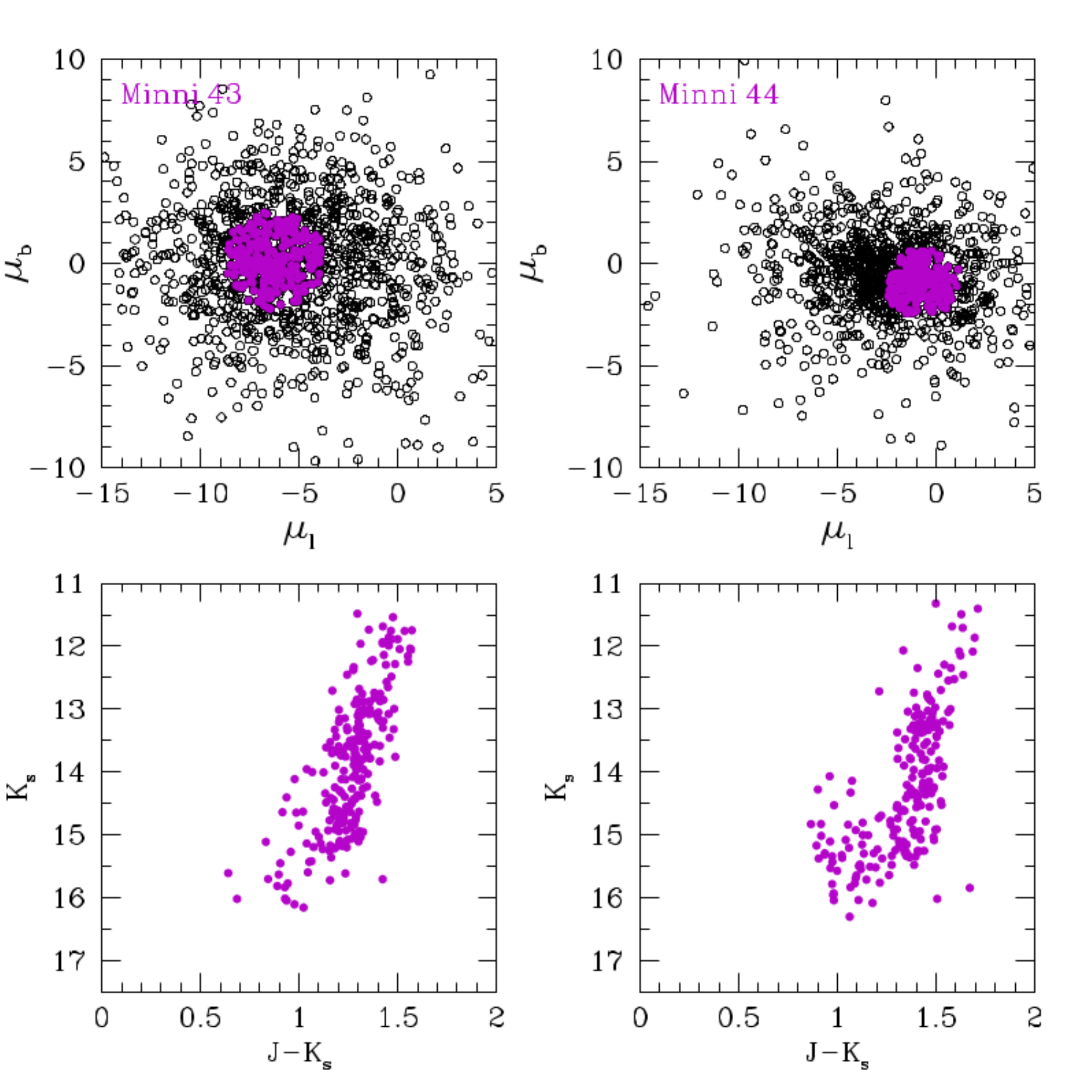}
\caption{Gaia-DR2 proper motion selections for each cluster. Upper panels show the PM distribution of the GC candidate regions (black circles) and the selected overdensities (purple filled circles). Lower panels show the corresponding PM-decontaminated CMDs.
\label{fig6}  }
\end{figure*}

As a second method to decontaminate the candidate CMDs we used PM data based on identifying overdensities in the ($\mu_l,\mu_b$) plane. For that purpose we search for Gaia-DR2 data \citep{gaia16,gaia18} which provide PMs across the whole sky for about 1.3 billion point sources, although is limited in fields with high extinction. The sources from Gaia-DR2 were first matched with the VVV-DR2 data at Vizier/CDS, i.e. the standard CASU aperture data. From the obtained catalogue a second match was done using the PSF data provided by \citet{alonsogarcia18}, in order to clean the data up using the following criteria: the sources should appear in 2 epochs in at least 3 of the 5 filters. PM in RA and DEC were transformed into Galactic coordinates. Thus the sources in the final catalogue are those appearing in Gaia-DR2, VVV-DR2 CASU data and VVV PSF photometry. No constrains were taken for the PMs or Parallaxes.  Figure \ref{fig6} show the results obtained within a radius of 2 arcmin of the observed GC candidate region. Upper panels show the selected overdensity region and its corresponding CMD in the lower panels. Selections in the ($\mu_l,\mu_b$) plane have ranges of $2.8 \leq \mu_l \leq 5.2$ mas/yr and $2.6 \leq \mu_b \leq 4.7$ mas/yr. Some GC candidates show an appropriate CMD in both decontamination methods, such as Minni\,23 and 28. Other candidates exhibit a good decontaminated CMD only by applying one of the two methods, such as Minni\,28, 32 and 41 in the PM selection procedure, and Minni\,30, 33, 34, 37, 38 and 40 in the statistical decontamination one. It looks more certain that the CMDs of the GC candidates Minni\,25, 27, 36, 43 and 44 show CMDs that do not correspond to a GC nature. \\

\section{Discussion } 

It is complicated to confirm the physical nature of small, poorly-populated GCs in the inner bulge regions, so in many cases alternative techniques should be used. In order to refine the physical parameters of these new objects, more powerful instruments to measure radial velocities and chemical composition are needed to follow them up. The majority of the likely good new GC candidates in the Galactic bulge appear to be small and intrinsically faint in comparison with the GCs that were already known \citep[e.g.][]{minniti17c,ryu18,camargo18}. It seems reasonable to speculate that these newly discovered GC candidates may be the debris of larger clusters that have been or are being eroded by the MW tidal field, which make them harder to pick kinematically and difficult our intent on proving or disproving their existence. \\ 

One interesting question is how far into the MW substructure a discrete, extended object like a GC can survive. We can use GCs as probes to explore this question. Tidal limitation is important and becomes dominant as the distance to the Galactic centre decreases. We are examining here the innermost candidates in which crowding is severe so that they are the most difficult to study. According to the catalogue of \cite{harris10}, the closest GCs to the Galactic centre are HP\,1, NGC\,6522 and NGC\,6528 at $R_G=0.5, 0.6$ and $0.64$ kpc, respectively. These GCs are located in the Galactic bulge, at heliocentric distances of $7.7, 7.9,$ and $8.2$ kpc, respectively.  Minni\,40 may be the new record holder for being the closest GC to the Galactic centre, at projected Galactocentric angular distance of $0.5$ deg. 
However, the GC distances have large associated errors, especially in the bulge region. In fact, their distances could be as uncertain as  $0.5$ up to $1.5$ kpc.  Therefore, it is preferable to inspect the projected angular distances. The previous record holders were Ter\,1, Ter\,6 and Pal\,6, located at $2.62, 2.59$ and $2.75$ deg, respectively, from the Galactic centre. Assuming $R_0=8.3$ kpc \citep{dekany13}, these angular distances are equivalent to projected Galactocentric distances of $R_G= 379, 375$ and $398$ pc, respectively. However, these three clusters are located in front of the Galactic bulge, at heliocentric distances of $6.7, 6.8,$ and $5.7$ kpc, respectively. Consequently,  their real Galactocentric distances are $R_G=1.6, 1.5,$ and $2.6$ kpc, respectively, according to the catalogue of \cite{harris10}. 
With a heliocentric distance D=6.1 kpc, Minni\,40 may be very close to being among the top record holders as the closest GC to the Galactic centre, with $R_G=2.2\pm 1.5$ kpc. The extinction values $A_{K_s}=0.46$ and $A_V=3.93$ mag from \cite{gonzalez12} using Nishiyama extinction law \citep{nishiyama09} are not so extreme in the field of this cluster so that its Galactocentric distance should be reliable. Spectroscopic parallax is a possibility for an independent distance estimate to confirm Minni\,40 as the closest GC to the Galactic centre (e.g. Dias et al. in prep.). \\

\begin{table*}
\centering
\caption{New globular cluster candidates discovered using NIR VVV Survey photometry. The table lists the GC identification, equatorial coordinates (J2000), tile ID, numbers of RR\,Lyrae and Mira stars within 5 arcmin of each cluster centre, $K_s$ band absorption, our outcome analysis, PMs in Galactic coordinates, the comparison GC, and estimated heliocentric distances in kpc. \label{tab1}}
\begin{threeparttable}
\begin{tabular}{cccccccccccc}
ID & RA2000 & DEC2000 & tile & NRR & NMira & ${A_{K_s}}^a$ & CMD$_{decont}$ & ($\mu_l,\mu_b$) & CMD$_{PM}$ & Comp. & Dist.$^a$ \\
\hline
Minni23   &   268.5595  &-36.1524  &b319    &4     &2      &0.11    &YES & (-6.1, -0.4)  &YES &N6642 & 8.4 kpc \\ 
Minni24   &   270.4500  &-28.3602  &b293    &13    &4      &0.17    &YES? & (-4.4, -0.5) &NO &N6642 & 7.9 kpc  \\ 
Minni25   &   265.9887  &-33.9416  &b302    &8     &3      &0.32    &NO &  (-6.9, +0.3) &NO &N6642 & ---  \\ 
Minni26   &   266.1200  &-34.8055  &b288    &8     &0      &0.26    &YES? & (-5.7, -0.3) &YES? &N6642 & 7.0 kpc  \\ 
Minni27   &  267.9595   &-33.8665  &b275    &12     &1      &0.29    &NO? & (-6.1, -0.3) &NO? &N6642 & ---  \\ 
Minni28   &   268.1345  &-33.4998  &b275    &10     &2      &0.20    &YES & (-6.0, +0.3) &YES &N6642 & 10.1 kpc  \\ 
Minni29   &   268.0991  &-32.2987  &b290    &6     &4      &0.30    &YES? & (-5.4, +0.2)  &YES? &N6642 & 9.6 kpc  \\ 
Minni30   &   268.5145  &-31.3104  &b291    &14     &7      &0.26    &YES & (-6.0, +0.2)   &YES? &N6642 & 9.9 kpc  \\ 
Minni31   &   269.6533  &-27.6393  &b307    &12    &4      &0.25    &YES? & (-4.1, -0.3) &YES? &N6642 & 9.1 kpc  \\ 
Minni32   &   271.6033  &-29.3081  &b279    &8     &1      &0.14    &NO & (-5.1, -0.2)  &YES &N6642 & ---   \\ 
Minni33   &  267.4658   &-30.7368  &b305    &25     &8      &0.43    &YES & (-3.0, -0.7)  &NO &N6642 & 10.5 kpc  \\ 
Minni34   &  268.5408   &-28.4309  &b320    &24    &0      &0.36    &YES & (-5.1, -0.2)  &NO? &N6642 & 8.8 kpc  \\ 
Minni35   &  268.0333   &-28.4206  &b320    &8     &0      &0.42    &YES? & (-6.4, -0.4) &NO &N6637 & 6.8 kpc   \\ 
Minni36   &  268.9833   &-29.9706  &b305    &9    &10      &0.24    &NO &  (-3.3, -0.4)  &NO? &N6642 & ---   \\ 
Minni37   &  269.0145   &-29.5806  &b306    &19    &8      &0.23   &YES & (-1.6, -0.4)  &YES? &N6642 & 8.8 kpc  \\ 
Minni38   &  268.4354   &-30.0209  &b305    &18     &5      &0.23    &YES & (-2.9, -0.6)  &YES? &N6642 & 8.5 kpc  \\ 
Minni39   &  268.0979   &-29.2945  &b306    &22     &9      &0.42    &YES? & (-1.6, -0.6) &NO &N6642 & 8.8 kpc  \\ 
Minni40   &  267.6770   &-29.6068  &b319    &24     &7      &0.46    &YES? & (-1.9, -0.3)  &NO &N6637 & 6.1 kpc   \\ 
Minni41   &   261.6104  &-28.7406  &b374    &8     &0      &0.28    &YES? & (-6.0, -0.5) &YES &N6642 & 8.8 kpc  \\ 
Minni42   &  264.1562   &-29.0379  &b347    &18     &0      &0.41    &YES? & (-2.1, -0.4) &NO? &N6642 & 10.4 kpc  \\ 
Minni43   &   263.4937  &-27.0084  &b375    &12     &0      &0.39    &NO & (-6.5, +0.5)  &NO &N6642 & ---   \\ 
Minni44   &  265.5895   &-26.5479  &b362    &11     &0      &0.45    &NO &  (-1.4, -0.8)  &NO &N6642 & ---   \\ 
\hline
\end{tabular}
\begin{tablenotes}
\item $^a$Typical distance, extinction and PMs errors are $\sigma_D \sim 1.5$ kpc, $\sigma _{A_{K_s}} \sim 0.10$ mag, $\sigma _{\mu_l} \sim 1.6$ mas/yr, and $\sigma _{\mu_b} \sim 0.8$ mas/yr, respectively. 
\end{tablenotes}
\end{threeparttable}
\end{table*}

\section{ Conclusions} 
We have presented the CMDs for 22 newly discovered GC candidates in the Galactic bulge. 
The deep NIR PSF photometry from the VVV Survey allowed us to differentially decontaminate the NIR CMDs of these GC candidates.  We used Gaia-DR2 PM dataset to search for overdensities in the PM plane in each of the GC candidate regions. Based on these analyses, we could discard in all certainty 5 out of the 22 objects as not real GCs, namely, Minni\,25, 27, 36, 43 and 44. On the other hand, our best candidates turned out to be Minni\,23 and Minni\,28.
We also compare the decontaminated CMDs with those of the well known GCs NGC\,6642 (metal-poor) and NGC\,6637 (metal-rich), in order to estimate their reddenings and distances. Many of the new GC candidates appear to be located within the Galactic bulge, at heliocentric distances $6.1<D<10$ kpc.
We have explored CMDs and LFs of the GC candidates, in order to identify which of them are most likely to be true GCs.
We conclude that it is very difficult to confirm the physical reality of bulge GCs, and in many cases other alternative techniques are needed to corroborate our findings.
There are at least two RR\,Lyrae stars clustering spatially for the GC candidates Minni\,24, Minni\,25, Minni\,28, Minni\,30, Minni\,33, Minni\,34, Minni\,37, Minni\,38, Minni\,39, Minni\,40 and Minni\,43. There appears to be at least two Mira variable stars clustering spatially for the GC candidates Minni\,30, Minni\,33, Minni\,36, Minni\,39 and Minni\,40. In addition, we find the closest GC in angular projection to the Galactic centre, Minni\,40, located at a projected angular distance of only $0.5$ deg from the Galactic centre. Those clusters that have been shown here as likely real, should be followed-up with other techniques in order to secure their classification and refine their physical parameters, as discussed by \cite{minniti18b}. The follow-up could be done with spectroscopy as well as in the future with long baseline VVVX data we might get astrometric signature for some of the clusters. \\

\section*{Acknowledgements}

We gratefully acknowledge data from the ESO Public Survey program ID 179.B-2002 taken with the VISTA telescope, and products from the Cambridge Astronomical Survey Unit (CASU). Support is provided by the BASAL Center for Astrophysics and Associated Technologies (CATA) through grant PFB-06, and the Ministry for the Economy, Development and Tourism, Programa Iniciativa Cienti\'fica Milenio grant IC120009, awarded to the Millennium Institute of Astrophysics (MAS). T.P. and J.J.C. acknowledge support from the Argentinian institutions CONICET and SECYT (Universidad Nacional de C\'ordoba). D.M. acknowledge support from FONDECYT Regular grants No. 1170121. J.A-G. acknowledges support by FONDECYT Iniciacion 11150916, and by the Ministry of Education through grant ANT-1656. R.K.S. acknowledges support from CNPq/Brazil through projects 308968/2016-6 and 421687/2016-9. R.K. acknowledges support from CNPq/Brazil. J.G.F.-T. is supported by FONDECYT No. 3180210. D.M., J.C. and H.N. express their deepest gratitude for the kind hospitality of the Vatican Observatory, where part of this project was carried out. D.G. gratefully acknowledges financial support from the Direcci\'on de Investigaci\'on y Desarrollo de
la Universidad de La Serena through the Programa de Incentivo a la Investigaci\'on de Acad\'emicos (PIA-DIDULS). \\




\bibliographystyle{mnras}
\bibliography{newGC} 

\begin{thebibliography}{}
\makeatletter
\relax
\def\mn@urlcharsother{\let\do\@makeother \do\$\do\&\do\#\do\^\do\_\do\%\do\~}
\def\mn@doi{\begingroup\mn@urlcharsother \@ifnextchar [ {\mn@doi@}
  {\mn@doi@[]}}
\def\mn@doi@[#1]#2{\def\@tempa{#1}\ifx\@tempa\@empty \href
  {http://dx.doi.org/#2} {doi:#2}\else \href {http://dx.doi.org/#2} {#1}\fi
  \endgroup}
\def\mn@eprint#1#2{\mn@eprint@#1:#2::\@nil}
\def\mn@eprint@arXiv#1{\href {http://arxiv.org/abs/#1} {{\tt arXiv:#1}}}
\def\mn@eprint@dblp#1{\href {http://dblp.uni-trier.de/rec/bibtex/#1.xml}
  {dblp:#1}}
\def\mn@eprint@#1:#2:#3:#4\@nil{\def\@tempa {#1}\def\@tempb {#2}\def\@tempc
  {#3}\ifx \@tempc \@empty \let \@tempc \@tempb \let \@tempb \@tempa \fi \ifx
  \@tempb \@empty \def\@tempb {arXiv}\fi \@ifundefined
  {mn@eprint@\@tempb}{\@tempb:\@tempc}{\expandafter \expandafter \csname
  mn@eprint@\@tempb\endcsname \expandafter{\@tempc}}}

\bibitem[\protect\citeauthoryear{{Alonso-Garc{\'{\i}}a}, {Mateo}, {Sen},
  {Banerjee}, {Catelan}, {Minniti}  \& {von Braun}}{{Alonso-Garc{\'{\i}}a}
  et~al.}{2012}]{alonsogarcia12}
{Alonso-Garc{\'{\i}}a} J.,  {Mateo} M.,  {Sen} B.,  {Banerjee} M.,  {Catelan}
  M.,  {Minniti} D.,   {von Braun} K.,  2012, \mn@doi [\aj]
  {10.1088/0004-6256/143/3/70}, \href
  {http://adsabs.harvard.edu/abs/2012AJ....143...70A} {143, 70}

\bibitem[\protect\citeauthoryear{{Alonso-Garc{\'{\i}}a}, {D{\'e}k{\'a}ny},
  {Catelan}, {Contreras Ramos}, {Gran}, {Amigo}, {Leyton}  \&
  {Minniti}}{{Alonso-Garc{\'{\i}}a} et~al.}{2015}]{alonsogarcia15}
{Alonso-Garc{\'{\i}}a} J.,  {D{\'e}k{\'a}ny} I.,  {Catelan} M.,  {Contreras
  Ramos} R.,  {Gran} F.,  {Amigo} P.,  {Leyton} P.,   {Minniti} D.,  2015,
  \mn@doi [\aj] {10.1088/0004-6256/149/3/99}, \href
  {http://adsabs.harvard.edu/abs/2015AJ....149...99A} {149, 99}

\bibitem[\protect\citeauthoryear{{Alonso-Garc{\'{\i}}a}
  et~al.,}{{Alonso-Garc{\'{\i}}a} et~al.}{2018}]{alonsogarcia18}
{Alonso-Garc{\'{\i}}a} J.,  et~al., 2018, \mn@doi [\aap]
  {10.1051/0004-6361/201833432}, \href
  {http://adsabs.harvard.edu/abs/2018A%26A...619A...4A} {619, A4}

\bibitem[\protect\citeauthoryear{{Barb{\'a}}, {Minniti}, {Geisler},
  {Alonso-Garc{\'{\i}}a}, {Hempel}, {Monachesi}, {Arias}  \&
  {G{\'o}mez}}{{Barb{\'a}} et~al.}{2019}]{barba19}
{Barb{\'a}} R.~H.,  {Minniti} D.,  {Geisler} D.,  {Alonso-Garc{\'{\i}}a} J.,
  {Hempel} M.,  {Monachesi} A.,  {Arias} J.~I.,   {G{\'o}mez} F.~A.,  2019,
  \mn@doi [\apjl] {10.3847/2041-8213/aaf811}, \href
  {http://adsabs.harvard.edu/abs/2019ApJ...870L..24B} {870, L24}

\bibitem[\protect\citeauthoryear{{Baumgardt} \& {Makino}}{{Baumgardt} \&
  {Makino}}{2003}]{Baumgardt03}
{Baumgardt} H.,  {Makino} J.,  2003, \mn@doi [\mnras]
  {10.1046/j.1365-8711.2003.06286.x}, \href
  {http://adsabs.harvard.edu/abs/2003MNRAS.340..227B} {340, 227}

\bibitem[\protect\citeauthoryear{{Bica}, {Minniti}, {Bonatto}  \&
  {Hempel}}{{Bica} et~al.}{2018}]{bica18}
{Bica} E.,  {Minniti} D.,  {Bonatto} C.,   {Hempel} M.,  2018, \mn@doi [\pasa]
  {10.1017/pasa.2018.24}, \href
  {http://adsabs.harvard.edu/abs/2018PASA...35...25B} {35, e025}

\bibitem[\protect\citeauthoryear{{Camargo}}{{Camargo}}{2018}]{camargo18}
{Camargo} D.,  2018, \mn@doi [\apjl] {10.3847/2041-8213/aacc68}, \href
  {http://adsabs.harvard.edu/abs/2018ApJ...860L..27C} {860, L27}

\bibitem[\protect\citeauthoryear{{Camargo} \& {Minniti}}{{Camargo} \&
  {Minniti}}{2019}]{camargo19}
{Camargo} D.,  {Minniti} D.,  2019, \mn@doi [\mnras] {10.1093/mnrasl/slz010},
  \href {http://adsabs.harvard.edu/abs/2019MNRAS.484L..90C} {484, L90}

\bibitem[\protect\citeauthoryear{{Catelan}, {Pritzl}  \& {Smith}}{{Catelan}
  et~al.}{2004}]{catelan04}
{Catelan} M.,  {Pritzl} B.~J.,   {Smith} H.~A.,  2004, \mn@doi [\apjs]
  {10.1086/422916}, \href {http://adsabs.harvard.edu/abs/2004ApJS..154..633C}
  {154, 633}

\bibitem[\protect\citeauthoryear{{Contreras Ramos} et~al.,}{{Contreras Ramos}
  et~al.}{2018}]{contreras18}
{Contreras Ramos} R.,  et~al., 2018, \mn@doi [\apj] {10.3847/1538-4357/aacd09},
  \href {http://adsabs.harvard.edu/abs/2018ApJ...863...78C} {863, 78}

\bibitem[\protect\citeauthoryear{{D{\'e}k{\'a}ny}, {Minniti}, {Catelan},
  {Zoccali}, {Saito}, {Hempel}  \& {Gonzalez}}{{D{\'e}k{\'a}ny}
  et~al.}{2013}]{dekany13}
{D{\'e}k{\'a}ny} I.,  {Minniti} D.,  {Catelan} M.,  {Zoccali} M.,  {Saito}
  R.~K.,  {Hempel} M.,   {Gonzalez} O.~A.,  2013, \mn@doi [\apjl]
  {10.1088/2041-8205/776/2/L19}, \href
  {http://adsabs.harvard.edu/abs/2013ApJ...776L..19D} {776, L19}

\bibitem[\protect\citeauthoryear{{Emerson} \& {Sutherland}}{{Emerson} \&
  {Sutherland}}{2010}]{emerson10}
{Emerson} J.,  {Sutherland} W.,  2010, The Messenger, \href
  {http://adsabs.harvard.edu/abs/2010Msngr.139....2E} {139, 2}

\bibitem[\protect\citeauthoryear{{Fall} \& {Rees}}{{Fall} \&
  {Rees}}{1977}]{fall77}
{Fall} S.~M.,  {Rees} M.~J.,  1977, \mn@doi [\mnras] {10.1093/mnras/181.1.37P},
  \href {http://adsabs.harvard.edu/abs/1977MNRAS.181P..37F} {181, 37P}

\bibitem[\protect\citeauthoryear{{Fall} \& {Rees}}{{Fall} \&
  {Rees}}{1985}]{fall85}
{Fall} S.~M.,  {Rees} M.~J.,  1985, \mn@doi [\apj] {10.1086/163585}, \href
  {http://adsabs.harvard.edu/abs/1985ApJ...298...18F} {298, 18}

\bibitem[\protect\citeauthoryear{{Fern{\'a}ndez-Trincado}
  et~al.,}{{Fern{\'a}ndez-Trincado} et~al.}{2017}]{trincado17}
{Fern{\'a}ndez-Trincado} J.~G.,  et~al., 2017, \mn@doi [\apjl]
  {10.3847/2041-8213/aa8032}, \href
  {http://adsabs.harvard.edu/abs/2017ApJ...846L...2F} {846, L2}

\bibitem[\protect\citeauthoryear{{Fern{\'a}ndez-Trincado}, {Beers}, {Tang},
  {Moreno}, {P{\'e}rez-Villegas}  \&
  {Ortigoza-Urdaneta}}{{Fern{\'a}ndez-Trincado} et~al.}{2019b}]{trincado19b}
{Fern{\'a}ndez-Trincado} J.~G.,  {Beers} T.~C.,  {Tang} B.,  {Moreno} E.,
  {P{\'e}rez-Villegas} A.,   {Ortigoza-Urdaneta} M.,  2019b, preprint
  (arXiv:1904.05369), \href {http://adsabs.harvard.edu/abs/2019arXiv190405369F}
  {}

\bibitem[\protect\citeauthoryear{{Fern{\'a}ndez-Trincado}
  et~al.,}{{Fern{\'a}ndez-Trincado} et~al.}{2019a}]{trincado19a}
{Fern{\'a}ndez-Trincado} J.~G.,  et~al., 2019a, preprint (arXiv:1902.10635),
  \href {http://adsabs.harvard.edu/abs/2019arXiv190210635F} {}

\bibitem[\protect\citeauthoryear{{Fern{\'a}ndez-Trincado}
  et~al.,}{{Fern{\'a}ndez-Trincado} et~al.}{2019c}]{trincado19c}
{Fern{\'a}ndez-Trincado} J.~G.,  et~al., 2019c, preprint (arXiv:1904.05884),
  \href {http://adsabs.harvard.edu/abs/2019arXiv190405884F} {}

\bibitem[\protect\citeauthoryear{{Froebrich}, {Meusinger}  \&
  {Scholz}}{{Froebrich} et~al.}{2007}]{froebrich07}
{Froebrich} D.,  {Meusinger} H.,   {Scholz} A.,  2007, \mn@doi [\mnras]
  {10.1111/j.1745-3933.2007.00302.x}, \href
  {http://adsabs.harvard.edu/abs/2007MNRAS.377L..54F} {377, L54}

\bibitem[\protect\citeauthoryear{{Gaia Collaboration} et~al.,}{{Gaia
  Collaboration} et~al.}{2016}]{gaia16}
{Gaia Collaboration} et~al., 2016, \mn@doi [\aap]
  {10.1051/0004-6361/201629272}, \href
  {http://adsabs.harvard.edu/abs/2016A%26A...595A...1G} {595, A1}

\bibitem[\protect\citeauthoryear{{Gaia Collaboration} et~al.,}{{Gaia
  Collaboration} et~al.}{2018}]{gaia18}
{Gaia Collaboration} et~al., 2018, \mn@doi [\aap]
  {10.1051/0004-6361/201833051}, \href
  {http://adsabs.harvard.edu/abs/2018A%26A...616A...1G} {616, A1}

\bibitem[\protect\citeauthoryear{{Gonzalez}, {Rejkuba}, {Zoccali}, {Valenti},
  {Minniti}, {Schultheis}, {Tobar}  \& {Chen}}{{Gonzalez}
  et~al.}{2012}]{gonzalez12}
{Gonzalez} O.~A.,  {Rejkuba} M.,  {Zoccali} M.,  {Valenti} E.,  {Minniti} D.,
  {Schultheis} M.,  {Tobar} R.,   {Chen} B.,  2012, \mn@doi [\aap]
  {10.1051/0004-6361/201219222}, \href
  {http://adsabs.harvard.edu/abs/2012A%26A...543A..13G} {543, A13}

\bibitem[\protect\citeauthoryear{{Harris}}{{Harris}}{2010}]{harris10}
{Harris} W.~E.,  2010, preprint (arXiv: 1012.3224), \href
  {http://adsabs.harvard.edu/abs/2010arXiv1012.3224H} {}

\bibitem[\protect\citeauthoryear{{Kim}, {Jerjen}, {Mackey}, {Da Costa}  \&
  {Milone}}{{Kim} et~al.}{2016}]{kim16}
{Kim} D.,  {Jerjen} H.,  {Mackey} D.,  {Da Costa} G.~S.,   {Milone} A.~P.,
  2016, \mn@doi [\apj] {10.3847/0004-637X/820/2/119}, \href
  {http://adsabs.harvard.edu/abs/2016ApJ...820..119K} {820, 119}

\bibitem[\protect\citeauthoryear{{Koposov}, {Belokurov}  \&
  {Torrealba}}{{Koposov} et~al.}{2017}]{koposov17}
{Koposov} S.~E.,  {Belokurov} V.,   {Torrealba} G.,  2017, \mn@doi [\mnras]
  {10.1093/mnras/stx1182}, \href
  {http://adsabs.harvard.edu/abs/2017MNRAS.470.2702K} {470, 2702}

\bibitem[\protect\citeauthoryear{{Laevens} et~al.,}{{Laevens}
  et~al.}{2015}]{laevens15}
{Laevens} B.~P.~M.,  et~al., 2015, \mn@doi [\apj] {10.1088/0004-637X/813/1/44},
  \href {http://adsabs.harvard.edu/abs/2015ApJ...813...44L} {813, 44}

\bibitem[\protect\citeauthoryear{{Lamers}, {Gieles}  \& {Portegies
  Zwart}}{{Lamers} et~al.}{2005}]{lamers05}
{Lamers} H.~J.~G.~L.~M.,  {Gieles} M.,   {Portegies Zwart} S.~F.,  2005,
  \mn@doi [\aap] {10.1051/0004-6361:20041476}, \href
  {http://adsabs.harvard.edu/abs/2005A%26A...429..173L} {429, 173}

\bibitem[\protect\citeauthoryear{{Luque} et~al.,}{{Luque}
  et~al.}{2016}]{luque16}
{Luque} E.,  et~al., 2016, \mn@doi [\mnras] {10.1093/mnras/stw302}, \href
  {http://adsabs.harvard.edu/abs/2016MNRAS.458..603L} {458, 603}

\bibitem[\protect\citeauthoryear{{Majewski} et~al.,}{{Majewski}
  et~al.}{2017}]{majewski17}
{Majewski} S.~R.,  et~al., 2017, \mn@doi [\aj] {10.3847/1538-3881/aa784d},
  \href {http://adsabs.harvard.edu/abs/2017AJ....154...94M} {154, 94}

\bibitem[\protect\citeauthoryear{{Minniti}}{{Minniti}}{1995}]{minniti95}
{Minniti} D.,  1995, \mn@doi [\aj] {10.1086/117393}, \href
  {http://adsabs.harvard.edu/abs/1995AJ....109.1663M} {109, 1663}

\bibitem[\protect\citeauthoryear{{Minniti} et~al.,}{{Minniti}
  et~al.}{2010}]{minniti10}
{Minniti} D.,  et~al., 2010, \mn@doi [\na] {10.1016/j.newast.2009.12.002},
  \href {http://adsabs.harvard.edu/abs/2010NewA...15..433M} {15, 433}

\bibitem[\protect\citeauthoryear{{Minniti} et~al.,}{{Minniti}
  et~al.}{2011}]{minniti11}
{Minniti} D.,  et~al., 2011, \mn@doi [\aap] {10.1051/0004-6361/201015795},
  \href {http://adsabs.harvard.edu/abs/2011A%26A...527A..81M} {527, A81}

\bibitem[\protect\citeauthoryear{{Minniti}, {Alonso-Garc{\'{\i}}a}, {Braga},
  {Contreras Ramos}, {Hempel}, {Palma}, {Pullen}  \& {Saito}}{{Minniti}
  et~al.}{2017a}]{minniti17a}
{Minniti} D.,  {Alonso-Garc{\'{\i}}a} J.,  {Braga} V.,  {Contreras Ramos} R.,
  {Hempel} M.,  {Palma} T.,  {Pullen} J.,   {Saito} R.~K.,  2017a, \mn@doi
  [Research Notes of the American Astronomical Society]
  {10.3847/2515-5172/aa9ab7}, \href
  {http://adsabs.harvard.edu/abs/2017RNAAS...1a..16M} {1, 16}

\bibitem[\protect\citeauthoryear{{Minniti} et~al.,}{{Minniti}
  et~al.}{2017b}]{minniti17b}
{Minniti} D.,  et~al., 2017b, \mn@doi [\apjl] {10.3847/2041-8213/838/1/L14},
  \href {http://adsabs.harvard.edu/abs/2017ApJ...838L..14M} {838, L14}

\bibitem[\protect\citeauthoryear{{Minniti} et~al.,}{{Minniti}
  et~al.}{2017c}]{minniti17c}
{Minniti} D.,  et~al., 2017c, \mn@doi [\apjl] {10.3847/2041-8213/aa95b8}, \href
  {http://adsabs.harvard.edu/abs/2017ApJ...849L..24M} {849, L24}

\bibitem[\protect\citeauthoryear{{Minniti} et~al.,}{{Minniti}
  et~al.}{2018a}]{minniti18b}
{Minniti} D.,  et~al., 2018a, \mn@doi [\apj] {10.3847/1538-4357/aadd06}, \href
  {http://adsabs.harvard.edu/abs/2018ApJ...866...12M} {866, 12}

\bibitem[\protect\citeauthoryear{{Minniti}, {Fern{\'a}ndez-Trincado}, {Ripepi},
  {Alonso-Garc{\'{\i}}a}, {Contreras Ramos}  \& {Marconi}}{{Minniti}
  et~al.}{2018b}]{minniti18a}
{Minniti} D.,  {Fern{\'a}ndez-Trincado} J.~G.,  {Ripepi} V.,
  {Alonso-Garc{\'{\i}}a} J.,  {Contreras Ramos} R.,   {Marconi} M.,  2018b,
  \mn@doi [\apjl] {10.3847/2041-8213/aaf1cd}, \href
  {http://adsabs.harvard.edu/abs/2018ApJ...869L..10M} {869, L10}

\bibitem[\protect\citeauthoryear{{Moni Bidin} et~al.,}{{Moni Bidin}
  et~al.}{2011}]{monibidin11}
{Moni Bidin} C.,  et~al., 2011, \mn@doi [\aap] {10.1051/0004-6361/201117488},
  \href {http://adsabs.harvard.edu/abs/2011A%26A...535A..33M} {535, A33}

\bibitem[\protect\citeauthoryear{{Mutlu-Pakdil} et~al.,}{{Mutlu-Pakdil}
  et~al.}{2018}]{mutlu18}
{Mutlu-Pakdil} B.,  et~al., 2018, \mn@doi [\apj] {10.3847/1538-4357/aacd0e},
  \href {http://adsabs.harvard.edu/abs/2018ApJ...863...25M} {863, 25}

\bibitem[\protect\citeauthoryear{{Nataf}, {Gould}, {Pinsonneault}  \&
  {Udalski}}{{Nataf} et~al.}{2013}]{nataf13}
{Nataf} D.~M.,  {Gould} A.~P.,  {Pinsonneault} M.~H.,   {Udalski} A.,  2013,
  \mn@doi [\apj] {10.1088/0004-637X/766/2/77}, \href
  {http://adsabs.harvard.edu/abs/2013ApJ...766...77N} {766, 77}

\bibitem[\protect\citeauthoryear{{Nishiyama}, {Tamura}, {Hatano}, {Kato},
  {Tanab{\'e}}, {Sugitani}  \& {Nagata}}{{Nishiyama}
  et~al.}{2009}]{nishiyama09}
{Nishiyama} S.,  {Tamura} M.,  {Hatano} H.,  {Kato} D.,  {Tanab{\'e}} T.,
  {Sugitani} K.,   {Nagata} T.,  2009, \mn@doi [\apj]
  {10.1088/0004-637X/696/2/1407}, \href
  {http://adsabs.harvard.edu/abs/2009ApJ...696.1407N} {696, 1407}

\bibitem[\protect\citeauthoryear{{Palma}, {Minniti}, {D{\'e}k{\'a}ny},
  {Clari{\'a}}, {Alonso-Garc{\'{\i}}a}, {Gramajo}, {Ram{\'{\i}}rez
  Alegr{\'{\i}}a}  \& {Bonatto}}{{Palma} et~al.}{2016}]{palma16}
{Palma} T.,  {Minniti} D.,  {D{\'e}k{\'a}ny} I.,  {Clari{\'a}} J.~J.,
  {Alonso-Garc{\'{\i}}a} J.,  {Gramajo} L.~V.,  {Ram{\'{\i}}rez Alegr{\'{\i}}a}
  S.,   {Bonatto} C.,  2016, \mn@doi [\na] {10.1016/j.newast.2016.05.008},
  \href {http://adsabs.harvard.edu/abs/2016NewA...49...50P} {49, 50}

\bibitem[\protect\citeauthoryear{{Pietrukowicz} et~al.,}{{Pietrukowicz}
  et~al.}{2015}]{pietru15}
{Pietrukowicz} P.,  et~al., 2015, \mn@doi [\apj] {10.1088/0004-637X/811/2/113},
  \href {http://adsabs.harvard.edu/abs/2015ApJ...811..113P} {811, 113}

\bibitem[\protect\citeauthoryear{{Recio-Blanco} et~al.,}{{Recio-Blanco}
  et~al.}{2017}]{recio17}
{Recio-Blanco} A.,  et~al., 2017, \mn@doi [\aap] {10.1051/0004-6361/201630220},
  \href {http://adsabs.harvard.edu/abs/2017A%26A...602L..14R} {602, L14}

\bibitem[\protect\citeauthoryear{{Rossi} \& {Hurley}}{{Rossi} \&
  {Hurley}}{2015}]{rossi15}
{Rossi} L.~J.,  {Hurley} J.~R.,  2015, \mn@doi [\mnras]
  {10.1093/mnras/stu2015}, \href
  {http://adsabs.harvard.edu/abs/2015MNRAS.446.3389R} {446, 3389}

\bibitem[\protect\citeauthoryear{{Ryu} \& {Lee}}{{Ryu} \& {Lee}}{2018}]{ryu18}
{Ryu} J.,  {Lee} M.~G.,  2018, \mn@doi [\apjl] {10.3847/2041-8213/aad8b7},
  \href {http://adsabs.harvard.edu/abs/2018ApJ...863L..38R} {863, L38}

\bibitem[\protect\citeauthoryear{{Saito} et~al.,}{{Saito}
  et~al.}{2012}]{saito12}
{Saito} R.~K.,  et~al., 2012, \mn@doi [\aap] {10.1051/0004-6361/201118407},
  \href {http://adsabs.harvard.edu/abs/2012A%26A...537A.107S} {537, A107}

\bibitem[\protect\citeauthoryear{{Schechter}, {Mateo}  \& {Saha}}{{Schechter}
  et~al.}{1993}]{schechter93}
{Schechter} P.~L.,  {Mateo} M.,   {Saha} A.,  1993, \mn@doi [\pasp]
  {10.1086/133316}, \href {http://adsabs.harvard.edu/abs/1993PASP..105.1342S}
  {105, 1342}

\bibitem[\protect\citeauthoryear{{Schiavon} et~al.,}{{Schiavon}
  et~al.}{2017}]{schiavon17}
{Schiavon} R.~P.,  et~al., 2017, \mn@doi [\mnras] {10.1093/mnras/stw2162},
  \href {http://adsabs.harvard.edu/abs/2017MNRAS.465..501S} {465, 501}

\bibitem[\protect\citeauthoryear{{Skrutskie} et~al.,}{{Skrutskie}
  et~al.}{2006}]{skrutskie06}
{Skrutskie} M.~F.,  et~al., 2006, \mn@doi [\aj] {10.1086/498708}, \href
  {http://adsabs.harvard.edu/abs/2006AJ....131.1163S} {131, 1163}

\bibitem[\protect\citeauthoryear{{Smolec}}{{Smolec}}{2005}]{smolec05}
{Smolec} R.,  2005, \actaa, \href
  {http://adsabs.harvard.edu/abs/2005AcA....55...59S} {55, 59}

\bibitem[\protect\citeauthoryear{{Soszy{\'n}ski} et~al.,}{{Soszy{\'n}ski}
  et~al.}{2011}]{soszynski11}
{Soszy{\'n}ski} I.,  et~al., 2011, \actaa, \href
  {http://adsabs.harvard.edu/abs/2011AcA....61....1S} {61, 1}

\bibitem[\protect\citeauthoryear{{Soszy{\'n}ski} et~al.,}{{Soszy{\'n}ski}
  et~al.}{2013}]{soszynski13}
{Soszy{\'n}ski} I.,  et~al., 2013, \actaa, \href
  {http://adsabs.harvard.edu/abs/2013AcA....63...21S} {63, 21}

\bibitem[\protect\citeauthoryear{{Trenti}, {Vesperini}  \& {Pasquato}}{{Trenti}
  et~al.}{2010}]{trenti10}
{Trenti} M.,  {Vesperini} E.,   {Pasquato} M.,  2010, \mn@doi [\apj]
  {10.1088/0004-637X/708/2/1598}, \href
  {http://adsabs.harvard.edu/abs/2010ApJ...708.1598T} {708, 1598}

\bibitem[\protect\citeauthoryear{{Valenti}, {Ferraro}  \& {Origlia}}{{Valenti}
  et~al.}{2007}]{valenti07}
{Valenti} E.,  {Ferraro} F.~R.,   {Origlia} L.,  2007, \mn@doi [\aj]
  {10.1086/511271}, \href {http://adsabs.harvard.edu/abs/2007AJ....133.1287V}
  {133, 1287}

\bibitem[\protect\citeauthoryear{{Wang} et~al.,}{{Wang} et~al.}{2016}]{wang16}
{Wang} L.,  et~al., 2016, \mn@doi [\mnras] {10.1093/mnras/stw274}, \href
  {http://adsabs.harvard.edu/abs/2016MNRAS.458.1450W} {458, 1450}

\bibitem[\protect\citeauthoryear{{Whitelock}, {Feast}  \& {Van
  Leeuwen}}{{Whitelock} et~al.}{2008}]{whitelock08}
{Whitelock} P.~A.,  {Feast} M.~W.,   {Van Leeuwen} F.,  2008, \mn@doi [\mnras]
  {10.1111/j.1365-2966.2008.13032.x}, \href
  {http://adsabs.harvard.edu/abs/2008MNRAS.386..313W} {386, 313}

\bibitem[\protect\citeauthoryear{{Zinn}}{{Zinn}}{1985}]{zinn85}
{Zinn} R.,  1985, \mn@doi [\apj] {10.1086/163249}, \href
  {http://adsabs.harvard.edu/abs/1985ApJ...293..424Z} {293, 424}

\makeatother
\end{thebibliography}








\bsp	
\label{lastpage}
\end{document}